\documentclass [a4paper, 11pt] {article}
\usepackage{amsmath}
\usepackage{amsfonts}
\usepackage{latexsym}
\usepackage{graphicx}
\usepackage[toc,page]{appendix}
\usepackage{multicol}
\usepackage{color}

\usepackage{amssymb}
\usepackage{mathtools}
\usepackage{empheq}
\usepackage{caption}
\usepackage{subcaption}
\usepackage{breqn} % For the dmath environment (multiline equations)
\usepackage{bm} % bold math

\begin{document}

\title{\textbf{The effect of magnetic twist on resonant absorption of slow sausage waves in magnetic flux tubes}}

\author{Mohammad Sadeghi$^{1}$\thanks{m.sadeghi@uok.ac.ir} ,
Kayoomars Karami$^{1,2}$\thanks{kkarami@uok.ac.ir}\\
\small{$^{1}$Department of Physics, University of Kurdistan, Pasdaran Street, P.O. Box 66177-15175, Sanandaj, Iran}\\
\small{$^{2}$Research Institute for Astronomy and Astrophysics of Maragha (RIAAM), P.O. Box 55134-441, Maragha, Iran}}

\maketitle

\begin{abstract}
Observations show that twisted magnetic flux tubes are present throughout the sun's atmosphere.
The main aim of this work is to obtain the damping rate of sausage modes in the presence of magnetic twist.
Using the connection formulae obtained by Sakurai et al. (1991), we investigate resonant absorption of the sausage modes in the slow continuum under photosphere conditions. We derive the dispersion relation and solve it numerically and consequently obtain the frequencies and damping rates of the slow surface sausage modes. We conclude that the magnetic twist can result in strong damping in comparison with the untwisted case.
\end{abstract}

%\noindent{\textit{Key words:} magnetohydrodynamics (MHD) --- Sun:
%corona --- Sun: magnetic topology --- Sun: oscillations}

%\noindent{\textit{Key words:} Sun: corona -- Sun: magnetic fields -- Sun: oscillations.}

%----------------------------------------------------------------------------------------------------
\section{Introduction}\label{secint}
One of important questions regarding the sun is that how the solar corona reaches temperatures exceeding 1 MK. There are different theories to justify this problem. One of these is propagation of magnetohydrodynamic (MHD) waves and their damping by the resonant absorbtion proposed for the first time by Ionson \cite{Ionson1978}. At the observational point of view, there are some evidences for propagating and damping of MHD waves (see e.g. Yu et al. \cite{TYu1}). Based on propagating speed of MHD waves, they are classified into fast, slow and Alfv\`{e}n waves. Here, we focus on the slow sausage MHD waves which has recently been observed by Dunn Solar Telescope \cite{grant2015}.

Dorotovi\^{c} et al. \cite{Dorotovic} observed the linear slow sausage waves in magnetic pore with the period from 20 minutes to 70 minutes in a
photospheric pore size by the Swedish Solar Telescope. Also, Morton et al. \cite{Morton} observed sausage modes in magnetic pores with the
period range from as short as 30 s up to 450 s. Grant et al. \cite{grant2015} reported the observational period of sausage mode as 181 s to 412 s. Moreover,
Keys et al. \cite{Keys} observationally found that in solar magnetic pores the number of surface modes  are more than body modes. They also pointed out that surface modes appear to carry more energy compared to body modes. The slow waves are usually generated by the motions of sunspots, magnetic pores, and granules and they play significant role in heating of the lower part of the sun's atmosphere \cite{TYu1}. Yu et al. (2017) showed that the resonant absorption plays an important role in the wave damping for the slow surface sausage modes in the slow continuum \cite{TYu}. \\

Besides, there is ample evidence on existence of magnetic twist in the solar atmosphere and below. For instance, it has been
suggested that magnetic flux tubes are twisted while rising
through the convection zone (e.g., Murray \& Hood  \cite{{Murray}};
Hood et al.  \cite{{Hood}}; Luoni et al. \cite{{Luoni}}). Also, the observations show that magnetic twist
is one of important mechanisms in energy transport from the
photosphere to the corona, see Wedemeyer-Böhm et al. \cite{{Wedemeyer}}.

So far, many studies have been done about the effect of twisted magnetic fields on the kink and sausage MHD waves in magnetic flux tubes. For instance,
%\textbf{Shafranov \cite{Shafranov} investigated the stability of magnetic flux tubes with the azimuthal
%component of magnetic field proportional to $r$ inside the cylinder and no magnetic twist outside. Shafranov \cite{Shafranov} concluded that if the magnetic
%twist value in the loop, which is defined as $\phi_{\rm twist} ≡
%(L/R)(B_{\phi} /B_z) = 2\pi N_{\rm twist}$ exceeds a critical value $\phi_c$, then the loop
%becomes kink unstable.}
the connection formulae (or conservation laws) were first used for studying surface waves in cylindrical plasmas in the
presence of magnetic twist by Goossens et al. \cite{Goossens}. Bennett et al. \cite{Bennett} studied the sausage modes in a magnetic flux tube containing a uniform magnetic twist. They pointed out that in the presence of magnetic twist, an infinite set of body modes is generated.

Erd\'{e}lyi \& Carter \cite{Erdelyi3} considered magnetic twist just in the annulus for surface
and hybrid modes. They found that when the magnetic twist increases the hybrid modes include a wide range of phase speeds for the sausage modes.

Erd\'{e}lyi \& Fedun \cite{Erdelyi1} investigated the propagation of MHD waves in an incompressible twisted magnetic flux tube. They showed that an increase in the twisted magnetic field from 0 to 0.3 could lead to an increase of 1 to 2 percent of the period of the sausage waves. Erd\'{e}lyi \& Fedun \cite{Erdelyi} extended their previous work \cite{Erdelyi1} to the compressibility condition and concluded that the period of the sausage waves increases 3 to 5 percent.

Karami \& Bahari \cite{Karami} considered the effect of twisted magnetic field on the resonant
absorption of MHD waves in coronal loops. They showed that the period ratio $P_1/P_2$ of the fundamental
and its first-overtone surface waves for kink ($m = 1$) and fluting ($m = 2, 3$) modes is lower
than 2 in the presence of twisted magnetic field. Ebrahimi \& Karami \cite{Ebrahimi} investigated resonant absorption of kink MHD waves by a magnetic
twist in coronal loops. They concluded that the resonant absorption by the magnetic twist can justify the rapid
damping of kink MHD waves observed in coronal loops.

Giagkiozis et al. \cite{Giagkiozis2016} studied resonant absorption of axisymmetric modes ($m=0$) in twisted magnetic flux tube. They showed that in the presence of twisted magnetic field,
both the longitudinal magnetic field and the density have crucial roles in the wave damping. Giagkiozis et al. \cite{Giagkiozis2015} also elaborated that the magnetic twist can remove the cut-off of fast body sausage modes. Besides the works mentioned above, there are some further studies on the effect of magnetic twist on the MHD waves in the literature, see e.g. Erd\'{e}lyi \& Fedun \cite{Erdelyi2}; Carter \& Erd\'{e}lyi \cite{Carter}, \cite{Carter1}; Ruderman
\cite{Ruderman},\cite{Ruderman1}; Karami \& Barin \cite{Karami2}; Terradas \& Goossens \cite{Terradas}; Karami \& Bahari \cite{Karami1}.

In the present work, our main goal is to study the effect of magnetic twist on resonant absorption of the slow surface sausage modes in the sun's atmosphere. To do so, we apply the magnetic twist to the model of Yu et al. \cite{TYu}. To achieve this aim, in Section 2 we introduce the model and solve the equations of motion governing the slow surface sausage modes. In Section 3, we obtain the
dispersion relation under magnetic pore conditions. In Section 4, using the connection formulae, we
derive the damping rate for the slow surface waves. The numerical results are shown in Section 5. Finally, we conclude the paper in Section 6.

%----------------------------------------------------------------------------------------------------
\section{Equations of motion and model}\label{sec1}

The linearized ideal MHD equations are as follows \cite{kadomtsev1966hydromagnetic}
 \begin{dgroup}
  \begin{dmath}\label{eqn:linearized:mhd1}
\rho \frac{\partial^{2}\bm{\xi}}{\partial t^{2}}=-\nabla\delta p-\frac{1}{\mu_{0}}\Big(\delta \bm{B} \times (\nabla\times \bm{B})+\bm{B}\times (\nabla\times\delta \bm{B})\Big),
  \end{dmath}
\begin{dmath}\label{eqn:linearized:mhd2}
\delta p =- \bm{\xi} \cdot \nabla p-\gamma p \nabla \cdot \bm{\xi},
\end{dmath}
\begin{dmath}\label{eqn:linearized:mhd3}
\delta \bm{B}=-\nabla \times (\bm{B}\times\bm{\xi}) ,
\end{dmath}
\end{dgroup}
where $\rho, p$ and $\bm{B}$ are the background density, kinetic pressure and magnetic field, respectively. Also $\bm{\xi}$ is the Lagrangian displacement vector, $\delta p$
and $\delta \bm{B}$ are the Eulerian perturbations of the pressure and magnetic field, respectively. Here, $\gamma$ is the ratio of specific heats (taken to be $5/3$ in this work), and $\mu_0$ is the permeability of free space.

In our model, we consider a magnetic field as follows
\begin{dmath}
  \bm{B} = \Big(0,B_{\phi}(r),B_{z}(r)\Big).
\end{dmath}
Now, the plasma pressure and magnetic field must satisfy the following magnetohydrostatic equation in the $r$-direction as
\begin{dmath}\label{eqn:pressure:r}
 \frac{{\rm d}}{{\rm d}r}\left(p + \frac{B_{\phi}^2 + B_{z}^2}{2 \mu_0}\right) + \frac{B_{\phi}^2}{\mu_0 r} = 0.
\end{dmath}
Here, following Yu et al. \cite{TYu} we consider the following profiles for the density and $z$-component of the background magnetic field as
\begin{eqnarray}\label{rho}
\rho(r)=\left\{\begin{array}{lll}
    \rho_{{\rm i}}, &r\leqslant r_i,&\\
     \rho_i+(\rho_e-\rho_i)\left(\frac{r-r_i}{r_e-r_i}\right), &r_i< r< r_e,&\\
    \rho_{{\rm e}}, &r\geqslant r_e,
      \end{array}\right.
\end{eqnarray}
\begin{eqnarray}\label{Bz}
B^2_z(r)=\left\{\begin{array}{lll}
    B^2_{{z\rm i}}, &r\leqslant r_i,&\\
     B^2_{{z\rm i}}+\left(B^2_{{z\rm e}}-B^2_{{z\rm i}}\right)\left(\frac{r-r_i}{r_e-r_i}\right), &r_i< r< r_e,&\\
    B^2_{{z\rm e}}, &r\geqslant r_e,
      \end{array}\right.
\end{eqnarray}
where $\rho_i$ and $\rho_e$ are the constant densities of the interior and exterior
regions of the flux tube, respectively. Also $B_{z\rm i}$ and $B_{z\rm e}$ are the interior and exterior constant longitudinal magnetic fields, respectively. We further assume the $\phi$-component of the background magnetic field takes the form
\begin{eqnarray}\label{Bphi}
B_{\phi}^2(r)=\left\{\begin{array}{lll}
  \frac{B^2_{{\phi\rm i}}r^2}{r^2_i}, &r\leqslant r_i,&\\
      B^2_{{\phi\rm i}}+\left(B^2_{{\phi\rm e}}-B^2_{{\phi\rm i}}\right)\left(\frac{r-r_i}{r_e-r_i}\right), &r_i< r< r_e,&\\
    B^2_{{\phi\rm e}}, &r\geqslant r_e,
      \end{array}\right.
\end{eqnarray}
where $B_{\phi\rm i}$ and $B_{\phi\rm e}$ are constant. Putting Eqs. (\ref{Bz}) and (\ref{Bphi}) into the magnetohydrostatic equation (\ref{eqn:pressure:r}), we obtain the background gas pressure as follows
%%%%%%%%%%%%%%%%%%%%%%%%%%%%%%%%%%%%%%%%%%%%%%%%%%%%%%%%%%%%%%%

\begin{eqnarray}\label{p}
p(r)=\left\{\begin{array}{lll}
   p_{{\rm i}}-\frac{B^2_{{\phi\rm i}}}{\mu_0}(r/r_i)^2, &r\leqslant r_i,&\\
   A_1+A_2 r+A_3\ln(r/r_i), &r_i< r< r_e,&\\
   p_{{\rm e}}-\frac{B^2_{{\phi\rm e}}}{\mu_0}\ln(r/r_e), &r\geqslant r_e,
      \end{array}\right.
\end{eqnarray}
where
\begin{align}\label{A123}
&A_1=\left( p_i-\frac{B^2_{{\phi\rm i}}}{\mu_0}\right)-A_2r_i,\nonumber\\
&A_2\equiv\frac{3\left(B^2_{{\phi\rm i}}-B^2_{{\phi\rm e}}\right)+\left(B^2_{{z\rm i}}-B^2_{{z\rm e}}\right)}{2\mu_0 (r_e-r_i)},\nonumber\\
&A_3\equiv\frac{r_i B^2_{{\phi\rm e}}-r_e B^2_{{\phi\rm i}}}{\mu_0 (r_e-r_i)},\nonumber\\
&p_e=\left(p_i-\frac{B^2_{{\phi\rm i}}}{\mu_0}\right)+A_2(r_e-r_i)+A_3\ln(r_e/r_i),
\end{align}
%\begin{align}\label{A123}
%&A_1= p_i-\frac{B^2_{{\phi\rm i}}}{\mu_0}-A_2r_i,\nonumber\\
%&A_2\equiv\frac{3\left(B^2_{{\phi\rm i}}-B^2_{{\phi\rm e}}\right)+\left(B^2_{{z\rm i}}-B^2_{{z\rm e}}\right)}{2\mu_0 (r_e-r_i)},\nonumber\\
%&A_3\equiv\frac{r_i B^2_{{\phi\rm e}}-r_e B^2_{{\phi\rm i}}}{\mu_0 (r_e-r_i)},\nonumber\\
%&p_e=p_i-\frac{B^2_{{\phi\rm i}}}{\mu_0}+A_2(r_e-r_i)+A_3\ln(r_e/r_i),\nonumber\\
%&~~~= p_i+\frac{\left(B^2_{{\phi\rm i}}-3B^2_{{\phi\rm e}}\right)+\left(B^2_{{z\rm i}}-B^2_{{z\rm e}}\right)}{2\mu_0}
%+\left[\frac{r_i B^2_{{\phi\rm e}}-r_e B^2_{{\phi\rm i}}}{\mu_0 (r_e-r_i)}\right]\ln(r_e/r_i),
%\end{align}
and $p_{i}$ is an arbitrary constant. Also the constants $A_1$ and $p_e$ have been obtained from the continuity of the gas pressure across the boundaries $r=r_i$ and $r=r_e$. Note that Eq. (\ref{p}) in the absence of twist, i.e. $B_{\phi\rm i}=B_{\phi\rm e}=0$ recovers the gas pressure profile given by Yu et al. \cite{TYu}.

\begin{figure}
\centering
        \includegraphics[width=0.7\textwidth]{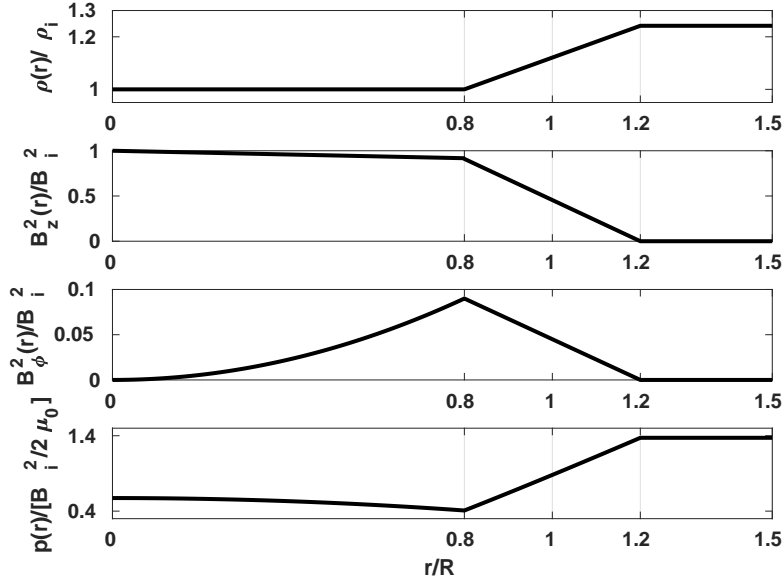}
         \caption{Variations of the background quantities of the flux tube including the density, the magnetic field components and the plasma pressure versus the fractional radius $r/R$ for the twist parameter $B_{\phi\rm i}/B_{z\rm i}=0.3$ and $r_i/R=0.8$ and $r_e/R=1.2$. For the photospheric conditions, the auxiliary parameters are taken from \cite{grant2015} as  $v_{Ai}=12~{\rm km~s}^{-1}$, $v_{Ae}=0 ~{\rm km~s}^{-1}$ (i.e. $B_{\phi e}=B_{ze}=0$), $v_{si}=7 ~{\rm km~s}^{-1}$, $v_{se}=11.5 ~{\rm km~s}^{-1}$. With the help of these values and using $\gamma=5/3$, from Eqs. (\ref{beta}) and (\ref{chi}) we find $\beta=0.41$ and $\rho_e/\rho_i=1.242$.}
    \label{profil}
\end{figure}

In addition, we define the following quantities
\begin{equation}
v^2_{Ai}\equiv\frac{B^2(r_i)}{\mu_0\rho_{i}},
\end{equation}
\begin{equation}
v^2_{Ae}\equiv\frac{B^2(r_e)}{\mu_0\rho_{e}},
\end{equation}
\begin{equation}\label{v2si}
v^2_{si}\equiv\gamma\frac{ p(r_i)}{\rho_i}=\gamma \left(p_i-\frac{B^2_{{\phi\rm i}}}{\mu_0}\right)/\rho_{i},
\end{equation}
\begin{equation}\label{v2se}
v^2_{se}\equiv\gamma\frac{ p(r_e)}{\rho_e}=\gamma p_e/\rho_e,
\end{equation}
\begin{equation}
v^2_{c(i,e)}\equiv\frac{v^2_{s(i,e)} v^2_{A(i,e)} }{v^2_{s(i,e)}+v^2_{A(i,e)}},
\end{equation}
where $B^2=B^2_{{\phi}}+B^2_{z}$. Also $v_{A(i,e)}$, $v_{s(i,e)}$ and $v_{c(i,e)}$ are the interior/exterior Alfv\'{e}n, sound and cusp velocities, respectively. Besides, we define the parameter $\beta$ as the ratio of the plasma pressure to the magnetic field pressure, inside the flux tube as
\begin{equation}\label{beta}
\beta\equiv\frac{p(r_i)}{B^2(r_i)/(2\mu_0)}=2v^2_{si}/(\gamma v^2_{Ai}).
\end{equation}
With the help of the parameter $\beta$, Eq. (\ref{beta}), the arbitrary constant $p_i$ can be determined as
\begin{equation}
\frac{p_i}{B_i^2/(2\mu_0)}=\beta+2\left(\frac{B_{\phi_i}}{B_i}\right)^2,
\end{equation}
where $B_i=B(r_i)$. Using Eqs. (\ref{v2si}), (\ref{v2se}) and (\ref{beta}), one can find the density ratio $\rho_e/\rho_i$ as
 \begin{align}\label{chi}
\frac{\rho_e}{\rho_i}&=\left(\frac{ v_{si}}{v_{se} }\right)^2\frac{p_e}{\left(p_i-\frac{B^2_{{\phi\rm i}}}{\mu_0}\right)},\nonumber\\
&=\frac{2}{\beta}\left(\frac{ v_{si}}{v_{se} }\right)^2\Bigg[\beta/2+(\widetilde{B}_i^2-\widetilde{B}_e^2)/2+\left(1-\frac{}{}\frac{r_e\ln(r_e/r_i) }{ r_e-r_i}\right)\widetilde{B}^2_{{\phi\rm i}}-\left(1-\frac{}{}\frac{r_i\ln(r_e/r_i) }{r_e-r_i}\right)\widetilde{B}^2_{{\phi\rm e}}\Bigg],
\end{align}
where $\widetilde{B}=B/B_i$ and $\widetilde{B}_i^2=1$.  Figure \ref{profil} shows the background quantities containing the density (\ref{rho}), the magnetic field components, Eqs. (\ref{Bz}) and (\ref{Bphi}), and the plasma pressure (\ref{p}) for the twist parameter $B_{\phi\rm i}/B_{z\rm i}=0.3$ under the magnetic pore conditions.

The set of Eqs. (\ref{eqn:linearized:mhd1})-(\ref{eqn:linearized:mhd3}) in cylindrical coordinates can be solved by Fourier decomposition of the perturbed quantities as follows
\begin{dmath}\label{eqn:fourier:xi:pt}
(\bm{\xi}, \delta P_{T}) \propto e^{i(m \phi + k_z z - \omega t)},
\end{dmath}
where $\omega$ is the angular frequency, $m$ is the azimuthal wavenumber for which only integer values are allowed and, $k_z$, is the longitudinal wavenumber in the $z$ direction. Also $\delta P_{T}= \delta p + {\bm B}. \delta{\bm B}/\mu_0$ is the Eulerian perturbation of total (gas and magnetic) pressure.  Putting Eq. (\ref{eqn:fourier:xi:pt}) into (\ref{eqn:linearized:mhd1})-(\ref{eqn:linearized:mhd3}), we obtain the two coupled first order differential equations
\begin{dgroup}\label{eqn:sakurai}
\begin{dmath}\label{eqn:sakurai:xidel}
D \frac{r {\rm d}\bm{\xi}}{{\rm d}r} = C_1 (r \bm{\xi}) - r C_2 \delta P_T,
\end{dmath}
\begin{dmath}\label{eqn:sakurai:ptdel}
D \frac{r{\rm d}\delta P_T}{{\rm d}r} =\frac{1}{r} C_3 (r \bm{\xi}) -  C_1 \delta P_T.
\end{dmath}
\end{dgroup}
The above equations derived earlier by Appert et al. \cite{Appert}
and later by Hain \& Lust, Goedbloed and Sakurai et al. \cite{hain1958normal,goedbloed1971stabilization,sakurai1991resonantI}. Here, the multiplicative factors are defined as
\begin{dgroup}\label{eqn:sakurai}
\begin{dmath}\label{eqn:sakurai:xidel1}
D\equiv\rho \left(\omega^2-\omega^2_A\right)C_4,
\end{dmath}
\begin{dmath}\label{eqn:sakurai:ptdel1}
C_1\equiv\frac{2 B_{\phi}}{\mu_0 r}\left(\omega^4 B_{\phi}-\frac{m}{r}f_BC_4\right),
\end{dmath}
\begin{dmath}\label{eqn:sakurai:ptdel1}
C_2\equiv n^2-k_z^2,
\end{dmath}
\begin{dmath}\label{eqn:sakurai:ptdel2}
C_3\equiv\rho D \left[\omega^2-\omega^2_A+\frac{2 B_{\phi}}{\mu_0 \rho}\frac{{\rm d}}{{\rm d}r}\left(\frac{B_{\phi}}{r}\right)\right]+4\omega^4\frac{ B^2_{\phi}}{\mu^2_0 r^2}-\rho C_4 \frac{4 B^2_{\phi}\omega^2_A}{\mu_0 r^2},
\end{dmath}
\begin{dmath}\label{eqn:sakurai:ptdel3}
C_4\equiv\left(v^2_s+v^2_A\right)\left(\omega^2-\omega^2_c\right),
\end{dmath}
\end{dgroup}
where
\begin{dgroup*}\label{eqn:sakurai}
\begin{dmath*}\label{eqn:sakurai:ptdel3}
f_B\equiv\frac{m}{r}B_{\phi}+k_z B_z,
\end{dmath*}
\begin{dmath*}\label{eqn:sakurai:ptdel1}
\omega^2_A\equiv\frac{f^2_B}{\mu_0 \rho},
\end{dmath*}
\begin{dmath*}\label{eqn:sakurai:xidel1}
\omega^2_c\equiv\left(\frac{v^2_s}{v^2_A+v^2_s}\right)\omega^2_A,
\end{dmath*}
\end{dgroup*}
and
\begin{equation}\label{n2}
n^2\equiv\frac{\omega^4}{(v^2_{A}+v^2_{s})(\omega^2- \omega^2_{c})}.
\end{equation}
Here $\omega_A(=k_zv_A)$ is the Alfv\'{e}n angular frequency and $\omega_c(=k_zv_c)$ is the cusp angular frequency. Also $v_A=\left|\textbf{B}\right|/\sqrt{\mu_0 \rho}$ is the Alfv\'{e}n speed, $v_s=\sqrt{\gamma p/\rho}$ is the sound speed, and $v_c=\frac{v_s v_A}{(v^2_s+v^2_A)^{1/2}}$  is the cusp velocity.

Combining Eqs. (\ref{eqn:sakurai:xidel}) and (\ref{eqn:sakurai:ptdel}), one can obtain a second-order ordinary differential equation for radial component of the Lagrangian displacement $\bm{\xi}$ as \cite{Erdelyi,Giagkiozis2016}
\begin{dmath}\label{eqn:secon-order}
\frac{{\rm d}}{{\rm d}r}\left[\frac{D}{r C_2}\frac{{\rm d}}{{\rm d}r}(r \bm{\xi}_r)\right]+\left[\frac{1}{D}\left(C_3-\frac{C^2_1}{C_2}\right)-r\frac{{\rm d}}{{\rm d}r}\left(\frac{C_1}{rC_2}\right)\right]\bm{\xi}_r=0.
\end{dmath}
For the sausage modes ($m=0$), solutions of Eq. (\ref{eqn:secon-order}) in the interior ($r\leqslant r_i$) and exterior ($r\geqslant r_e$) regions are given by \cite{Giagkiozis2015}
\begin{dgroup}
\begin{dmath}\label{xiri}
\bm{\xi}_{ri}(s) = A_{i} \frac{s^{1/2}}{E^{1/4}} e^{-s/2} M(a,b;s),
\end{dmath}
\begin{dmath}\label{xire}
\bm{\xi}_{re}(r) = A_{e } K_\nu(k_{re} r),
\end{dmath}
\end{dgroup}
where $A_i$ and $A_e$ are constant. Also $M(.)$ is the Kummer function, and $K(.)$ is the modified Bessel function of the second kind \cite{Abramowitz}. Replacing the solutions (\ref{xiri}) and (\ref{xire}) into Eq. (\ref{eqn:sakurai:ptdel}) read
\begin{dgroup}
\begin{dmath}\label{Pi}
\delta P_{Ti}(s) = A_{i}e^{-s/2}\left( \frac{k_a D_i}{n_i^2 - k_z^2}\right) \left[\left( \frac{n_i+k_z}{k_z}\right) s M(a,b;s) - 2M(a,b-1;s)\right],
\end{dmath}
\begin{dmath}\label{Pe}
\delta P_{Te}(r) = A_{e} \left[\left( \frac{\mu_0(1-\nu)D_e-2B^2_{\phi e}n^2_e}{\mu_0 r (k^2_z- n^2_e)} \right)K_{\nu}(k_{re}r) - \frac{D_e}{k_{re}}K_{\nu-1}(k_{re}r)\right].
\end{dmath}
\end{dgroup}
 The parameters appeared in Eqs. (\ref{xiri})-(\ref{Pe}) are defined as
\begin{equation}\label{s}
s\equiv k^2_a E^{1/2}r^2,~~~E\equiv\frac{4 B^4_{\phi i}n^2_{i}}{\mu^2_0 r^4_i D^2_i k^2_z(1-\alpha^2)^2},
\end{equation}
\begin{equation}\label{ka}
k_a\equiv k_z(1-\alpha^2)^{1/2},~~~\alpha^2\equiv\frac{4 B^2_{\phi i}\omega^2_{Ai}}{\mu_0 r^2_i\rho_i(\omega^2-\omega^2_{Ai})^2},
\end{equation}
%\begin{equation}\label{n2}
%n^2\equiv\frac{\omega^4}{(v^2_{A}+v^2_{s})(\omega^2- \omega^2_{c})},
%\end{equation}
\begin{equation}\label{a}
a\equiv 1+\frac{k^2_{ri}}{4 k^2_z E^{1/2}},~~~b=2,
\end{equation}

\begin{equation}\label{k2}
k^2_{r}\equiv\frac{(\omega^2_{s}-\omega^2)(\omega^2_{A}-\omega^2)}{(v^2_{A}+v^2_{s})(\omega_c^2-\omega^2)},
\end{equation}

\begin{equation}\label{D}
D_i\equiv\rho_i \left(\omega^2-\omega^2_{Ai}\right),~~~ D_e\equiv\rho_e \left(\omega^2-\omega^2_{Ae}\right),
\end{equation}

\begin{dmath}\label{nu}
\nu^2(0,r)\equiv 1+2\frac{B^2_{\phi e}}{\mu^2_0 D^2_e}\Big(2B^2_{\phi e}n^2_e k^2_z+\mu_0\rho_e\left[\omega^2_{Ae}\left(3n^2_e-k^2_z\right)-\omega^2\left(n^2_e+k^2_z\right)\right]\Big).
\end{dmath}
Note that in the annulus region ($r_i< r< r_e$), we don't solve the MHD equations. Instead, we relate the interior solutions to exterior ones by using the connection formula introduced in section \ref{sec3}.

%----------------------------------------------------------------------------------------------------
\section{Dispersion relation for the case of no inhomogeneous layer}\label{sec2}

Here, we are interested in obtaining the dispersion relation for the sausage mode in the case of no annulus region. The solutions (\ref{xiri})-(\ref{Pe}) for inside and outside of the flux tube must satisfy the following boundary conditions
\begin{dgroup}\label{boundaryconditions}
\begin{dmath}\label{boundaryconditions1}
\bm{\xi}_{ri}\Big|_{r=R}=\bm{\xi}_{re}\Big|_{r=R},
\end{dmath}
\begin{dmath}\label{boundaryconditions2}
\Big(\delta P_{Ti}-\frac{B^2_{\phi i}}{\mu_0 r}~\bm{\xi}_{ri}\Big)\Big|_{r=R}=\Big(\delta P_{Te}-\frac{B^2_{\phi e}}{\mu_0 r}~\bm{\xi}_{re}\Big)\Big|_{r=R},
\end{dmath}
\end{dgroup}
where $R=r_i=r_e$ is the tube radius. The above relations show continuity conditions for the Lagrangian displacement and Lagrangian changes of the total pressure across the tube boundary, respectively. Inserting the solutions (\ref{xiri})-(\ref{Pe}) into the boundary conditions (\ref{boundaryconditions1}) and (\ref{boundaryconditions2}), after some algebra one can find the following dispersion relation

\begin{dmath}\label{dispersionrelation}
-\frac{\mu_0D_i}{ k^2_{ri}}\left[\left(\frac{n_i+k_z}{k_z}\right)s-2\frac{M(a,b-1,s)}{M(a,b,s)}\right]=
\frac{\mu_0 D_e }{ k^2_{re}}\left(1-\nu- \frac{k_{re}R~K_{\nu-1}(k_{re} R)}{K_{\nu}(k_{re} R)}\right)-\left(1+\frac{ 2  n^2_e}{ k^2_{re}}\right)B_{\phi e}^2
+
B^2_{\phi i},
\end{dmath}
where $s=k^2_a E^{1/2}R^2$.

Now, we are interested in investigating the dispersion relation (\ref{dispersionrelation}) in the limit of no twist inside and outside the tube, i.e. $B_{\phi i}=B_{\phi e}=0$. For the small twist, from Eq. (\ref{s}) we have $E\ll 1$ and then from the first relation of Eq. (\ref{a}) we get $E^{1/2}\simeq k^2_{ri}/(4ak_z^2)$. Also from Eq. (\ref{ka}) we obtain $\alpha^2\ll 1$ and then $k_a\simeq k_z$. Consequently, from the first relation of Eq. (\ref{s}) we find $s\simeq k^2_{ri}R^2/(4a)$. Using these approximations for the small twist, the Kummer functions appeared in the dispersion relation (\ref{dispersionrelation}) behave as
\begin{equation}\label{dislim1}
\lim_{a\to\infty}M(a,b-1,s)=\lim_{a\to\infty}M\left(a,1,\frac{k_{ri}^2 R^2}{4 a}\right)=\Gamma(1)I_{0}\left(2\sqrt{\frac{k_{ri}^2 R^2}{4}}\right)=I_{0}\left(k_{ri} R\right),
\end{equation}
\begin{equation}\label{dislim2}
\lim_{a\to\infty}M(a,b,s)=\lim_{a\to\infty}M\left(a,2,\frac{k_{ri}^2 R^2}{4 a}\right)=\Gamma(2)\left(\frac{k_{ri}^2 R^2}{4}\right)^{-1/2}I_{1}\left(2\sqrt{\frac{k_{ri}^2 R^2}{4}}\right)=\frac{2}{k_{ri} R}I_{1}\left(k_{ri} R\right),
\end{equation}
where we have used the following relation \cite{Abramowitz}
\begin{equation}
\lim_{a\to\infty}M(a,b,z/a)=\Gamma(b)z^{(1-b)/2}I_{b-1}(2\sqrt{z}),
\end{equation}
in which $\Gamma(b)$ is the Gamma function and $I(.)$ is the modified Bessel function of the first kind.

Now, putting Eqs. (\ref{dislim1}) and (\ref{dislim2}) in the dispersion relation (\ref{dispersionrelation}) and using $B_{\phi i}\rightarrow0$, $B_{\phi e}\rightarrow0$, $s\rightarrow 0$, and $\nu\rightarrow 1$ (see Eq. (\ref{nu})), one can find
\begin{equation}
\frac{D_i}{k_{ri}}\frac{I_{0}\left(k_{ri} R\right)}{I_{1}\left(k_{ri} R\right)}=-\frac{D_e}{k_{re}}\frac{K_{0}\left(k_{re} R\right)}{K_{1}\left(k_{re} R\right)}.
\end{equation}
Replacing $D_i$ and $D_e$ from Eq. (\ref{D}) into the above relation, we get
\begin{equation}\label{dispER}
\rho_i \left(\omega^2-\omega_{Ai}^2\right)+\frac{k_{ri}}{k_{re}}\rho_e \left(\omega^2-\omega_{Ae}^2\right)Q_0=0,
\end{equation}
where $Q_0=-\frac{I_{1}\left(k_{ri} R\right)K_{0}\left(k_{re} R\right)}{I_{0}\left(k_{ri} R\right)K_{1}\left(k_{re} R\right)}$. Finally, Eq. (\ref{dispER}) can be recast as
 \begin{equation}\label{dislim}
\omega^2=\frac{\rho_i \omega_{Ai}^2 - \rho_e \frac{k_{ri}}{k_{re}}\omega_{Ae}^2 Q_0 }{\rho_i - \rho_e \frac{k_{ri}}{k_{re}} Q_0},
\end{equation}
which recovers exactly the same result obtained by Edwin \& Robertes \cite{Edwin1983Robertes} and Yu et al. \cite{TYu} in the absence of magnetic twist.

In addition, we turn to solve the dispersion relation (\ref{dispersionrelation}), numerically, in the presence of twist. To do this, we use its dimensionless form (see Eq. (\ref{dispersionrelation-dimless}) in Appendix D). In Fig. \ref{5bl0tn}, we plot the results obtained for the phase speed $v/v_{si}=\omega/\omega_{si}$ of the slow surface sausage mode ($m=0$) versus $k_z R$ for different twist parameters $B_{\phi i}/B_{zi}$. The figure shows that (i) for a given $k_zR$, when the twist increases the phase speed decreases. (ii) For a given twist $B_{\phi i}/B_{zi}$, the phase speed decreases for larger $k_zR$ values. (iii) For the case of no twist, the result of  \cite{TYu} is recovered.

In the next section, we turn to obtain the dispersion relation of sausage modes in the presence of annulus region.

%---------------------------------------------------------------------------------------------------
\begin{figure}
\centering
        \includegraphics[width=0.6\textwidth]{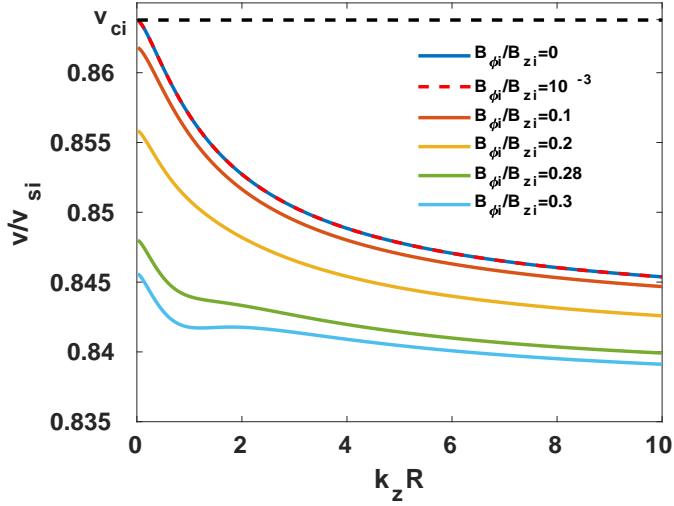}
         \caption{The phase speed $v/v_{si}(=\omega/\omega_{si})$, Eq. (\ref{dispersionrelation}), of the slow surface sausage modes ($m=0$) versus $k_z R$ for different twist parameters $B_{\phi i}/B_{zi}$. Under the magnetic pore conditions, following  \cite{grant2015} the auxiliary parameters are taken as $v_{Ai}=12 ~{\rm km~s}^{-1}$, $v_{Ae}=0 ~{\rm km~s}^{-1}$ (i.e. $B_{\phi e}=B_{ze}=0$), $v_{si}=7 ~{\rm km~s}^{-1}$, $v_{se}=11.5 ~{\rm km~s}^{-1}$, $v_{ci}=6.0464~{\rm km~s}^{-1}(\simeq 0.8638~ v_{si})$ and $v_{ce}=0 ~{\rm km~s}^{-1}$. Here, the results for $B_{\phi_i}/B_{zi}=0$ and $10^{-3}$ overlap with each other.}
    \label{5bl0tn}
\end{figure}

%----------------------------------------------------------------------------------------------------
%----------------------------------------------------------------------------------------------------
\section{Dispersion relation in the presence of inhomogeneous layer and resonant absorption}\label{sec3}

Considering an inhomogeneous layer, where the background density, pressure and magnetic field change continuously from inside to outside the flux tube, see Eqs. (\ref{rho})-(\ref{Bphi}),
the differential equation (\ref{eqn:secon-order}) may become singular at $\omega=\omega_c(r)$ and $\omega=\omega_A(r)$.
The resonant absorption that may occur in these singular points causes damping of the wave amplitude. Because the presence of the inhomogeneous layer, the values of $\omega_c(r)$ and
$\omega_A(r)$ change continuously from inside to outside the flux tube. These processes are called slow (cusp) and Alfv\'{e}n continua, respectively. Note that following Yu et al. \cite{TYu} under the magnetic pore conditions, only the cusp singularity occurs where the phase speed of the slow surface sausage (sss) mode lies in the range of $v_{ce}<v_{sss}<v_{ci}$.

Following Sakurai et al. \cite{sakurai1991resonantI}, in the resonant layer where the singularity occurs because of magnetic twist one does not need to solve Eq. (\ref{eqn:secon-order}). Instead, the solutions inside and outside of the flux tube are related to each other via the following connection formula given by \cite{sakurai1991resonantI}
\begin{dgroup}\label{conection}
\begin{dmath}\label{conection1}
[\bm{\xi}_r]\equiv\bm{\xi}_{re}(r_e)-\bm{\xi}_{ri}(r_i),=\frac{-i \pi \mu \omega_{c}^4}{|\Delta_c|r B^2 \omega_{A}^2}\Big|_{r=r_c}\Big(\delta P_{Ti}-\frac{2 B_{\phi i}^2 \bm{\xi}_{ri}}{\mu_0 r}\Big)\Big|_{r=r_i},
\end{dmath}
\begin{dmath}\label{conection2}
[\delta P_T]\equiv\delta P_{Te}(r_e)-\delta P_{Ti}(r_i),=\frac{-i 2 \pi \omega_{c}^4 B_{\phi}^2}{|\Delta_c|r B^2 \omega_{A}^2}\Big|_{r=r_c}\Big(\delta P_{Ti}-\frac{2 B_{\phi i}^2 \bm{\xi}_{ri}}{\mu_0 r}\Big)\Big|_{r=r_i},
\end{dmath}
\end{dgroup}
where $[\bm{\xi}_r]$ and $[\delta P_T]$ are the jump conditions in the Lagrangian radial displacement and total pressure
perturbation, respectively, across the inhomogeneous (resonant) layer.  The subscript $c$ denotes the position of the slow resonance $(r=r_c)$ and $|\Delta_c|\equiv\left |\frac{d(\omega^2-\omega^2_c)}{dr}\right|_{r=r_c}$. Note that we will determine the cusp resonance point $r_c$ later, see Eq. (\ref{rc}).

Substituting the solutions (\ref{xiri})-(\ref{Pe}) into the connection formula (\ref{conection1}) and (\ref{conection2}), one can find the dispersion relation governing the slow surface sausage modes in the presence of magnetic twist as
\begin{dmath}\label{dispersionrelationthin}
\frac{D_i}{k_{ri}^{2}} \left[\frac{n_i+k_z}{k_z}s-2\frac{M(a,b-1;s)}{M(a,b;s)}\right]+r_i
  \left(\frac{\mu_0(1-v)D_e-2B_{\phi e}^{2}n_{e}^{2}}{\mu_0 r_e k_{re}^{2}} -\frac{D_e}{k_{re}}\frac{K_{v-1}(k_{re}r_e)}{K_{v}(k_{re}r_e)}\right)\\
 +i \frac{ \pi \mu_0 \omega_{c}^4}{|\Delta_c| B^2 \omega_{A}^2}\Big|_{r=r_c}  \left(\frac{D_i}{k_{ri}^{2}}\left[\frac{n_i+k_z}{k_z}s-2\frac{M(a,b-1;s)}{M(a,b;s)}\right]+\frac{2 B_{\phi i}^2}{\mu_0}\right)\times \left(-\frac{2 B_{\phi }^2}{\mu_0 r}\Big|_{r=r_c}+\frac{\mu_0(1-v)D_e-2B_{\phi e}^{2}n_{e}^{2}}{\mu_0 r_e k_{re}^{2}} -\frac{D_e}{k_{re}}\frac{K_{v-1}(k_{re}r_e)}{K_{v}(k_{re}r_e)}\right)=0.
\end{dmath}

For the case of no twist, i.e. $ B_{\phi i}=B_{\phi e}=0$, the dispersion relation (\ref{dispersionrelationthin}) following the same approach that was used in the previous section takes the form
\begin{dmath}\label{lim6}
\rho_i\left(\omega^2-\omega_{Ai}^2\right)-\rho_e\left(\omega^2-\omega_{Ae}^2\right)\dfrac{k_i}{k_e}Q_0+\frac{i \pi k_{z}^2 }{\rho|\Delta_c|  }\left(\dfrac{v_{s}^2}{v_{s}^2+v_{A}^2}\right)^2\rho_i\rho_e\left(\omega^2-\omega_{Ai}^2\right)\left(\omega^2-\omega_{Ae}^2\right)\dfrac{G_0}{k_e}=0,~~
\end{dmath}
where $G_0=-\frac{K_{0}(k_{re}r_e)}{K_{1}(k_{re}r_e)}$. The above relation is same as that obtained by Yu et al. \cite{TYu} in the absence of magnetic twist.

Using Eqs. (\ref{rho}) to (\ref{v2se}), one can obtain the quantities  $v_s=\sqrt{\gamma p/\rho}$ , $v_A=\left|\textbf{B}\right|/\sqrt{\mu_0 \rho}$ and the cusp velocity $v_c\equiv\frac{v_s v_A}{(v^2_s+v^2_A)^{1/2}}$ in the inhomogeneous layer ($r_i< r< r_e$) as
\begin{dmath}\label{vs2}
v_s^2=v_{si}^2\left[\frac{1+\delta(\chi v^2_{sei}-1)+\zeta}{1+\delta(\chi -1)}\right],
\end{dmath}
\begin{dmath}\label{vA2}
v_A^2=v_{Ai}^2\left[\frac{1+\delta(\chi v^2_{Aei}-1)}{1+\delta(\chi -1)}\right],
\end{dmath}
\begin{dmath}\label{v2C}
v^2_c=\frac{v^2_{si}v^2_{Ai}\Big[1+\delta(\chi v^2_{sei}-1)+\zeta\Big]\Big[1+\delta(\chi v^2_{Aei}-1)\Big]}{\Big[1+\delta(\chi -1)\Big]\Big[v^2_{si}\Big(1+\delta(\chi v^2_{sei}-1)+\zeta\Big)+v^2_{Ai}\Big(1+\delta(\chi v^2_{Aei}-1)\Big)\Big]},
\end{dmath}
where $\delta\equiv\frac{r-r_i}{r_e-r_i}$ , $\chi\equiv\rho_e/\rho_i$, $v_{sei}\equiv v_{se}/v_{si}$, $v_{Aei}\equiv v_{Ae}/v_{Ai}$ and
\begin{align}\label{V}
&\zeta\equiv\frac{\gamma v^2_{A i}}{v^2_{si}}\left[\frac{r_iB^2_{\phi e}-r_eB^2_{\phi i}}{(r_e-r_i)B^2_{i}}\right]\Big(\ln(r/r_i)-\delta \ln(r_e/r_i)\Big).
\end{align}
Notice that in the absence of twist, i.e. $ B_{\phi i}=B_{\phi e}=0$, Eqs. (\ref{vs2}), (\ref{vA2}) and (\ref{v2C}) transform to the corresponding relations in Yu et al. \cite{TYu}.

\begin{figure}[h]
\centering
\includegraphics[width=0.6\textwidth]{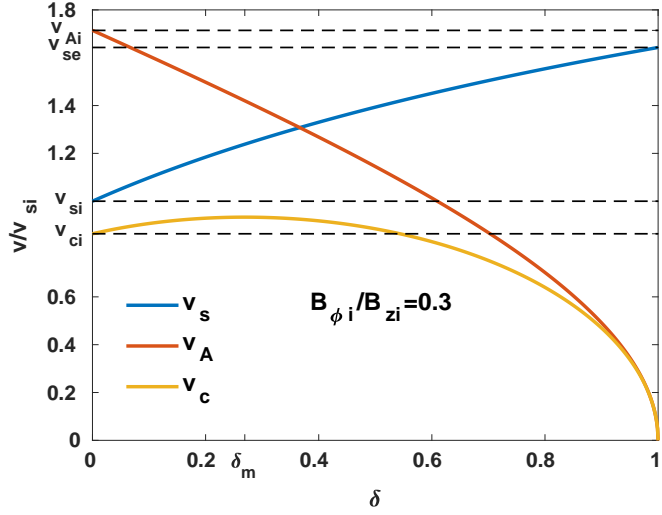}
\caption{Variations of the dimensionless velocity $v/v_{si}$, Eqs. (\ref{vs2}), (\ref{vA2}) and (\ref{v2C}), as a function of $\delta\equiv\frac{r-r_i}{r_e-r_i}$ in the nonuniform
(transitional) layer for the twist parameter $B_{\phi i}/B_{zi}=0.3$ under magnetic pore conditions. The auxiliary parameters are $v_{Ai}=12 ~{\rm km~s}^{-1}$, $v_{Ae}=0 ~{\rm km~s}^{-1}$ (i.e. $B_{\phi e}=B_{ze}=0$), $v_{si}=7 ~{\rm km~s}^{-1}$, $v_{se}=11.5 ~{\rm km~s}^{-1}$, $v_{ci}=6.0464~{\rm km~s}^{-1}(\simeq 0.8638~ v_{si})$ and $v_{ce}=0 ~{\rm km~s}^{-1}$ \cite{grant2015}. }
    \label{vcp}
\end{figure}

In Fig. \ref{vcp}, using Eqs. (\ref{vs2}), (\ref{vA2}) and (\ref{v2C}) we plot the sound, Alfv\'{e}n and cusp velocities for the twist parameters $B_{\phi i}/B_{zi}=0.3$ under magnetic pore conditions. Figure \ref{vcp} shows that for $v_c< v_{ci}$ and $v_{ci}<v_c<v_{c_{\rm max}}$, respectively, the surface and body sausage modes can
resonantly damp in the slow continuum. Here, $v_{c_{\rm max}}$ is the maximum value of the cusp velocity.

Note that according to Yu et al. \cite{{TYu}}, the position of the cusp resonance point $r_c$ is obtained by setting $\omega^2=\omega_c^2\Big|_{r=r_c}\equiv k_z^2v^2_c\Big|_{r=r_c}$ in Eq. (\ref{v2C}). Consequently, the resulting equation
in terms of the variable $\delta_c\equiv\delta\Big|_{r=r_c}=\frac{r_c-r_i}{r_e-r_i}$ yields the following second order equation
\begin{align}\label{vc1}
A\delta_c^2+B\delta_c+C=0,
\end{align}
with
\begin{eqnarray}
A&\equiv&1-\frac{v_{c}^2}{v_{ci}^2}-\frac{\gamma v^2_{A i}}{v^2_{si}}\left(\frac{r_iB^2_{\phi e}-r_eB^2_{\phi i}}{(r_e-r_i)B^2_{i}}\right)\left(1-\frac{v_{c}^2}{v_{Ai}^2}\right)\left[\frac{1}{2}\left(\frac{r_e-r_i}{r_i}\right)^2+\frac{r_e-r_i}{r_i}+\ln(r_e/r_i)\right]\nonumber\\
&&+\chi\left[\frac{2v_{c}^2}{v_{ci}^2}-(v_{sei}^2+v_{Aei}^2)-
\frac{\gamma v^2_{A i}}{v^2_{si}}\left(\frac{r_iB^2_{\phi e}-r_eB^2_{\phi i}}{(r_e-r_i)B^2_{i}}\right)\left(v_{Aei}^2-\frac{v_{c}^2}{v_{Ai}^2}\right)\left(\frac{r_e-r_i}{r_i}-\ln(r_e/r_i)\right)\right]\nonumber\\
&&-\chi^2\left(\frac{v_{c}^2}{v_{ci}^2}-v_{sei}^2v_{Aei}^2\right),\label{eq:d8}\\
B&\equiv&2\left(\frac{v_{c}^2}{v_{ci}^2}-1\right)+\frac{\gamma v^2_{A i}}{v^2_{si}}\left(\frac{r_iB^2_{\phi e}-r_eB^2_{\phi i}}{(r_e-r_i)B^2_{i}}\right)\left(1-\frac{v_{c}^2}{v_{Ai}^2}\right)\left(\frac{r_e-r_i}{r_i}+\ln(r_e/r_i)\right)\nonumber\\&&-\chi\left[\frac{v_{c}^2}{v_{ci}^2}
\left(1+\frac{v_{se}^2+v_{Ae}^2}{v_{si}^2+v_{Ai}^2}\right)-(v_{sei}^2+v_{Aei}^2)\right],\label{eq:d9}\\
C&\equiv&1-\frac{v_{c}^2}{v_{ci}^2},\label{eq:d10}
\end{eqnarray}
where we have used the following approximation
\begin{align}\label{ap}
\ln(r/r_i)=\ln\left[1+\left(\frac{r_e-r_i}{r_i}\right)\delta_c\right]\simeq\left(\frac{r_e-r_i}{r_i}\right)\delta_c-\frac{1}{2}\left(\frac{r_e-r_i}{r_i}\right)^2\delta_c^2,
\end{align}
for $\left(\frac{r_e-r_i}{r_i}\right)\delta_c<1$. We checked that keeping the higher order terms $O(\delta_c^3)$ does not affect the results. Equation (\ref{vc1}) has two roots for $\delta_c$ as
\begin{align}\label{sol}
\delta_{c_1}=-\frac{B}{2A}+\frac{\sqrt{B^2-4AC}}{2A},~~~ 0<\delta\leqslant\delta_m,
\end{align}
\begin{align}\label{sol1}
\delta_{c_2}=-\frac{B}{2A}-\frac{\sqrt{B^2-4AC}}{2A},~~~ \delta_m\leqslant\delta\leqslant1,
\end{align}
where $\delta_m$ is the value of $\delta$ when $v_c=v_{c_{\rm max}}$. For instance, in Fig. \ref{vcp} for $B_{\phi i}/B_{zi}=0.3$ we have $v_{c_{\rm max}}=0.93v_{si}$ and $\delta_m=0.27$.

Under the magnetic pore conditions, for slow surface mode we have only one root denoted by $\delta_{c_2}$ . Consequently, the cusp resonance position $r_c$ reads
\begin{equation}\label{rc}
r_c=r_i\left[\left(\frac{r_e}{r_i}-1\right)\delta_{c_2}+1\right].
\end{equation}
Next, we turn to calculate the parameter $\Delta_c$ appeared in the dispersion relation (\ref{dispersionrelationthin}). To this aim, using Eq. (\ref{v2C}) and  $\omega_c^2(r_c)=k_z^2v^2_c\Big|_{r=r_c}$ we obtain
\begin{eqnarray}\label{delta}
\Delta_{c}&\equiv&\left[\frac{d}{dr}(\omega^2-\omega_c^2)\right]_{r=r_c}=-2\left(\omega_{c}\frac{d\omega_{c}}{dr}\right)_{r=r_c},\nonumber\\
&=&-\Bigg(\frac{\omega_{c}^2(r_c)}{l}\Bigg)
\Bigg\{\frac{\left(\chi v_{sei}^2-1\right)+l\zeta'}{1+\delta\left(\chi v_{sei}^2-1\right)+\zeta}-\frac{(\chi-1)}{1+\delta(\chi-1)}\nonumber\\&&+\frac{(\chi v_{Aei}^2-1)}{1+\delta\left(\chi v_{Aei}^2-1\right)}\\&&-\frac{v_{si}^2\Big(\chi v_{sei}^2-1+l\zeta'\Big)+v_{Ai}^2\Big(\chi v_{Aei}^2-1\Big)}{v_{si}^2\left[1+\delta\Big(\chi v_{sei}^2-1\right)+\zeta\Big]+v_{Ai}^2\Big[1+\delta\left(\chi v_{Aei}^2-1\right)\Big]}\Bigg\}_{r=r_c},\nonumber
\end{eqnarray}
where
\begin{align}\label{V11}
\zeta'\equiv\frac{d\zeta}{dr}=\frac{\gamma v^2_{A i}}{v^2_{si}}\left[\frac{r_iB^2_{\phi e}-r_eB^2_{\phi i}}{l^2 B^2_{i}}\right]\Big(l/r-\ln(r_e/r_i)\Big).
\end{align}
Note that in both Eqs. (\ref{dispersionrelationthin}) and (\ref{delta}) due to having the cusp resonance, $\delta(r=r_c)$ should be replaced by $\delta_{c_2}$, Eq. (\ref{sol1}).
%----------------------------------------------------------------------------------------------------
\subsection{Weak damping limit - slow continuum}

Here, we are interested in investigating the dispersion relation (\ref{dispersionrelationthin}) in the limit of weak damping. To this aim, we first rewrite Eq. (\ref{dispersionrelationthin}) as follows
\begin{dmath}\label{dispersionrelationthin1}
D_i + k_{ri}^{2}
  \frac{\left(\frac{\mu_0 D_e(1-v)-2B_{\phi e}^{2}n_{e}^{2}}{\mu_0 r_e k_{re}^2} -\frac{D_e}{k_{re}}\frac{K_{v-1}(k_{re}r_e)}{K_{v}(k_{re}r_e)}\right)}{\left[\frac{n_i+k_z}{k_z}s-2\frac{M(a,b-1;s)}{M(a,b;s)}\right]}\\
 +i \frac{ \pi \mu_0 \omega_{c}^4}{|\Delta_c| B^2 \omega_{A}^2}\Big|_{r=r_c}  \frac{\left(D_i\left[\frac{n_i+k_z}{k_z}s-2\frac{M(a,b-1;s)}{M(a,b;s)}\right]+\frac{2 B_{\phi i}^2}{\mu_0}k_{ri}^{2}\right)}{\left[\frac{n_i+k_z}{k_z}s-2\frac{M(a,b-1;s)}{M(a,b;s)}\right]}\times \left(-\frac{2 B_{\phi }^2}{\mu_0 r}\Big|_{r=r_c}+\frac{\mu_0(1-v)D_e-2B_{\phi e}^{2}n_{e}^{2}}{\mu_0 r_e k_{re}^{2}} -\frac{D_e}{k_{re}}\frac{K_{v-1}(k_{re}r_e)}{K_{v}(k_{re}r_e)}\right)=0.
\end{dmath}
This can also be recast in the following compact form
\begin{equation}
D_{{\rm AR}}+iD_{\rm {AI}}=0,
\end{equation}
where $D_{{\rm AR}}$ and $D_{\rm {AI}}$, respectively, are the real and imaginary parts of Eq. (\ref{dispersionrelationthin1}) given by
\begin{dmath}\label{limw1}
D_{{\rm AR}}=\rho_i(\omega^2-\omega_{Ai}^2) -r_i \rho_e(\omega^2-\omega_{Ae}^2) \frac{k_{ri}}{k_{re}}Q,
\end{dmath}
\begin{dmath}\label{limw2}
D_{{\rm AI}}=\frac{ \pi \rho_i \rho_e k_z^2}{k_{re}\rho_c|\Delta_c| }\Big|_{r=r_c}\Big(\frac{v_{sc}^2}{v_{Ac}^2+v_{sc}^2}\Big)^2 \left((\omega^2-\omega_{Ai}^2)+Z\right)(\omega^2-\omega_{Ae}^2) G,
\end{dmath}
where
\begin{dmath}\label{limw110}
Q\equiv-k_{ri}r_i
  \frac{\left(\frac{\mu_0 D_e (1-v)-2B_{\phi e}^{2}n_{e}^{2}}{D_e \mu_0 r_e k_{re}} -\frac{K_{v-1}(k_{re}r_e)}{K_{v}(k_{re}r_e)}\right)}{\left[\frac{n_i+k_z}{k_z}s-2\frac{M(a,b-1;s)}{M(a,b;s)}\right]},
\end{dmath}
\begin{dmath}\label{limw111}
G\equiv\left(-\frac{2k_{re} B_{\phi }^2}{D_e \mu_0 r}\Big|_{r=r_c}+\frac{\mu_0(1-v)D_e-2B_{\phi e}^{2}n_{e}^{2}}{\mu_0 r_e k_{re} D_e} -\frac{K_{v-1}(k_{re}r_e)}{K_{v}(k_{re}r_e)}\right),
\end{dmath}
\begin{dmath}\label{limw112}
Z\equiv\frac{2 B_{\phi i}^2k_{ri}^{2}}{\mu_0\rho_i\left[\frac{n_i+k_z}{k_z}s-2\frac{M(a,b-1;s)}{M(a,b;s)}\right]}.
\end{dmath}
Note that in Eqs. (\ref{limw1}) and (\ref{limw2}) we have the complex frequency $\omega=\omega_r+i\gamma$, in which $\omega_r$ and $\gamma$ are the cusp (slow) frequency and the damping rate, respectively.
In the limit of weak damping, i.e. $\gamma \ll \omega_r$, the damping rate $\gamma$ is given by Goossens et al. \cite{{Goossens}}
\begin{equation}\label{gammac1}
\gamma=-D_{{\rm AI}}(\omega_r)\left(\frac{\partial D_{{\rm AR}}}{\partial \omega}\Big|_{\omega_r}\right)^{-1}.
\end{equation}
With the help of Eq. (\ref{gammac1}), one can obtain an analytical expression for $\gamma$ (see Appendix \ref{ap1}) in the slow
(cusp) continuum as follows
\begin{align}\label{gammacwd1}
\gamma=-\frac{\frac{ \pi \rho_e k_z^2}{k_{re}\rho_c|\Delta_c| }\Big|_{r=r_c}\left(\frac{v_{s}^2}{v_{A}^2+v_{s}^2}\right)^2 \left((\omega_{r}^2-\omega_{Ai}^2)+Z\right)(\omega_{r}^2-\omega_{Ae}^2) G}{2\omega_r\left(1-\chi\frac{k_{ri}}{k_{re}}Q\right)-\omega_r \chi T},
\end{align}
where the quantity $T$ is given by Eq. (\ref{di11}). Note that in the limit of no twist, i.e. $B_{\phi i}=B_{\phi e}=0$, Eqs. (\ref{limw110}), (\ref{limw111}), (\ref{limw112}), (\ref{div64}) and (\ref{div65}) reduce to
\begin{align}\label{div55}
&Q=Q_0\equiv\frac{I'_0(x) K_0(y)}{I_0(x)K'_0(y)},\nonumber\\
&G=G_0\equiv\frac{K_0(y)}{K'_0(y)},\nonumber\\
&Z=0,\nonumber\\
&P=P_0\equiv\left(\frac{I''_0(x)}{I_0(x)}-\frac{{I'_0(x)}^2}{I_0(x)^2}\right)\frac{K_0(y)}{K_0'(y)},\nonumber\\
&S=S_0\equiv\left(1-\frac{K''_0(y) K_0(y)}{{K'_0(y)}^2}\right)\frac{I'_0(x)}{I_0(x)},
\end{align}
where $x=k_{ri}r_i$ and $y=k_{re}r_e$ and we have also used the relation
\begin{equation}\label{dislim11}
\lim_{a\to\infty}\frac{M(a,b-1,s)}{M(a,b,s)}=\frac{x}{2}\frac{I_{0}(x)}{I_{1}(x)}.
\end{equation}
Replacing the relations (\ref{div55}) into Eq. (\ref{gammacwd1}), the damping rate $\gamma$ in the absence of twist takes the form
\begin{align}\label{gammacwd11}
\gamma=-\frac{\frac{ \pi \rho_e k_z^2}{k_{re}\rho_c|\Delta_c| }\Big|_{r=r_c}\left(\frac{v_{s}^2}{v_{A}^2+v_{s}^2}\right)^2 (\omega_r^2-\omega_{Ai}^2)(\omega_r^2-\omega_{Ae}^2)G_0}{2\omega_r\left(1-\chi\frac{k_{ri}}{k_{re}}Q_0\right)-\omega_r \chi T_0},
\end{align}
where
\begin{align}\label{di12}
&T_0=\omega_{r}^2 \left(\omega_{r}^2-\omega_{Ae}^2\right)\frac{k_{ri}}{k_{re}}\left(\frac{(Q_0+xP_0)(\omega_{r}^2-2\omega_{ci}^2)}{(\omega_{si}^2-\omega_{c}^2)(\omega_{Ai}^2-\omega_{r}^2)(\omega_{r}^2-\omega_{ci}^2)}-\frac{ (Q_0-yS_0)(\omega_{r}^2-2\omega_{ce}^2)}{(\omega_{se}^2-\omega_{r}^2)(\omega_{Ae}^2-\omega_{r}^2)(\omega_{r}^2-\omega_{ce}^2) }\right).
\end{align}
Notice that Eq. (\ref{gammacwd11}) is the same as the dispersion relation (28) in Yu et al.  \cite{TYu} for the slow surface sausage modes when the twist is absent.

%----------------------------------------------------------------------------------------------------
\subsection{Weak damping rate in long wavelength limit - slow continuum}

Here, we try to examine the damping rate (\ref{gammacwd1}) in the long wavelength limit, i.e. ${k_zR}\ll 1$ ($\omega_r\approx \omega_{ci}=k_zv_{ci}$). In this limit, one can show that Eq. (\ref{gammacwd1}) takes the form (see Eq. (\ref{gammacwd22}) in Appendix \ref{ap2})
\begin{align}\label{gammacwd12}
\gamma=-\frac{ \pi \rho_e k_z}{|\Delta_c| }\frac{\omega_{ci}^4}{\omega_{Ai}^4}\frac{ \left((\omega_{ci}^2-\omega_{Ai}^2)+Z\right)(\omega_{ci}^2-\omega_{Ae}^2) G}{2\omega_{ci} \left(\rho_i-\rho_e\frac{k_{ri}}{k_{re}}Q\right)-\omega_{ci} \rho_e T},
\end{align}
where the quantities $T$, $Q$, $G$ and $Z$, are given by Eqs. (\ref{di22}), (\ref{limw1101}), (\ref{limw1111}) and (\ref{limw1121}), respectively. Under photospheric (magnetic pore) conditions, i.e. $v_{Ae}=v_{A\phi e}=v_{ce}\simeq 0$, one can show that Eq. (\ref{gammacwd12}) reduces to (see Eq. (\ref{gammacwd3}) in Appendix \ref{ap2})
\begin{align}\label{gammacwd3}
\gamma=-\frac{ \pi \rho_e k_zR}{|\Delta_c| }\frac{\omega_{ci}^5}{\omega_{Ai}^4}\frac{ \left((\omega_{ci}^2-\omega_{Ai}^2)+Z\right)G}{2\left(\rho_i-\rho_e\frac{k_{ri}}{k_{re}}Q\right)- \rho_e T},
\end{align}
where now the quantities $T$, $Q$, $G$ and $Z$ are given by Eqs. (\ref{di2210}), (\ref{limw11011}), (\ref{limw1112}) and (\ref{limw11211}), respectively.

Note that in the absence of twist (i.e. $B_{\phi i}=B_{\phi e}=0$), the weak damping rate $\gamma$ in long wavelength limit, Eq. (\ref{gammacwd12}), reduces to (see Eq. (\ref{gammacwd21121}) in Appendix \ref{ap3})
\begin{align}\label{gammacwd2112}
\gamma=\frac{ 2\pi \chi^3 }{|\Delta_c| R}\left[\frac{ \omega_{ci}^7\omega_{si}^2\left(\omega_{ci}^2-\omega_{Ae}^2\right)^3 }{3\omega_{Ai}^{10} \omega_{ci}^2+8\chi \omega_{Ai}^8 \omega_{si}^2\left(\omega_{ci}^2-\omega_{Ae}^2\right)\ln(k_zR)}\right](k_zR)^4\ln^3(k_zR).
\end{align}
For the photosphere conditions (i.e. $\omega_{Ae}=\omega_{ce}=0$), Eq. (\ref{gammacwd2112}) reads
\begin{align}\label{gammacwd223}
\gamma=\frac{ 2\pi \chi^3 }{|\Delta_c| R}\left[\frac{ \omega_{ci}^{11}\omega_{si}^2 }{3\omega_{Ai}^{10}+8\chi \omega_{Ai}^8 \omega_{si}^2\ln(k_zR) }\right](k_zR)^4\ln^3(k_zR).
\end{align}
It should be noted that Eqs. (\ref{gammacwd2112}) and (\ref{gammacwd223}) without the terms $8\chi \omega_{Ai}^8 \omega_{si}^2\left(\omega_{ci}^2-\omega_{Ae}^2\right)\ln(k_zR)$ and $8\chi \omega_{Ai}^8 \omega_{si}^2\ln(k_zR)$ appeared in their denominator are same as Eqs. (36) and (37) in  \cite{TYu}. This difference is because of wrong minus sign appeared in Eq. (A.7) in  \cite{TYu1}.
%----------------------------------------------------------------------------------------------------
\section{Numerical results}

Here, we solve numerically the dispersion relation (\ref{dispersionrelationthin}) to obtain the frequencies and damping rates of the slow surface sausage modes. To this aim, it is convenient to recast Eq. (\ref{dispersionrelationthin}) in dimensionless form (see Eq. (\ref{dim}) in Appendix \ref{ap4}). Under the magnetic pore
conditions, following  \cite{grant2015} we set again the model parameters as $v_{Ai}=12 ~{\rm km~s}^{-1}$, $v_{Ae}=0 ~{\rm km~s}^{-1}$ (i.e. $B_{\phi e}=B_{ze}=0$), $v_{si}=7 ~{\rm km~s}^{-1}$, $v_{se}=11.5 ~{\rm km~s}^{-1}$, $v_{ci}=6.0464~{\rm km~s}^{-1}(\simeq 0.8638~ v_{si})$ and $v_{ce}=0 ~{\rm km~s}^{-1}$. Our numerical results are shown in Figs. \ref{5abcdl0.1t} to \ref{5abcphil0.20t}.

Figures \ref{5abcdl0.1t}, \ref{5abcdl0.2t}, \ref{5abcdl0.3t} and \ref{5abcdl0.4t} present variations of the phase speed (or normalized frequency) $v/v_{si}\equiv\omega_r/\omega_{si}$, the damping rate to frequency ratio $|\gamma|/\omega_r$ and the damping time to period ratio  $\tau_D/T=\omega_r/(2\pi|\gamma|)$ of the slow surface sausage modes versus $k_z R$ for different twist parameters $B_{\phi_i}/B_{zi}=(0,10^{-3},0.1,0.2,0.3)$ and different thickness of the inhomogeneous layer $l/R=(0.1,0.2,0.3,0.4)$. The figures clear that i) the minimum value of the phase speed $v/v_{si}$ decreases and shifts to smaller $k_zR$ with increasing the twist parameter $B_{\phi_i}/B_{zi}$. ii) For a given $l/R$, in the small wavelength limit ($k_z R \gg 1$), we have asymptotically $v/v_{si}\rightarrow v_{ci}/v_{si}=0.8638$ and $|\gamma|/\omega_r\rightarrow 0$. This shows that the effect of magnetic twist for larger $k_zR$ is negligible. iii) The maximum value of $\left|\gamma\right|/\omega_r$ increases and its position moves to smaller $k_zR$ when $B_{\phi_i}/B_{zi}$ increases. iv) In the absence/presnt of twist,  the maximum value of $\left|\gamma\right|/\omega_r$ decreases and its position moves to smaller $k_zR$ when $l/R$ increases (see also Fig. \ref{5abcphi0.2t}). v) For a given $l/R$, the minimum value of $\tau_D/T$ decreases with increasing $B_{\phi_i}/B_{zi}$. For instance, for the case of $l/R=0.1$, the minimum value of $\tau_D/T$ for $B_{\phi_i}/B_{zi}=0.3$ changes $\sim38\%$ less than the case of no twist. vi) The dashed-line curves in Figs. \ref{5abcdl0.1t}, \ref{5abcdl0.2t}, \ref{5abcdl0.3t} and \ref{5abcdl0.4t} present the analytical results of the damping rate to frequency ratio $|\gamma|/\omega_{r}$ evaluated by Eq. (\ref{gammacwd1}). These curves show that for the weak damping (i.e. $\gamma \ll \omega_r$) and in the long wavelength limit (i.e. ${k_zR}\ll 1$), our numerical results are in good agreement with analytical ones.

%In Fig. \ref{5abct}, we plot $V/V_{si}$, $-\left|\gamma_0\right|/\omega_r$ and $\tau_D/T$ as a function of the twist parameter $B_{\phi_i}/B_{zi}$ for different values of $k_zR$ and $l/R$. The figure shows that for given $k_zR$ and $l/R$, the quantities $V/V_{si}$, $\mid\gamma_0\mid/\omega_r$ and $\tau_D/T$ decrease with increasing the twist parameter.

In Figs. \ref{5abckz0.5t}, \ref{5abckz1t}, \ref{5abckz2t}, \ref{5abckz4t} and \ref{5abckz8t}, we plot variations of $v/v_{si}$, $|\gamma|/\omega_r$ and $\tau_D/T$ versus the thickness of the inhomogeneous layer $l/R$ for different $B_{\phi_i}/B_{zi}=(0,10^{-3},0.1,0.2,0.3)$ and $k_zR=(0.5,1,2,4,8)$. Figures show that i) the maximum value of $\left|\gamma\right|/\omega_r$ increases and moves to smaller $l/R$ when $B_{\phi_i}/B_{zi}$ increases. ii) In the absence/presnt of twist, the maximum value of $\left|\gamma\right|/\omega_r$ increases and its position moves to smaller $l/R$ when $k_zR$ increases (see also Fig. \ref{5abcphil0.20t}). iii) For a given $k_zR$, the phase speed and the damping rate to frequency ratio, respectively, approach $v/v_{si}\rightarrow v_{ci}/v_{si}=0.8638$ and $|\gamma|/\omega_r\rightarrow 0$ in the limit of larger values of $l/R$. iv) For a given $k_zR$, the minimum value of $\tau_D/T$ decreases with increasing $B_{\phi_i}/B_{zi}$. For instance, for the case of $k_zR=1$, the minimum value of $\tau_D/T$ for $B_{\phi_i}/B_{zi}=0.3$ decreases $\sim63.5\%$ in comparison to the case of no twist.

It is worth to mention that for the case of no twist $B_{\phi_i}/B_{zi}=0$, the results of Figs. \ref{5abcdl0.1t} to \ref{5abcphil0.20t} recover those obtained in Yu et al. \cite{{TYu}}. Note that the results for $B_{\phi_i}/B_{zi}=0$ and $10^{-3}$ overlap with each other.

%----------------------------------------------------------------------------------------------------
\section{Conclusions}

Here, we investigated the effect of magnetic twist on resonant absorption of slow sausage waves in magnetic flux tubes under the solar photospheric (or magnetic poor) conditions. We considered a
straight cylindrical flux tube with different magnetic twist profiles in the interior, annulus and exterior regions. Besides, we assumed the density and longitudinal magnetic field to be constant inside and outside of the flux tube, but to be inhomogeneous in the annulus layer. We presented the solutions
of ideal MHD equations for the interior and exterior regions of the
flux tube. In the case of no inhomogeneous (annulus) layer, we derived the dispersion relation which recovers the result obtained by Edwin \& Robertes \cite{Edwin1983Robertes} and Yu et al. \cite{TYu} in the absence of magnetic twist. In the presence of inhomogeneous layer, with the help of the appropriate connection formula of resonant absorption introduced by Sakurai et al.
 \cite{sakurai1991resonantI}, we obtained the jump conditions governing the solutions inside and outside of
the flux tube. Consequently, we derived the dispersion
relation of the slow surface sausage modes in the presence of magnetic twist. Using this, we first obtained an analytical relation for damping rate of the slow surface sausage modes in the limit of weak damping and long wavelength limit. Then, we showed that our analytical expression for the damping rate in the absence of twist recover the result obtained by Yu et al. \cite{{TYu}}. In addition, we solved the dispersion relation, numerically, and obtain the phase speed (or normalized frequency) $v/v_{si}\equiv\omega_r/\omega_{si}$, the damping rate to frequency ratio $|\gamma|/\omega_r$ and the damping time to period ratio  $\tau_D/T=\omega_r/(2\pi|\gamma|)$ of the slow surface sausage modes under the photospheric (magnetic poor) conditions. Our results show the following:

\begin{itemize}

\item For a given thickness of the inhomogeneous layer $l/R$, with increasing the twist parameter $B_{\phi_i}/B_{zi}$ (i) the minimum values of both the phase speed $v/v_{si}$ and the damping time to period ratio  $\tau_D/T=\omega_r/(2\pi|\gamma|)$ decrease and shift to smaller $k_zR$; (ii) the maximum value of $\left|\gamma\right|/\omega_r$ increases and moves to smaller $k_zR$.

\item For a given $l/R$, the phase speed and the damping rate to frequency ratio, approach $v/v_{si}\rightarrow v_{ci}/v_{si}=0.8638$ and $|\gamma|/\omega_r\rightarrow 0$, respectively, in the small wavelength limit ($k_z R \gg 1$). This asymptotic behaviours also hold for a given $k_zR$ in the limit of larger values of $l/R$.

\item  For a given $k_zR$, the maximum value of $\left|\gamma\right|/\omega_r$ (or minimum value of $\tau_D/T$) increases (or decreases) and moves to smaller $l/R$ when the twist parameter increases.

%\item For a given $k_zR$, the phase speed and the damping rate to frequency ratio, respectively, approach $v/v_{si}\rightarrow v_{ci}/v_{si}=0.8638$ and $|\gamma|/\omega_r\rightarrow 0$ in the limit of larger values of $l/R$.

\item For the case of $l/R=0.1$, the minimum value of $\tau_D/T$ for $B_{\phi_i}/B_{zi}=0.3$, for instance, changes $\sim38\%$ less than the case of no twist. Also for $k_zR=1$, the minimum value of $\tau_D/T$ for $B_{\phi_i}/B_{zi}=0.3$, for example, decreases $\sim63.5\%$ in comparison to the case of no twist. These results show that the magnetic twist can considerably affect the resonant absorption of the slow surface sausage modes in magnetic flux tubes under the photospheric conditions.

\end{itemize}

%----------------------------------------------------------------------------------------------------
\section*{Acknowledgements}
The authors thank Tom Van Doorsselaere and Marcel Goossens for reading the manuscript and useful discussions. The work of K. Karami has been supported financially
by Research Institute for Astronomy and Astrophysics of Maragha (RIAAM) under research project No. 1/5440-61.\\
%----------------------------------------------------------------------------------------------------

%----------------------------------------------------------------------------
\section*{Appendix}
\begin{appendix}
\section{Weak damping rate for the surface sausage mode}\label{ap1}

Here, with the help of Eq. (\ref{gammac1}) we obtain the damping rate of surface sausage modes in the weak damping limit, i.e. $\gamma \ll \omega_r$. To this aim, we first calculate  $\frac{\partial D_{AR}}{\partial \omega}$ from Eq. (\ref{limw1})  as follows
\begin{align}\label{div}
&\frac{\partial D_{AR}}{\partial \omega}=2 \rho_i \omega -2 \rho_e \omega \frac{k_{ri}}{k_{re}}Q- \rho_e \left(\omega-\omega_{Ae}^2\right)\left(\frac{1}{k_{re}}\frac{dk_{ri}}{d\omega}-\frac{k_{ri}}{k_{re}^2}\frac{dk_{re}}{d\omega}\right)Q-\rho_e \left(\omega-\omega_{Ae}^2\right)\frac{k_{ri}}{k_{re}}\frac{dQ}{d\omega}.
\end{align}
Now from Eq. (\ref{k2}), one can obtain
\begin{equation}\label{kk}
 \frac{dk_{ri}}{dw}=\frac{- \omega^3(\omega^2-2\omega_{ci}^2)}{(v_{si}^2+v_{Ai}^2)(\omega^2-\omega_{ci}^2)^2 k_{ri}},
\end{equation}
\begin{equation}\label{kk2}
 \frac{dk_{re}}{dw}=\frac{- \omega^3(\omega^2-2\omega_{ce}^2)}{(v_{se}^2+v_{Ae}^2)(\omega^2-\omega_{ce}^2)^2 k_{re}}.
\end{equation}
Also from Eq. (\ref{limw110}) for $\frac{dQ}{d\omega}$, we get
\begin{align}\label{div51}
&\frac{dQ}{d\omega}=\frac{Q}{x}\frac{dx}{dw}-x\frac{\frac{d}{d\omega}\left(\frac{\mu_0 D_e (1-v)-2B_{\phi e}^{2}n_{e}^{2}}{\mu_0 D_e  y} -\frac{K_{v-1}(y)}{K_{v}(y)}\right)}{\left[\frac{n_i+k_z}{k_z}s-2\frac{M(a,b-1;s)}{M(a,b;s)}\right]}\nonumber\\
&+x\frac{\left(\frac{\mu_0 D_e (1-v)-2B_{\phi e}^{2}n_{e}^{2}}{\mu_0 D_e  y} -\frac{K_{v-1}(y)}{K_{v}(y)}\right) \frac{d}{d\omega}\left[\frac{n_i+k_z}{k_z}s-2\frac{M(a,b-1;s)}{M(a,b;s)}\right]}{\left[\frac{n_i+k_z}{k_z}s-2\frac{M(a,b-1;s)}{M(a,b;s)}\right]^2}.
\end{align}
After some algebra, we obtain
\begin{align}\label{div52}
&\frac{dQ}{d\omega}=\nonumber\\
&\frac{Q}{x}\frac{dx}{dw}+x\frac{\left(\frac{1-\nu}{y^2}+\frac{2 v_{A\phi_e}^2 n_e^2}{\left(\omega^2-\omega_{Ae}^2\right) y^2}+\left(\frac{K_{v-1}'}{K_{v}}-\frac{K_{v}^' K_{v-1}}{K_{v}^2}\right)\right)\frac{dy}{d\omega}+\frac{1}{y}\frac{d\nu}{d\omega}-\frac{4 v_{A\phi_e}^2 }{\left(\omega^2-\omega_{Ae}^2\right) y}\left(n_e\frac{dn_e}{d\omega}- \frac{\omega n_e^2}{\left(\omega^2-\omega_{Ae}^2\right)} \right)}{\left[\frac{n_i+k_z}{k_z}s-2\frac{M(a,b-1;s)}{M(a,b;s)}\right]}\nonumber\\
&+x\frac{\left[\frac{s}{k_z}\frac{dn_i}{d\omega}+\frac{n_i+k_z}{k_z}ds-2\frac{d}{d\omega}\left(\frac{M(a,b-1;s)}{M(a,b;s)}\right)\right]\left(\frac{\mu_0 D_e (1-v)-2B_{\phi e}^{2}n_{e}^{2}}{\mu_0  D_e y} -\frac{K_{v-1}(y)}{K_{v}(y)}\right)}{\left[\frac{n_i+k_z}{k_z}s-2\frac{M(a,b-1;s)}{M(a,b;s)}\right]^2},
\end{align}
where $x=k_{ri}r_i$ and $y=k_{re}r_e$. This can be rewritten as
\begin{align}\label{div53}
&\frac{dQ}{d\omega}= P\frac{dx}{dw}+S\frac{dy}{d\omega},
\end{align}
where
\begin{align}\label{div64}
P=\frac{Q}{x}+x\frac{\left[\frac{s}{k_z}\frac{dn_i}{d\omega}+\frac{n_i+k_z}{k_z}ds-2\frac{d}{d\omega}\left(\frac{M(a,b-1;s)}{M(a,b;s)}\right)\right]\left(\frac{\mu_0 D_e (1-v)-2B_{\phi e}^{2}n_{e}^{2}}{\mu_0  D_e y} -\frac{K_{v-1}(y)}{K_{v}(y)}\right)}{\frac{dx}{dw}\left[\frac{n_i+k_z}{k_z}s-2\frac{M(a,b-1;s)}{M(a,b;s)}\right]^2},
\end{align}
\begin{align}\label{div65}
S=x\frac{\left(\frac{1-\nu}{y^2}+\frac{2 v_{A\phi_e}^2 n_e^2}{\left(\omega^2-\omega_{Ae}^2\right) y^2}+\left(\frac{K_{v-1}'}{K_{v}}-\frac{K_{v}^' K_{v-1}}{K_{v}^2}\right)\right)\frac{dy}{d\omega}+\frac{1}{y}\frac{d\nu}{d\omega}-\frac{4 v_{A\phi_e}^2 }{\left(\omega^2-\omega_{Ae}^2\right) y}\left(n_e\frac{dn_e}{d\omega}- \frac{\omega n_e^2}{\left(\omega^2-\omega_{Ae}^2\right)} \right)}{\frac{dy}{d\omega}\left[\frac{n_i+k_z}{k_z}s-2\frac{M(a,b-1;s)}{M(a,b;s)}\right]}.
\end{align}
In addition, from Eq. (\ref{n2}) we have
\begin{equation}\label{kk1}
 \frac{dn_{i}}{dw}=\frac{\omega^3(\omega^2-2\omega_{ci}^2)}{(v_{si}^2+v_{Ai}^2)(\omega^2-\omega_{ci}^2)^2n_{i}},
\end{equation}
\begin{equation}\label{kk3}
 \frac{dn_{e}}{dw}=\frac{\omega^3(\omega^2-2\omega_{ce}^2)}{(v_{se}^2+v_{Ae}^2)(\omega^2-\omega_{ce}^2)^2n_{e}}.
\end{equation}
With the help of Eqs. (\ref{kk}) and (\ref{kk2}), Eq. (\ref{div53}) takes the form
\begin{align}\label{div57}
&\frac{dQ}{d\omega}= xP\frac{\omega^3(\omega^2-2\omega_{ci}^2)}{(\omega_{si}^2-\omega^2)(\omega_{Ai}^2-\omega^2)(\omega^2-\omega_{ci}^2)}+yS\frac{ \omega^3(\omega^2-2\omega_{ce}^2)}{(\omega_{se}^2-\omega^2)(\omega_{Ae}^2-\omega^2)(\omega^2-\omega_{ce}^2) }.
\end{align}
Replacing this into Eq. (\ref{div}) yields
\begin{align}\label{di}
&\frac{\partial D_{AR}}{\partial \omega}=2 \rho_i \omega -2 \rho_e \omega \frac{k_{ri}}{k_{re}}Q-\rho_e \omega^3 \left(\omega-\omega_{Ae}^2\right)\frac{k_{ri}}{k_{re}}\nonumber\\
&\left(\frac{(Q+xP)(\omega^2-2\omega_{ci}^2)}{(\omega_{si}^2-\omega^2)(\omega_{Ai}^2-\omega^2)(\omega^2-\omega_{ci}^2)}-\frac{ (Q-yS)(\omega^2-2\omega_{ce}^2)}{(\omega_{se}^2-\omega^2)(\omega_{Ae}^2-\omega^2)(\omega^2-\omega_{ce}^2) }\right).
\end{align}
Finally, substituting Eqs. (\ref{limw2}) and (\ref{di}) into Eq. (\ref{gammac1}) one can get the damping rate $\gamma$ in the limit of weak damping for the surface sausage modes in the slow continuum as
\begin{align}\label{gammacwd}
\gamma\Big|_{\omega=\omega_{r}}=-\frac{\frac{ \pi \rho_e k_z^2}{k_{re}\rho_c|\Delta_c| }\Big|_{r=r_c}\left(\frac{v_{s}^2}{v_{A}^2+v_{s}^2}\right)^2 \left((\omega^2-\omega_{Ai}^2)+Z\right)(\omega^2-\omega_{Ae}^2) G}{2\omega\left(1-\chi\frac{k_{ri}}{k_{re}}Q\right)-\omega \chi T},
\end{align}
where
\begin{align}\label{di11}
&T=\omega_{r}^2 \left(\omega_{r}^2-\omega_{Ae}^2\right)\frac{k_{ri}}{k_{re}}\left(\frac{(Q+xP)(\omega_{r}^2-2\omega_{ci}^2)}{(\omega_{si}^2-\omega_{r}^2)(\omega_{Ai}^2-\omega_{r}^2)(\omega_{r}^2-\omega_{ci}^2)}-\frac{ (Q-yS)(\omega_{r}^2-2\omega_{ce}^2)}{(\omega_{se}^2-\omega_{r}^2)(\omega_{Ae}^2-\omega_{r}^2)(\omega_{r}^2-\omega_{ce}^2) }\right).
\end{align}

%----------------------------------------------------------------------------------------------------
\section{Weak damping rate in long wavelength limit}\label{ap2}

Here, we turn to examine the dispersion relation (\ref{gammacwd}) in the long wavelength limit, i.e. ${k_zR}\ll 1$. In this limit, Eq. (\ref{nu}) yields $\nu^2=1+O(k_z^2R^2)\simeq 1$. Also following  \cite{{Abramowitz}} we have
\begin{equation}\label{kum}
\lim_{k_zR\to 0} \frac{M(a,b-1 ; s)}{M(a,b ; s)}=1+\frac{a}{b}~s+O(s^2),
\end{equation}
\begin{equation}\label{bes}
\lim_{k_zR\to 0} \frac{K_0(y)}{K_1(y)}=-y \ln(y).
\end{equation}
In the limit ${k_zR}\ll 1$ ($\omega_r\approx \omega_{ci}$), one should note that the damping rate (\ref{gammacwd}) at $\omega=\omega_{r}\approx \omega_{ci}$ becomes singular. To avoid of this singularity, we follow the approach of  \cite{TYu} in which one can assume $\omega^2_r=\omega_{ci}^2-\alpha$, where $\alpha\ll \omega_{ci}^2$. Substituting $\omega_{ci}^2-\omega^2_r=\alpha$ into Eq. (\ref{gammacwd}) and using Eqs. (\ref{kum}) and (\ref{bes}), one can obtain
\begin{align}\label{gammacwd22}
\gamma=-\frac{ \pi \rho_e k_z}{|\Delta_c| }\frac{\omega_{ci}^4}{\omega_{Ai}^4}\frac{ \left((\omega_{ci}^2-\omega_{Ai}^2)+Z\right)(\omega_{ci}^2-\omega_{Ae}^2) G}{2\omega_{ci} \left(\rho_i-\rho_e\frac{k_{ri}}{k_{re}}Q\right)-\omega_{ci} \rho_e T},
\end{align}
where we have used $r_i\approx R$ and $k_{re}=k_z$. Also
\begin{align}\label{di22}
&T=\omega_{ci}^2 \left(\omega_{ci}^2-\omega_{Ae}^2\right)\frac{k_{ri}}{k_{re}}\left(\frac{(Q+xP)\omega_{ci}^2}{(\omega_{si}^2-\omega_{ci}^2)(\omega_{Ai}^2-\omega_{ci}^2)\alpha}-\frac{ (Q-yS)(\omega_{ci}^2-2\omega_{ce}^2)}{(\omega_{se}^2-\omega_{ci}^2)(\omega_{Ae}^2-\omega_{ci}^2)(\omega_{ci}^2-\omega_{ce}^2) }\right),
\end{align}
\begin{dmath}\label{limw1101}
Q=-x
  \frac{\left(\frac{-2B_{\phi e}^{2}n_{e}^{2}}{\mu_0 D_e  y}+y\ln(y)\right)}{\left[\frac{n_i}{k_z}s-2\left(1+\frac{a}{2}s\right)\right]},
\end{dmath}
\begin{dmath}\label{limw1111}
G=\left(-\frac{2k_{re} B_{\phi }^2}{D_e \mu_0 r}\Big|_{r=r_c}-\frac{2B_{\phi e}^{2}n_{e}^{2}}{\mu_0 y D_e} +y\ln(y)\right),
\end{dmath}
\begin{dmath}\label{limw1121}
Z=\frac{2 B_{\phi i}^2k_{ri}^{2}}{\mu_0\rho_i\left[\frac{n_i}{k_z}s-2\left(1+\frac{a}{2}s\right)\right]},
\end{dmath}
\begin{align}\label{div641}
P=\frac{Q}{x}+x\frac{\left[\frac{s}{k_z}\frac{dn_i}{d\omega}+\frac{n_i}{k_z}ds-2\frac{d}{d\omega}\left(1+\frac{a}{2}s\right)\right]\left(\frac{-2B_{\phi e}^{2}n_{e}^{2}}{\mu_0  D_e y}+ y\ln(y)\right)}{\frac{dx}{dw}\left[\frac{n_i}{k_z}s-2\left(1+\frac{a}{2}s\right)\right]^2},
\end{align}
\begin{align}\label{div651}
S=x\frac{\left(\frac{2 v_{A\phi_e}^2 n_e^2}{\left(\omega_{ci}^2-\omega_{Ae}^2\right) y^2}-\frac{x}{2}\left(1+\ln(y)\right)\right)\frac{dy}{d\omega}-\frac{4 v_{A\phi_e}^2 }{\left(\omega_{ci}^2-\omega_{Ae}^2\right) y}\left(n_e\frac{dn_e}{d\omega}- \frac{\omega_{ci} n_e^2}{\left(\omega_{ci}^2-\omega_{Ae}^2\right)} \right)}{\frac{dy}{d\omega}\left[\frac{n_i}{k_z}s-2\left(1+\frac{a}{b}s\right)\right]}.
\end{align}
Now from Eqs. (\ref{s}), (\ref{a}) and using (\ref{ka}), one can obtain
\begin{align}\label{s1}
s=2\frac{v_{A\phi i}^2 n_i k_z}{\left(\omega_{ci}^2-\omega_{Ai}^2\right)},
\end{align}
\begin{align}\label{a1}
a=1+\frac{x^2\left[r_i^2\left(\omega_{ci}^2-\omega_{Ai}^2\right)^2-4v_{A\phi i}^2\omega_{Ai}^2\right]}{8v_{A\phi i}^2 n_i k_zr_i^2\left(\omega_{ci}^2-\omega_{Ai}^2\right)},
\end{align}
where $v^2_{A\phi}\equiv\frac{B^2_{\phi }}{\mu_0 \rho}$. With the help of Eqs. (\ref{s1}) and (\ref{a1}), one can get
\begin{align}\label{as}
&as=\frac{x^2}{4}+v_{A\phi i}^2\left(\frac{2n_i k_z\left(\omega_{ci}^2-\omega_{Ai}^2\right)-x^2r_i^{-2}\omega_{Ai}^2}{\left(\omega_{ci}^2-\omega_{Ai}^2\right)^2}\right),
\end{align}
\begin{align}
&\frac{da}{d\omega}=\frac{1}{8v_{A\phi i}^2 n_i k_z}\Bigg[2x\left(\omega_{ci}^2-\omega_{Ai}^2\right)\frac{dx}{d\omega}+x^2\left(2\omega_{ci}-\frac{\left(\omega_{ci}^2-\omega_{Ai}^2\right)}{n_i}\frac{dn_i}{d\omega}\right)\nonumber\\
&+4v_{A\phi i}^2\omega_{Ai}^2x^2r_i^{-2}\left(-\frac{2}{  \left(\omega_{ci}^2-\omega_{Ai}^2\right)}\frac{dx}{d\omega}+\frac{2\omega_{ci}}{\left(\omega_{ci}^2-\omega_{Ai}^2\right)^2}+\frac{1}{n_i \left(\omega_{ci}^2-\omega_{Ai}^2\right)}\frac{dn_i}{d\omega}\right)\Bigg],
\end{align}
\begin{align}
&\frac{ds}{d\omega}=2v_{A\phi i}^2 n_i k_z\left(\frac{1 }{n_i \left(\omega_{ci}^2-\omega_{Ai}^2\right)}\frac{dn_i}{d\omega}-\frac{2\omega _{ci}}{\left(\omega^2-\omega_{Ai}^2\right)^2}\right),
\end{align}
\begin{align}\label{sdadw}
&s\frac{da}{d\omega}=\frac{1}{4}\Bigg[2x\frac{dx}{d\omega}+\frac{x^2}{\left(\omega_{ci}^2-\omega_{Ai}^2\right)}\left(2\omega_{ci}-\frac{\left(\omega_{ci}^2-\omega_{Ai}^2\right)}{n_i}\frac{dn_i}{d\omega}\right)\nonumber\\
&+4v_{A\phi i}^2\omega_{Ai}^2x^2r_i^{-2}\left(-\frac{2}{  \left(\omega_{ci}^2-\omega_{Ai}^2\right)^2}\frac{dx}{d\omega}+\frac{2\omega_{ci}}{\left(\omega_{ci}^2-\omega_{Ai}^2\right)^3}+\frac{1}{n_i \left(\omega_{ci}^2-\omega_{Ai}^2\right)^2}\frac{dn_i}{d\omega}\right)\Bigg],
\end{align}
\begin{align}\label{adsdw}
&a\frac{ds}{d\omega}=x^2\left(\frac{1}{4n_i}\frac{dn_i}{d\omega}-\frac{\omega_{ci}}{2\left(\omega_{ci}^2-\omega_{Ai}^2\right)}\right)+2v_{A\phi i}^2 n_i k_z\left(\frac{1 }{n_i \left(\omega_{ci}^2-\omega_{Ai}^2\right)}\frac{dn_i}{d\omega}-\frac{2\omega_{ci} }{\left(\omega_{ci}^2-\omega_{Ai}^2\right)^2}\right)\nonumber\\
&-v_{A\phi i}^2\omega_{Ai}^2x^2r_i^{-2}\left(\frac{1 }{n_i \left(\omega_{ci}^2-\omega_{Ai}^2\right)^2}\frac{dn_i}{d\omega}-\frac{2\omega_{ci} }{\left(\omega_{ci}^2-\omega_{Ai}^2\right)^3}\right).
\end{align}
Combining Eqs. (\ref{sdadw}) and (\ref{adsdw}) yields
\begin{align}\label{sdadw-adsdw}
&s\frac{da}{d\omega}+a\frac{ds}{d\omega}=\frac{x}{2}\frac{dx}{d\omega}+4v_{A\phi i}^2\omega_{Ai}^2x^2r_i^{-2}\left(-\frac{2}{  \left(\omega_{ci}^2-\omega_{Ai}^2\right)^2}\frac{dx}{d\omega}+\frac{9\omega_{ci}}{4\left(\omega_{ci}^2-\omega_{Ai}^2\right)^3}+\frac{3}{4n_i \left(\omega_{ci}^2-\omega_{Ai}^2\right)^2}\frac{dn_i}{d\omega}\right)\nonumber\\
&+2v_{A\phi i}^2 n_i k_z\left(\frac{1 }{n_i \left(\omega_{ci}^2-\omega_{Ai}^2\right)}\frac{dn_i}{d\omega}-\frac{2\omega_{ci} }{\left(\omega_{ci}^2-\omega_{Ai}^2\right)^2}\right).
\end{align}
In addition, we need to evaluate the quantity $\alpha$. To this aim, following  \cite{TYu} we first replace $\omega^2=\omega_{ci}^2-\alpha$ into Eq. (\ref{k2}) and get
\begin{equation}\label{ki}
k_{ri}^2\simeq \frac{k_z^2}{\alpha} \frac{\left(\omega_{ci}^2-\omega_{si}^2\right)\left(\omega_{ci}^2-\omega_{Ai}^2\right)}{\left(\omega_{Ai}^2+\omega_{si}^2\right)}
=\frac{k_z^2}{\alpha}\frac{\omega_{ci}^6}{\omega_{si}^2\omega_{Ai}^2},
\end{equation}
where we have used the definition $\omega_c^2\equiv \frac{\omega_s^2\omega_A^2}{\omega_s^2+\omega_A^2}$ in obtaining the second equality of the above relation.
In the next, the dispersion relation (\ref{dispersionrelation}) in long wavelength limit (${k_zR}\ll 1$) reads
\begin{dmath}\label{disp}
\frac{(\omega_{ci}^2-\omega_{Ai}^2)}{ k^2_{ri}}\left[\left(\frac{n_i}{k_z}\right)s-2\left(1+\frac{a}{b}s\right)\right]=-
\chi(\omega_{ci}^2-\omega_{Ae}^2) R^2 \ln(y)+\left(1+\frac{ 2  n^2_e}{ k^2_{re}}\right)\frac{B_{\phi e}^2}{\mu_0\rho_i}
-
\frac{B^2_{\phi i}}{\mu_0\rho_i}.
\end{dmath}
Now, replacing $k_{ri}^2$ from Eq. (\ref{ki}) into (\ref{disp}), the quantity $\alpha$ can be obtained as follows
\begin{dmath}\label{alpha}
\alpha=-\frac{k_z^2\omega_{ci}^4}{\omega_{Ai}^4}\left(\frac{\chi(\omega^2-\omega_{Ae}^2) R^2 \ln(y)-\left(1+\frac{ 2  n^2_e}{ k^2_{re}}\right)v_{A\phi e}^2
+
v_{A\phi i}^2}{\left[\left(\frac{n_i}{k_z}\right)s-2\left(1+\frac{a}{b}s\right)\right]}\right).
\end{dmath}
Substituting this into Eq. (\ref{ki}) yields
\begin{dmath}\label{kii}
k_{ri}^2=\frac{\omega_{ci}^2\omega_{Ai}^2}{\omega_{si}^2}\frac{\left[\left(\frac{n_i}{k_z}\right)s-2\left(1+\frac{a}{b}s\right)\right]}{\left(\chi(\omega^2-\omega_{Ae}^2) R^2 \ln(y)-\left(1+\frac{ 2  n^2_e}{ k^2_{re}}\right)v_{A\phi e}^2
+
v_{A\phi i}^2\right)}.
\end{dmath}
Under photospheric (magnetic pore) conditions, i.e. $v_{Ae}=v_{A\phi e}=v_{ce}\simeq 0$, the weak damping rate $\gamma$, Eq. (\ref{gammacwd22}), in long wavelength limit reduces to
\begin{align}\label{gammacwd3}
\gamma=-\frac{ \pi \rho_e k_zR}{|\Delta_c| }\frac{\omega_{ci}^5}{\omega_{Ai}^4}\frac{ \left((\omega_{ci}^2-\omega_{Ai}^2)+Z\right)G}{2\left(\rho_i-\rho_e\frac{k_{ri}}{k_{re}}Q\right)- \rho_e T}.
\end{align}
Besides, Eqs. (\ref{di22}), (\ref{limw1101}), (\ref{limw1111}), (\ref{limw1121}), (\ref{div641}), (\ref{div651}), (\ref{alpha}) and (\ref{kii}) take the forms
\begin{align}\label{di2210}
&T=\omega_{ci}^2\frac{x}{y}\left(\frac{(Q+xP)}{\alpha}-\frac{ (Q-yS)}{(\omega_{ci}^2-\omega_{se}^2)}\right),
\end{align}
\begin{dmath}\label{limw11011}
Q=
  \frac{-xy\ln(y)}{\left[\frac{n_i}{k_z}s-2\left(1+\frac{a}{2}s\right)\right]}=\frac{xy\ln(y)}{2}\left(1-\frac{a}{2}s+\frac{v_{A\phi i}^2 n_i^2}{\omega_{ci}^2-\omega_{Ai}^2}\right)=\frac{xy\ln(y)}{2}\left(1-v_{A\phi i}^2\left(\frac{2\left(n_i k_z-n_i^2\right)\left(\omega_{ci}^2-\omega_{Ai}^2\right)-x^2r_i^{-2}\omega_{Ai}^2}{2\left(\omega_{ci}^2-\omega_{Ai}^2\right)^2}\right)\right),
\end{dmath}
\begin{dmath}\label{limw1112}
G=\left(-\frac{2k_{re} v_{A\phi }^2}{\chi \omega_{ci}^2 R}+y\ln(y)\right),
\end{dmath}
\begin{dmath}\label{limw11211}
Z=\frac{2 v_{A\phi i}^2k_{ri}^{2}}{\left[\frac{n_i}{k_z}s-2\left(1+\frac{a}{2}s\right)\right]}=-v_{A\phi i}^2k_{ri}^{2}\left(1-\frac{a}{2}s+\frac{v_{A\phi i}^2 n_i^2}{\omega_{ci}^2-\omega_{Ai}^2}\right)=-v_{A\phi i}^2k_{ri}^{2}\left(1-v_{A\phi i}^2\left(\frac{2\left(n_i k_z-n_i^2\right)\left(\omega_{ci}^2-\omega_{Ai}^2\right)-x^2r_i^{-2}\omega_{Ai}^2}{2\left(\omega_{ci}^2-\omega_{Ai}^2\right)^2}\right)\right),
\end{dmath}
\begin{dmath}\label{div6411}
P=\frac{Q}{x}+x\frac{\left[\frac{s}{k_z}\frac{dn_i}{d\omega}+\frac{n_i}{k_z}ds-2\frac{d}{d\omega}\left(1+\frac{a}{2}s\right)\right] y\ln(y)}{\frac{dx}{dw}\left[\frac{n_i}{k_z}s-2\left(1+\frac{a}{2}s\right)\right]^2}\nonumber\\
=\frac{y\ln(y)}{2}\left(1-\frac{a}{2}s+\frac{v_{A\phi i}^2 n_i^2}{(\omega_{ci}^2-\omega_{Ai}^2)}\right)+\frac{xy\ln(y)}{2}\left(1-as+\frac{2v_{A\phi i}^2 n_i^2}{(\omega_{ci}^2-\omega_{Ai}^2)}\right)\frac{d}{d\omega}\left(\frac{v_{A\phi i}^2 n_i^2}{(\omega_{ci}^2-\omega_{Ai}^2)}-\frac{a}{2}s\right)=\frac{y\ln(y)}{2}\left(1-\frac{x^2}{8}-v_{A\phi i}^2\left(\frac{2\left(n_i k_z-n_i^2\right)\left(\omega_{ci}^2-\omega_{Ai}^2\right)-x^2r_i^{-2}\omega_{Ai}^2}{2\left(\omega_{ci}^2-\omega_{Ai}^2\right)^2}\right)\right)+\frac{xy\ln(y)}{2}\left(1-v_{A\phi i}^2\left(\frac{2\left(n_i k_z-n_i^2\right)\left(\omega_{ci}^2-\omega_{Ai}^2\right)-x^2r_i^{-2}\omega_{Ai}^2}{2\left(\omega_{ci}^2-\omega_{Ai}^2\right)^2}\right)\right)\\
\left(\frac{2v_{A\phi i}^2 n_i}{(\omega_{ci}^2-\omega_{Ai})}\frac{dn_i}{d\omega}-\frac{2v_{A\phi i}^2 \omega_{ci} n_i^2}{(\omega_{ci}^2-\omega_{Ai}^2)}-\frac{sda+ads}{2}\right),
\end{dmath}
\begin{align}\label{div6511}
S&=-x\frac{1+\ln(y)}{\left[\frac{n_i}{k_z}s-2\left(1+\frac{a}{2}s\right)\right]}\nonumber\\
&=x\left(1+\ln(y)\right)\left(\frac{1}{2}-\frac{a}{4}s+\frac{v_{A\phi i}^2 n_i^2}{2(\omega_{ci}^2-\omega_{Ai}^2)}\right)\nonumber\\
&=\frac{x\left(1+\ln(y)\right)}{2}\left(1-v_{A\phi i}^2\left(\frac{2\left(n_i k_z-n_i^2\right)\left(\omega_{ci}^2-\omega_{Ai}^2\right)-x^2r_i^{-2}\omega_{Ai}^2}{2\left(\omega_{ci}^2-\omega_{Ai}^2\right)^2}\right)\right),
\end{align}
\begin{align}\label{alpha7}
&\alpha=-\frac{k_z^2\omega_{ci}^4}{\omega_{Ai}^4}\left(\frac{\chi\omega_{ci}^2 R^2 \ln(y)
+
v_{A\phi i}^2}{\left[\left(\frac{n_i}{k_z}\right)s-2\left(1+\frac{a}{b}s\right)\right]}\right)\nonumber\\
&=\frac{k_z^2\omega_{ci}^4}{\omega_{Ai}^4}\left(\chi\omega_{ci}^2 R^2 \ln(y)
+
v_{A\phi i}^2\right)\left(\frac{1}{2}-\frac{a}{4}s+\frac{v_{A\phi i}^2 n_i^2}{2(\omega_{ci}^2-\omega_{Ai}^2)}\right)\nonumber\\
&=\frac{k_z^2\omega_{ci}^4}{\omega_{Ai}^4}\left(\chi\omega_{ci}^2 R^2 \ln(y)
+
v_{A\phi i}^2\right)\left(1-v_{A\phi i}^2\left(\frac{2\left(n_i k_z-n_i^2\right)\left(\omega_{ci}^2-\omega_{Ai}^2\right)-x^2r_i^{-2}\omega_{Ai}^2}{2\left(\omega_{ci}^2-\omega_{Ai}^2\right)^2}\right)\right),
\end{align}
\begin{align}\label{kii7}
k_{ri}^2&=\frac{\omega_{ci}^2\omega_{Ai}^2}{\omega_{si}^2}\frac{\left[\left(\frac{n_i}{k_z}\right)s-2\left(1+\frac{a}{b}s\right)\right]}{\left(\chi\omega_{ci}^2 R^2 \ln(y)
+
v_{A\phi i}^2\right)}\nonumber\\
&=-\frac{2\omega_{ci}^2\omega_{Ai}^2}{\omega_{si}^2}\frac{1}{\left(\chi\omega_{ci}^2 R^2 \ln(y)
+
v_{A\phi i}^2\right)}\left(1+\frac{a}{2}s-\frac{v_{A\phi i}^2 n_i^2}{(\omega_{ci}^2-\omega_{Ai}^2)}\right)\nonumber\\
&=-\frac{2\omega_{ci}^2\omega_{Ai}^2}{\omega_{si}^2}\frac{1}{\left(\chi\omega_{ci}^2 R^2 \ln(y)
+
v_{A\phi i}^2\right)}\left(1+v_{A\phi i}^2\left(\frac{2\left(n_i k_z-n_i^2\right)\left(\omega_{ci}^2-\omega_{Ai}^2\right)-x^2r_i^{-2}\omega_{Ai}^2}{2\left(\omega_{ci}^2-\omega_{Ai}^2\right)^2}\right)\right).
\end{align}
Finally, substituting Eqs. (\ref{di2210}) to (\ref{kii7}) into (\ref{gammacwd3}) gives a long analytical expression for the weak damping $\gamma$ in long wavelength limit for the  photospheric conditions.\\

%----------------------------------------------------------------------------------------------------

\section{Weak damping rate in long wavelength limit with no twist}\label{ap3}

In the limit of no twist, i.e. $B_{\phi i}=B_{\phi e}=0$, Eqs. (\ref{as}) and (\ref{sdadw-adsdw}) read
\begin{align}
&as=\frac{x^2}{4},
\end{align}
\begin{align}
&s\frac{da}{d\omega}+a\frac{ds}{d\omega}=\frac{x}{2}\frac{dx}{d\omega}.
\end{align}
Substituting the above relations into Eqs. (\ref{limw1101}) to (\ref{div651}), (\ref{alpha}) and (\ref{kii}) one can get
\begin{dmath}\label{limw110111}
Q=\frac{xy}{2}\ln(y),
\end{dmath}
\begin{dmath}\label{limw11111}
G=y\ln(y),
\end{dmath}
\begin{dmath}\label{limw11212}
Z=0,
\end{dmath}
\begin{align}\label{div6411}
P=\left(\frac{1}{2}-\frac{3x^2}{16}\right)y\ln(y),
\end{align}
\begin{align}\label{div6511}
S=\frac{x}{2}\left(1+\ln(y)\right).
\end{align}
\begin{dmath}\label{alpha1}
\alpha=\frac{\chi \omega_{ci}^4}{2\omega_{Ai}^4}(\omega_{ci}^2-\omega_{Ae}^2) k_z^2 R^2 \ln(y),
\end{dmath}
\begin{dmath}\label{ki1}
k_{ri}^2=-2\frac{\omega_{ci}^2\omega_{Ai}^2}{\chi\omega_{si}^2(\omega_{ci}^2-\omega_{Ae}^2) R^2 \ln(y)}.
\end{dmath}
Inserting Eqs. (\ref{limw110111}), (\ref{limw11111}), (\ref{limw11212}) into (\ref{gammacwd22}) gives
\begin{align}\label{gammacwd221}
\gamma=-\frac{ \pi \rho_e k_z^2}{|\Delta_c|  }\frac{\omega_{ci}^3}{\omega_{Ai}^4}\frac{ \left((\omega_{ci}^2-\omega_{Ai}^2)\right)(\omega_{ci}^2-\omega_{Ae}^2)R \ln(y)}{2\left(\rho_i-\rho_e\frac{x^2}{2}\ln(y)\right)-\rho_e T}.
\end{align}
Putting Eqs. (\ref{div6411}) and (\ref{div6511}) into (\ref{di22}), one can get
\begin{align}\label{di221}
&T=\omega_{ci}^2 \left(\omega_{ci}^2-\omega_{Ae}^2\right)\Bigg(\frac{x^2 \ln(y)\omega_{ci}^2}{(\omega_{si}^2-\omega_{ci}^2)(\omega_{Ai}^2-\omega_{ci}^2)\alpha}-\frac{3x^4 \ln(y)\omega_{ci}^2}{16(\omega_{si}^2-\omega_{ci}^2)(\omega_{Ai}^2-\omega_{ci}^2)\alpha}\nonumber\\
&+\frac{ x^2(\omega_{ci}^2-2\omega_{ce}^2)}{2(\omega_{se}^2-\omega_{ci}^2)(\omega_{Ae}^2-\omega_{ci}^2)(\omega_{ci}^2-\omega_{ce}^2) }\Bigg).
\end{align}
Replacing Eqs. (\ref{alpha1}) and (\ref{ki1}) into (\ref{di221}) yields
\begin{align}\label{di2213}
&T=-\frac{4\omega_{Ai}^6}{\chi^2 \omega_{ci}^2\omega_{si}^2\left(\omega_{ci}^2-\omega_{Ae}^2\right)k_z^2R^2\ln(k_zR)}-\frac{3\omega_{Ai}^8}{2\chi^3 \omega_{si}^4\left(\omega_{ci}^2-\omega_{Ae}^2\right)^2k_z^2R^2\ln^2(k_zR)}\nonumber\\
&-\frac{ \omega_{ci}^4\omega_{Ai}^2(\omega_{ci}^2-2\omega_{ce}^2)}{\chi \omega_{si}^2(\omega_{se}^2-\omega_{ci}^2)(\omega_{Ae}^2-\omega_{ci}^2)(\omega_{ci}^2-\omega_{ce}^2) \ln(k_zR)}.
\end{align}
Note that in long wavelength limit (${k_zR}\ll 1$), the third term appeared in Eq. (\ref{di2213}) is small in comparison to the first two ones. Hence, Eq. (\ref{di2213}) reduces to
\begin{align}\label{di2214}
&T=-\frac{4\omega_{Ai}^6}{\chi^2 \omega_{ci}^2\omega_{si}^2\left(\omega_{ci}^2-\omega_{Ae}^2\right)k_z^2R^2\ln(k_zR)}-\frac{3\omega_{Ai}^8}{2\chi^3 \omega_{si}^4\left(\omega_{ci}^2-\omega_{Ae}^2\right)^2k_z^2R^2\ln^2(k_zR)}.
\end{align}
Finally, substituting Eq. (\ref{di2214}) into (\ref{gammacwd221}) gives the weak damping rate in long wavelength limit with no twist as
\begin{align}\label{gammacwd21121}
\gamma=\frac{ 2\pi \chi^3 }{|\Delta_c| R}\left[\frac{ \omega_{ci}^7\omega_{si}^2\left(\omega_{ci}^2-\omega_{Ae}^2\right)^3 }{3\omega_{Ai}^{10} \omega_{ci}^2+8\chi \omega_{Ai}^8 \omega_{si}^2\left(\omega_{ci}^2-\omega_{Ae}^2\right)\ln(k_zR)}\right](k_zR)^4\ln^3(k_zR).
\end{align}

\section{Dimensionless dispersion relation}\label{ap4}
In order to numerical solving the dispersion relation (\ref{dispersionrelationthin}), we recast it in the following dimensionless form

\begin{align}\label{dim}
&\frac{v_F^2-v_{AI}^2}{k_{rI}^2}\left[\left(N_I+1\right)s-2\frac{M(a,b-1;s)}{M(a,b;s)}\right]\nonumber\\
&+r_i \left[\frac{\left(1-\nu\right)\chi \left(v_F^2-v_{AE}^2\right)-2 \chi v_{A \phi E}^2 N_E^2}{r_e k_{rE}^2}-\frac{\chi k_z \left(v_F^2-v_{AE}^2\right)K_{\nu-1}\left(k_{rE}~k_zr_e\right)}{k_{rE}K_{\nu}\left(k_{rE}~k_zr_e\right)}\right]\nonumber\\
&+\frac{i \pi k_z^2 v_c^4 v_{Si}^2}{\left|\Delta_c\right|v_A^4 }\left(\frac{v_F^2-v_{AI}^2}{k_{rI}^2}\left[\left(N_I+1\right)s-2\frac{M(a,b-1;s)}{M(a,b;s)}\right]+2 v_{A \phi i}^2\right)\nonumber\\
&\times \left(2\frac{(1+\delta(\chi-1))v_{A \phi c}^2}{r_c} +\frac{\left(1-\nu\right)\chi \left(v_F^2-v_{AE}^2\right)-2 \chi v_{A \phi E}^2 N_E^2}{r_e k_{rE}^2}-\frac{\chi k_z \left(v_F^2-v_{AE}^2\right)K_{\nu-1}\left(k_{rE}~ k_zr_e\right)}{k_{rE}K_{\nu}\left(k_{rE}~k_z r_e\right)}\right)=0,
\end{align}
where
\begin{align}
&a=1+\frac{k_{rI}^2\left[k_{z}^2\left(v_F^2-v_{AI}^2\right)^2-4v_{A\phi I}^2v_{AI}^2\right]}{8v_{A\phi I}^2 N_I\left(v_F^2-v_{AI}^2\right)},~~~~~s=2\frac{v_{A\phi I}^2 N_I}{\left(v_F^2-v_{AI}^2\right)},\nonumber\\
&v_F=\frac{\omega_r}{\omega_{si}}=\frac{v}{v_{si}},~~~~~ v_{AE}=\frac{v_{Ae}}{v_{si}},~~~~~ v_{AI}=\frac{v_{Ai}}{v_{si}},~~~~~b=2,~~~~~\chi=\frac{\rho_e}{\rho_i},\nonumber\\
&v_{SE}=\frac{v_{se}}{v_{si}} ,~~~~~ v_{A\phi E}=\frac{v_{A \phi e}}{v_{si}},~~~~~ v_{A\phi I}=\frac{v_{A \phi i}}{v_{si}},~~~~~{v_{SI}}=1,~~~~~v^2_{A\phi}=\frac{B_{\phi}^2}{\mu_0\rho},\nonumber\\
&k_{re}^2=k_z^2 k_{rE}^2,~~~~~k_{ri}^2=k_z^2 k_{rI}^2,~~~~~n_{e}^2=k_z^2 N_{E}^2,~~~~~n_{i}^2=k_z^2 N_{I}^2,
\end{align}
and
\begin{align}
&N_I^2=\frac{v_F^4}{v_F^2 v_{AI}^2+\left(v_F^2-1\right)},~~~~~N_E^2=\frac{v_F^4}{v_F^2 v_{AE}^2+v_{SE}^2\left(v_F^2-1\right)},\nonumber\\
&k_{rI}^2=\frac{\left(1-v_F^2\right)\left(v_{AI}^2-v_F^2\right)}{\left(1+v_{AI}^2\right)\left(v_{cI}^2-v_{v_{F}}^2\right)},~~~~~
k_{rE}^2=\frac{\left(v_{SE}^2-v_F^2\right)\left(v_{AE}^2-v_F^2\right)}{\left(v_{SE}^2+v_{AE}^2\right)\left(v_{cE}^2-v_{v_{F}}^2\right)},\nonumber\\
&v_{cE}=\frac{v_{ce}}{v_{si}},~~~~~v_{cI}=\frac{v_{ci}}{v_{si}},\nonumber\\
&\nu^2=1+2\frac{v_{A\phi E}^2}{\left(v_{F}^2-v_{AE}^2\right)}\left[2v_{A\phi E}^2 N_E^2+ \left(v_{AE}^2\left(3N_E^2-1\right)-v_F^2\left(N_E^2+1\right)\right)\right].
\end{align}
Note that in the absence of inhomogeneous layer, Eq. (\ref{dim}) reduces to
\begin{align}\label{dispersionrelation-dimless}
&\frac{v_F^2-v_{AI}^2}{k_{rI}^2}\left[\left(N_I+1\right)s-2\frac{M(a,b-1;s)}{M(a,b;s)}\right]\nonumber\\
&= \frac{\left(1-\nu\right)\chi \left(v_F^2-v_{AE}^2\right)-2 \chi v_{A \phi i}^2 N_E^2}{r_e k_{rE}^2}-\frac{\chi k_z \left(v_F^2-v_{AE}^2\right)R K_{\nu-1}\left(k_{rE}~k_zR\right)}{k_{rE}K_{\nu}\left(k_{rE}~k_zR\right)}+v_{A \phi I}^2+v_{A \phi E}^2.
\end{align}
\end{appendix}
%----------------------------------------------------------------------------------------------------
%----------------------------------------------------------------------------------------------------

%----------------------------------------------------------------------------------------------------
\begin{figure}
    \centering
    \begin{subfigure}[b]{0.5\textwidth}
        \centering
        \includegraphics[width=\textwidth]{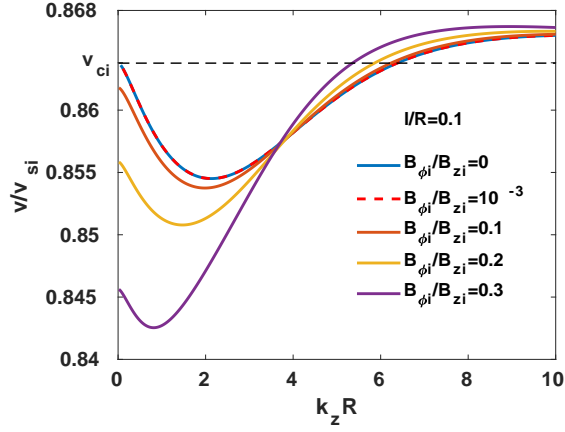}
      \caption{}
        \label{5al0.1t}
    \end{subfigure}
    \vfill
    \begin{subfigure}[b]{0.5\textwidth}
        \centering
        \includegraphics[width=\textwidth]{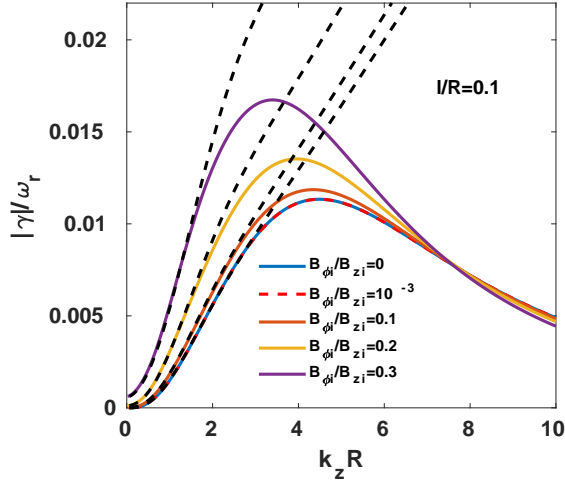}
       \caption{}
        \label{5bl0.1t}
    \end{subfigure}
    \vfill
    \begin{subfigure}[b]{0.5\textwidth}
        \centering
        \includegraphics[width=\textwidth]{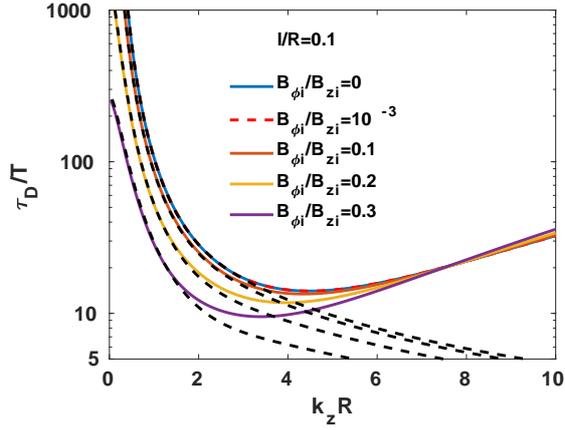}
        \caption{}
        \label{5cl0.1t}
    \end{subfigure}
    \vfill

    \caption{(a) The phase speed $v/v_{si}\equiv\omega_r/\omega_{si}$, (b) the damping rate to frequency ratio $\left|\gamma\right|/\omega_r$, and (c) the damping time to period ratio $\tau_D/T=\omega_r/(2\pi|\gamma|)$ of the slow surface sausage modes versus $k_zR$ for $l/R=0.1$ and different twist parameters $B_{\phi_i}/B_{zi}=(0,10^{-3},0.1,0.2,0.3)$. For comparison, the analytical results obtained by Eq. (\ref{gammacwd1}) are shown by the dashed-line curves. Auxiliary parameters are as in Fig. \ref{5bl0tn}. The results for $B_{\phi_i}/B_{zi}=0$ and $10^{-3}$ overlap with each other.}
    \label{5abcdl0.1t}
\end{figure}
%----------------------------------------------------------------------------------------------------
\begin{figure}
    \centering
    \begin{subfigure}[b]{0.5\textwidth}
        \centering
        \includegraphics[width=\textwidth]{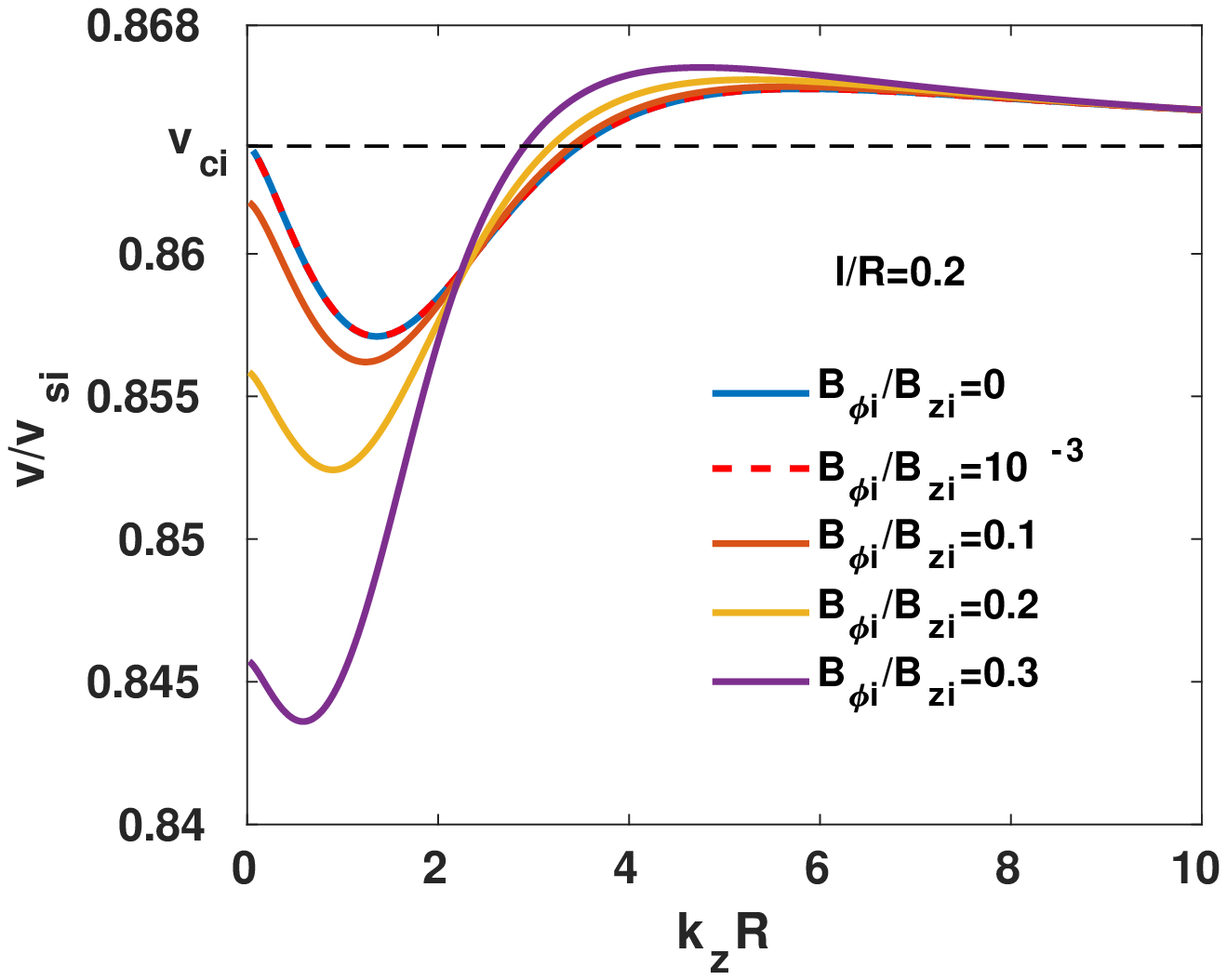}
      \caption{}
        \label{5al0.2t}
    \end{subfigure}
    \vfill
    \begin{subfigure}[b]{0.5\textwidth}
        \centering
        \includegraphics[width=\textwidth]{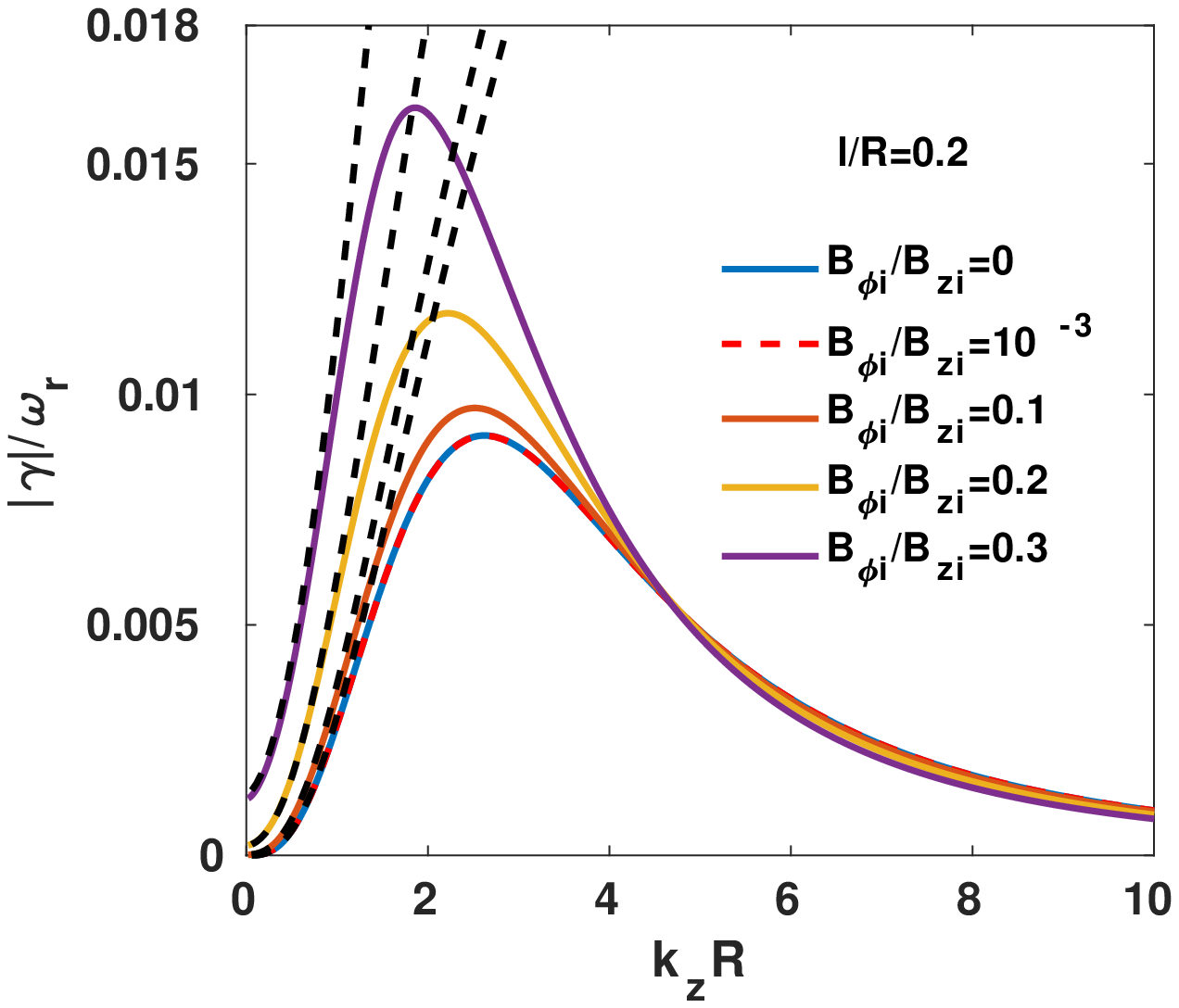}
       \caption{}
        \label{5bl0.2t}
    \end{subfigure}
    \vfill
    \begin{subfigure}[b]{0.5\textwidth}
        \centering
        \includegraphics[width=\textwidth]{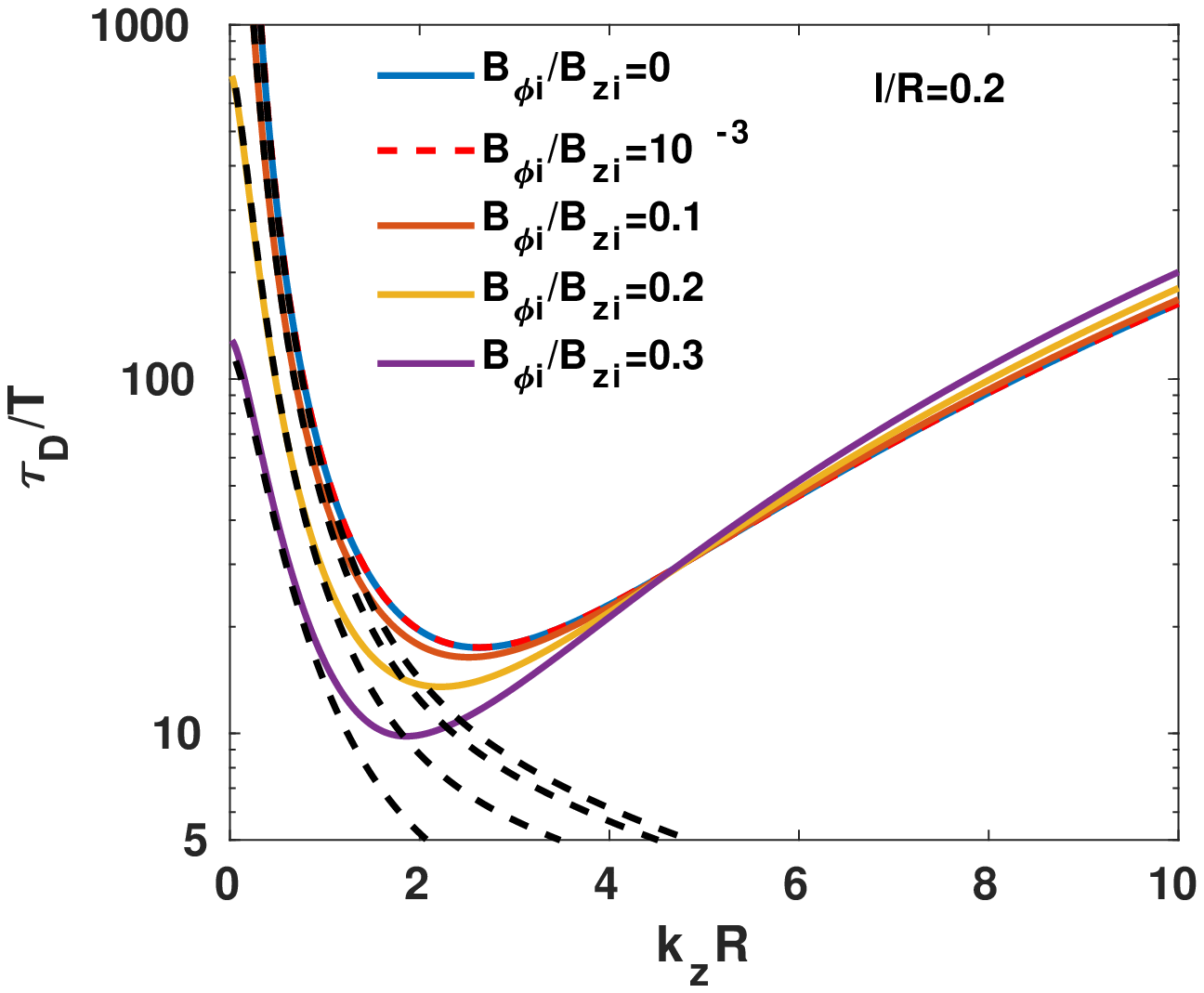}
        \caption{}
        \label{5cl0.2t}
    \end{subfigure}
    \vfill

    \caption{Same as Fig. \ref{5abcdl0.1t}, but for $l/R=0.2$.}
    \label{5abcdl0.2t}
\end{figure}
%----------------------------------------------------------------------------------------------------
\begin{figure}
    \centering
    \begin{subfigure}[b]{0.5\textwidth}
        \centering
        \includegraphics[width=\textwidth]{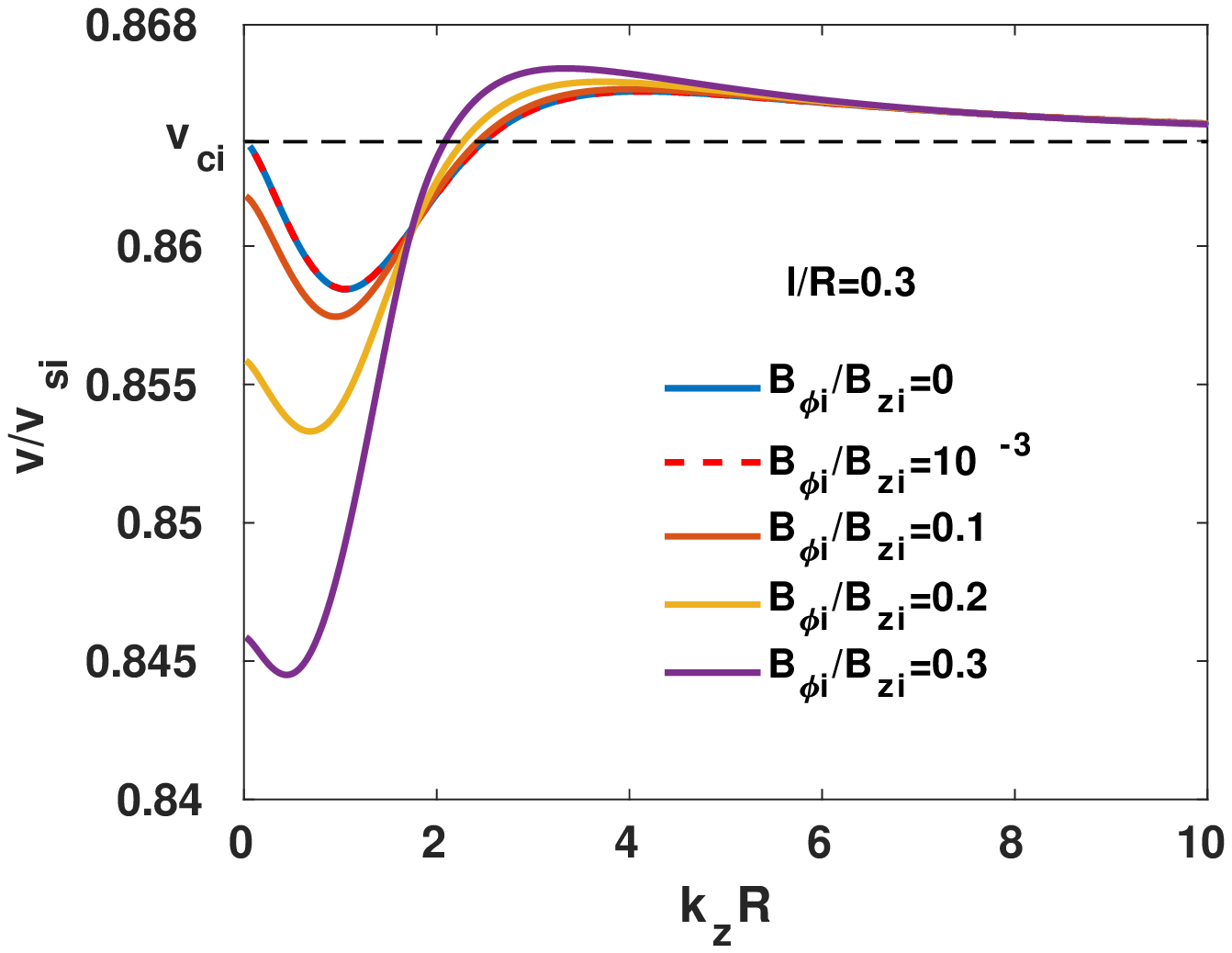}
      \caption{}
        \label{5al0.3t}
    \end{subfigure}
    \vfill
    \begin{subfigure}[b]{0.5\textwidth}
        \centering
        \includegraphics[width=\textwidth]{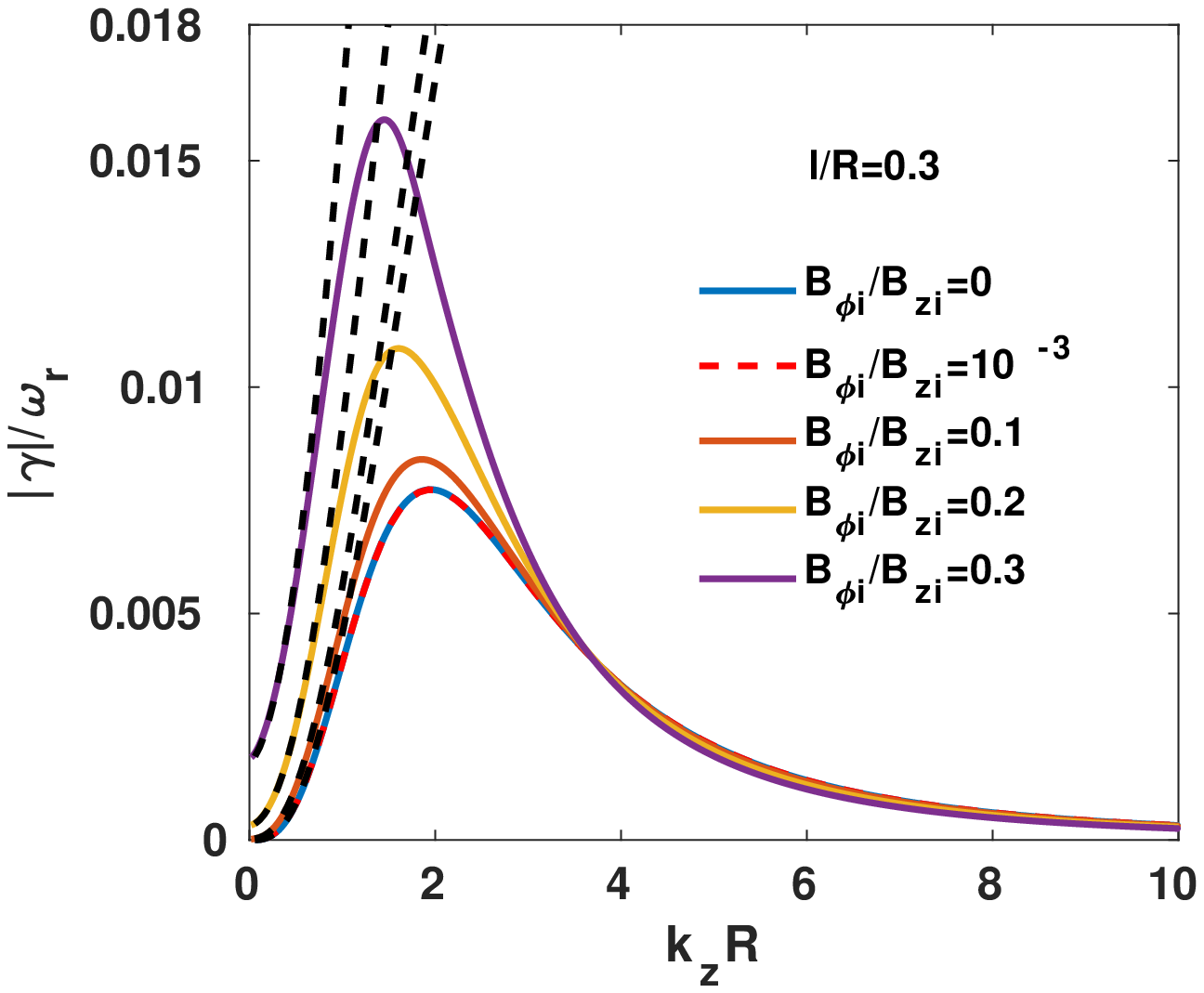}
       \caption{}
        \label{5bl0.3t}
    \end{subfigure}
    \vfill
    \begin{subfigure}[b]{0.5\textwidth}
        \centering
        \includegraphics[width=\textwidth]{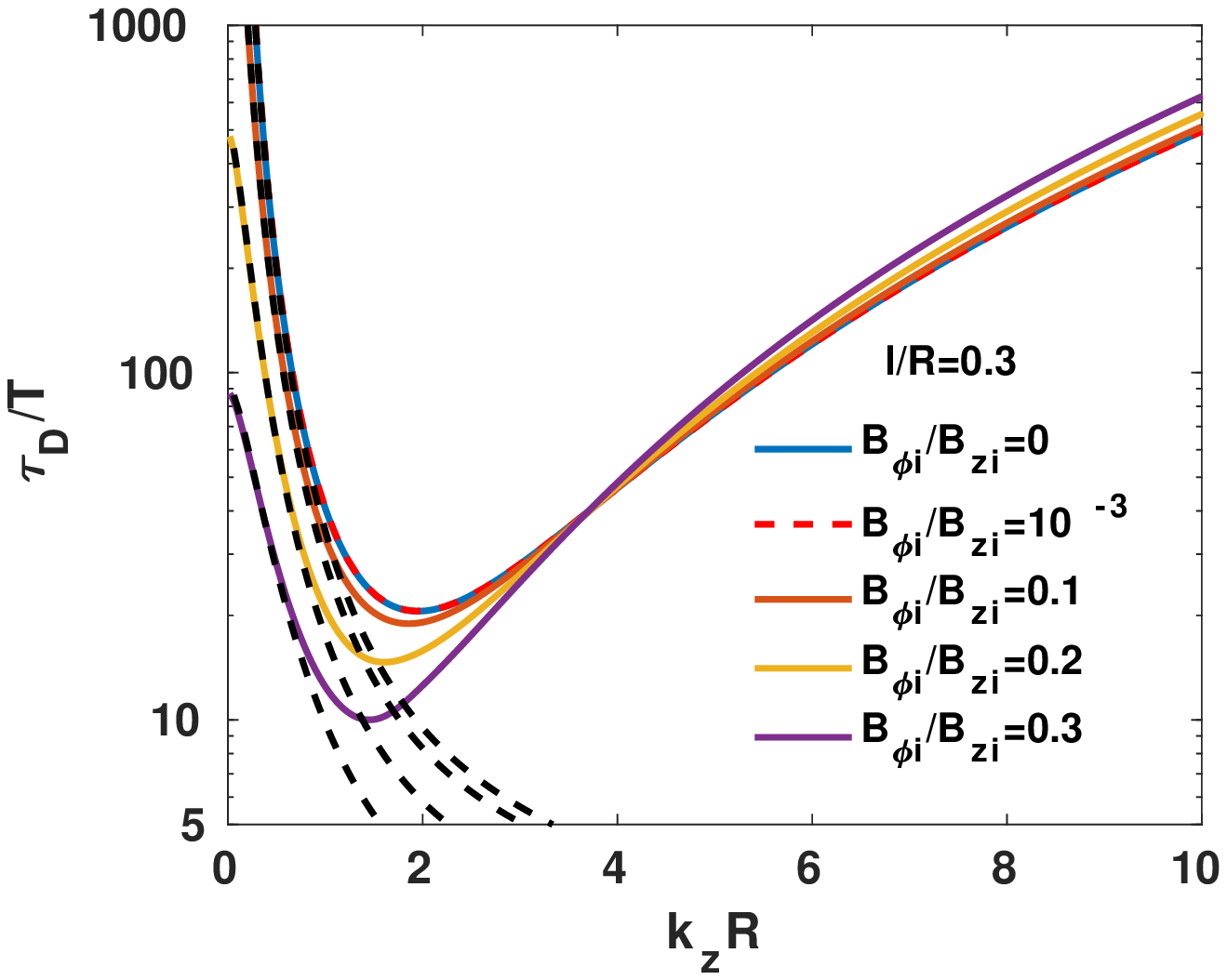}
        \caption{}
        \label{5cl0.3t}
    \end{subfigure}
    \vfill

    \caption{Same as Fig. \ref{5abcdl0.1t}, but for $l/R=0.3$.}
    \label{5abcdl0.3t}
\end{figure}
%----------------------------------------------------------------------------------------------------
\begin{figure}
    \centering
    \begin{subfigure}[b]{0.5\textwidth}
        \centering
        \includegraphics[width=\textwidth]{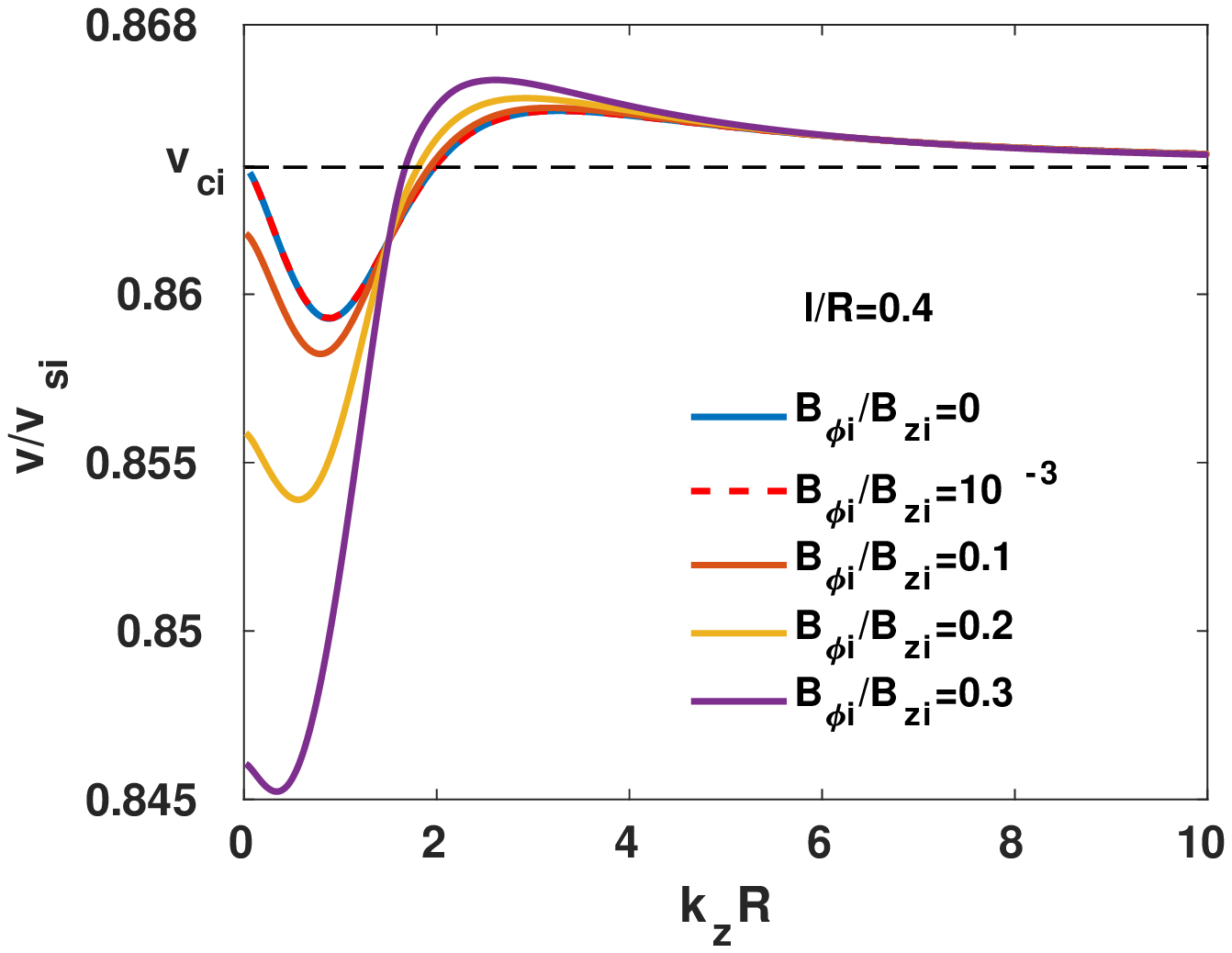}
      \caption{}
        \label{5al0.4t}
    \end{subfigure}
    \vfill
    \begin{subfigure}[b]{0.5\textwidth}
        \centering
        \includegraphics[width=\textwidth]{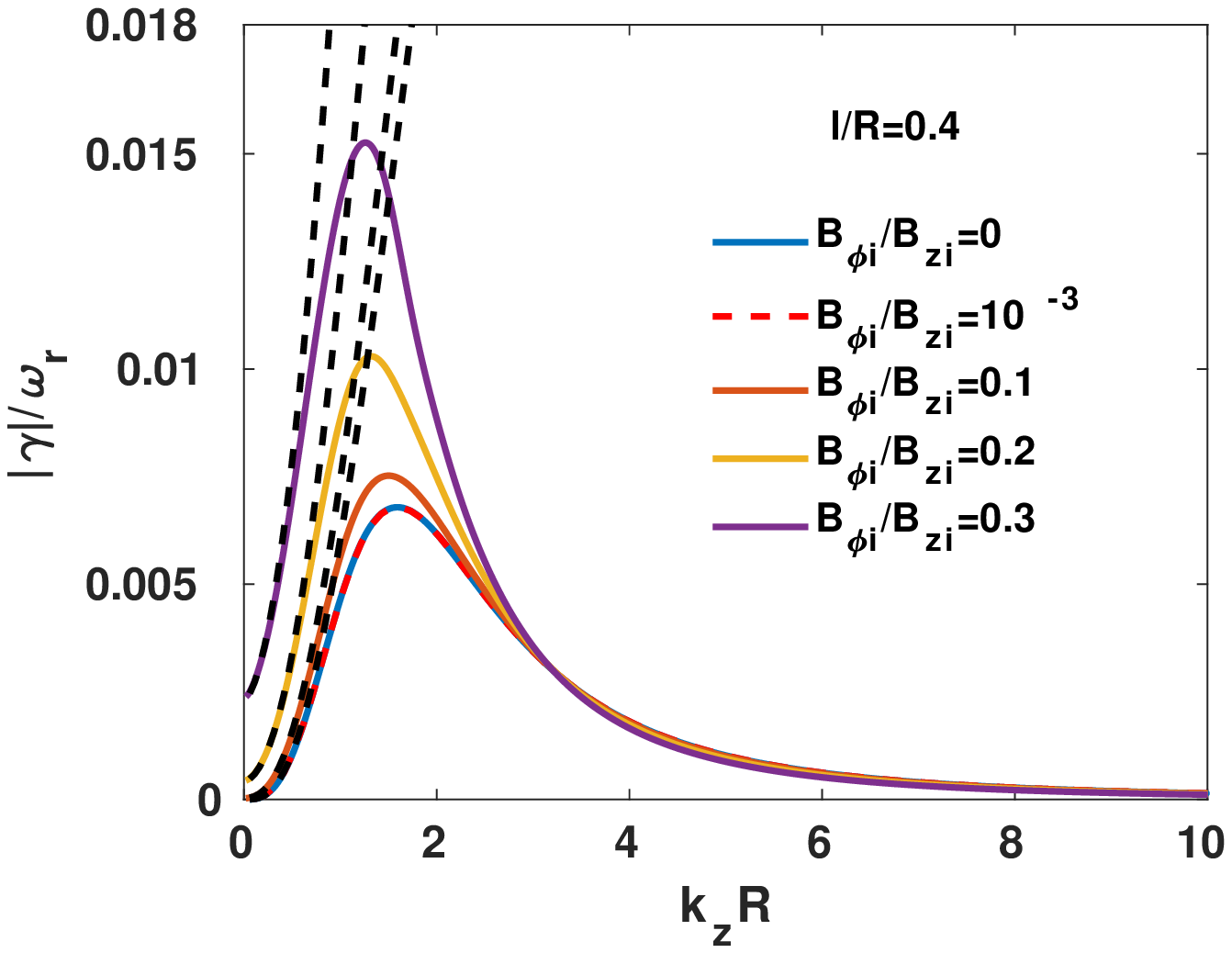}
       \caption{}
        \label{5bl0.4t}
    \end{subfigure}
    \vfill
    \begin{subfigure}[b]{0.5\textwidth}
        \centering
        \includegraphics[width=\textwidth]{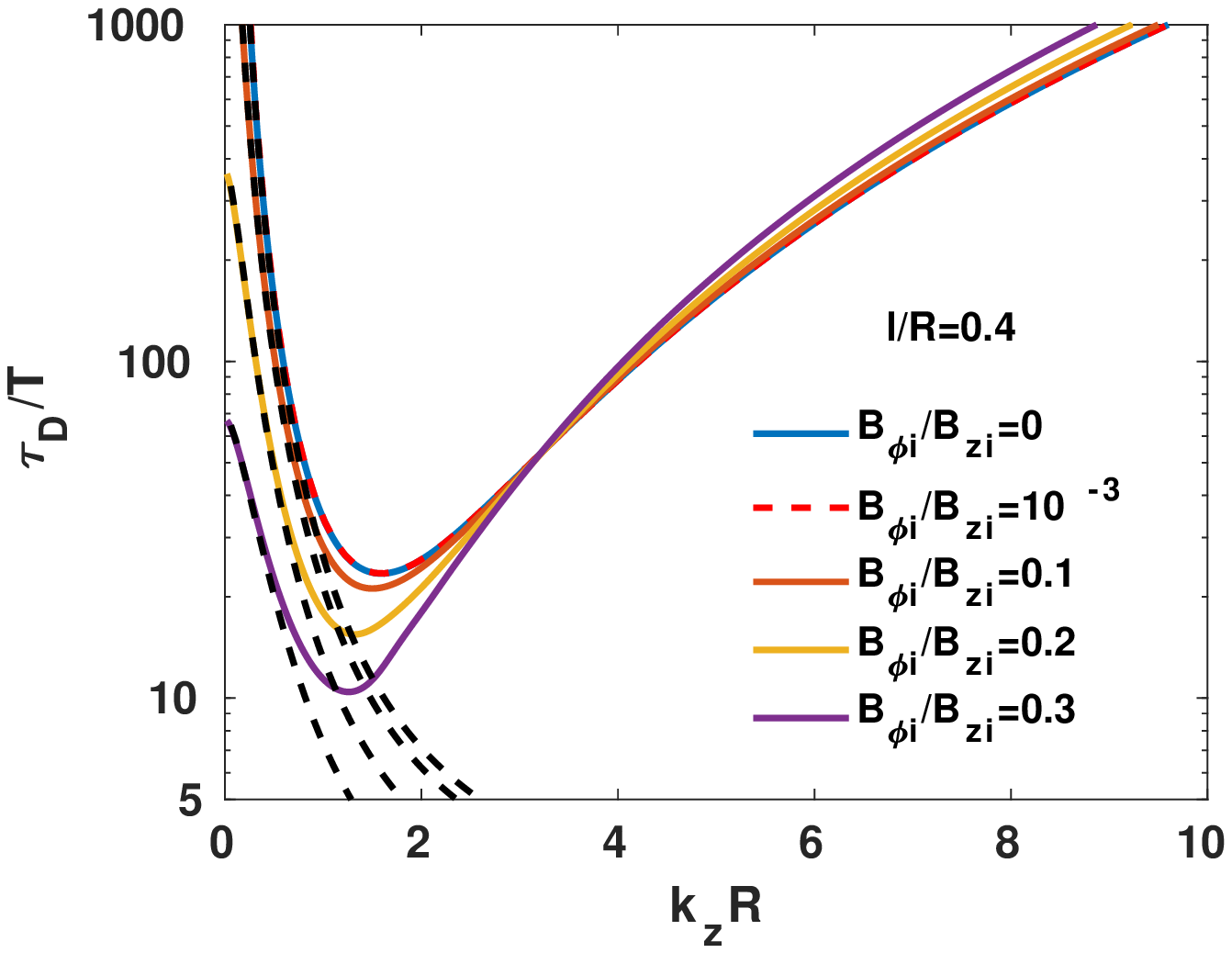}
        \caption{}
        \label{5cl0.4t}
    \end{subfigure}
    \vfill

    \caption{Same as Fig. \ref{5abcdl0.1t}, but for $l/R=0.4$.}
    \label{5abcdl0.4t}
\end{figure}
\begin{figure}
    \centering
    \begin{subfigure}[b]{0.5\textwidth}
        \centering
        \includegraphics[width=\textwidth]{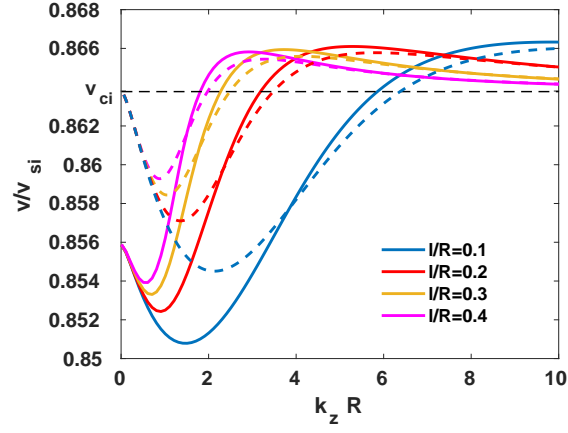}
      \caption{}
        \label{5aphi0.2t}
    \end{subfigure}
    \vfill
    \begin{subfigure}[b]{0.5\textwidth}
        \centering
        \includegraphics[width=\textwidth]{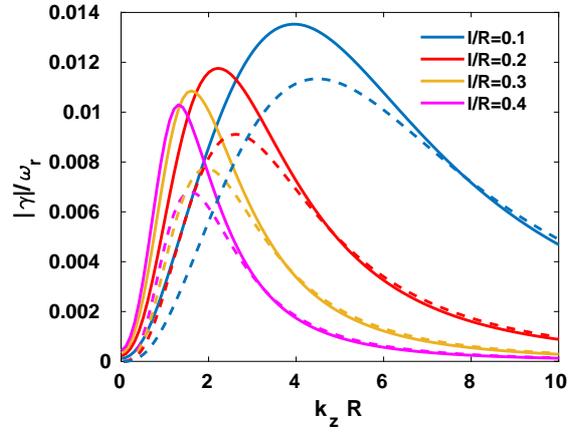}
       \caption{}
        \label{5bphi0.2t}
    \end{subfigure}
    \vfill
    \begin{subfigure}[b]{0.5\textwidth}
        \centering
        \includegraphics[width=\textwidth]{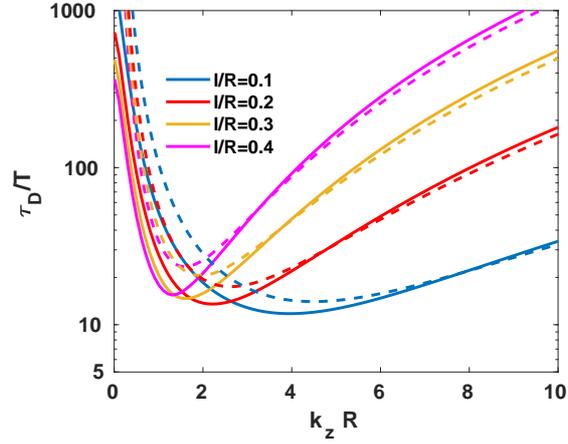}
        \caption{}
        \label{5cphi0.20t}
    \end{subfigure}
    \caption{(a) The phase speed $v/v_{si}\equiv\omega_r/\omega_{si}$, (b) the damping rate to frequency ratio $\left|\gamma\right|/\omega_r$, and (c) the damping time to period ratio $\tau_D/T=\omega_r/(2\pi|\gamma|)$ of the slow surface sausage modes versus $k_zR$ for different thickness of the inhomogeneous layer $l/R=(0.1,0.2,0.3,0.4)$. Here, the dashed and solid line curves, respectively, are related to $B_{\phi_i}/B_{zi}=0$ and $B_{\phi_i}/B_{zi}=0.2$.  Auxiliary parameters are as in
Fig. \ref{5bl0tn}.}
    \label{5abcphi0.2t}
\end{figure}
%-----------------------------------------------------------------------------------------------------------------------------------------
\begin{figure}
    \centering
    \begin{subfigure}[b]{0.5\textwidth}
        \centering
        \includegraphics[width=\textwidth]{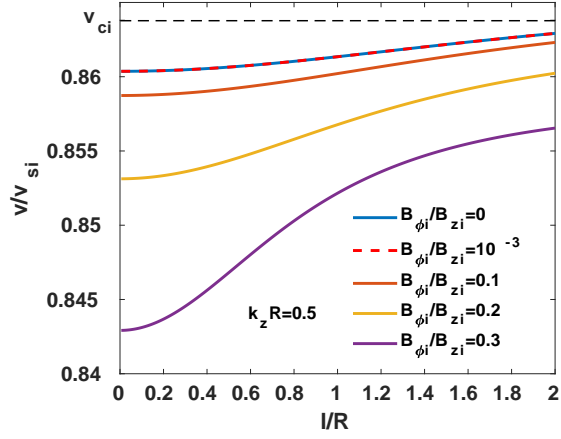}
      \caption{}
        \label{5akz0.5t}
    \end{subfigure}
    \vfill
    \begin{subfigure}[b]{0.5\textwidth}
        \centering
        \includegraphics[width=\textwidth]{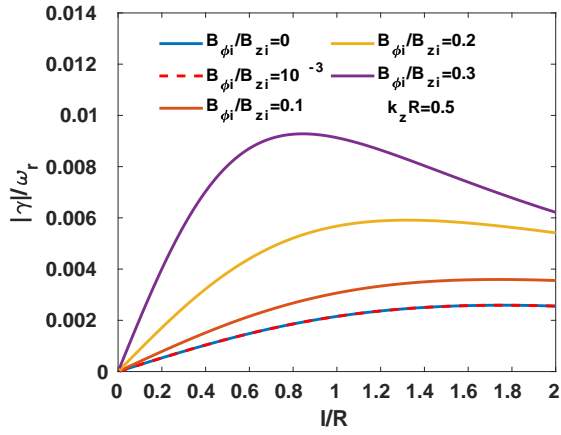}
       \caption{}
        \label{5bkz0.5t}
    \end{subfigure}
    \vfill
    \begin{subfigure}[b]{0.5\textwidth}
        \centering
        \includegraphics[width=\textwidth]{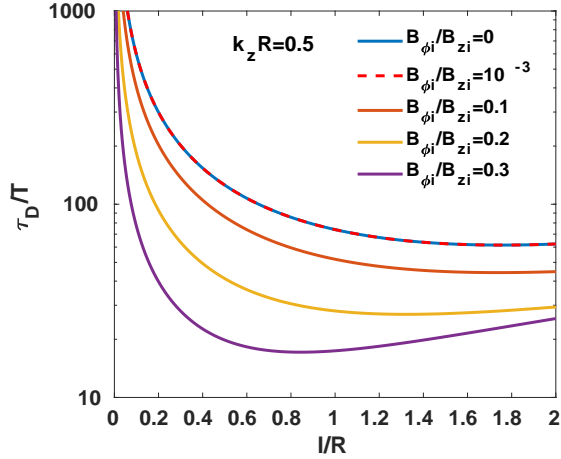}
        \caption{}
        \label{5ckz0.5t}
    \end{subfigure}
    \caption{(a) The phase speed $v/v_{si}\equiv\omega_r/\omega_{si}$, (b) the damping rate to frequency ratio $\left|\gamma\right|/\omega_r$, and (c) the damping time to period ratio $\tau_D/T=\omega_r/(2\pi|\gamma|)$ of the slow surface sausage modes versus $l/R$ for $k_zR=0.5$ and different twist parameters $B_{\phi_i}/B_{zi}=(0,10^{-3},0.1,0.2,0.3)$. Auxiliary parameters are as in
Fig. \ref{5bl0tn}. The results for $B_{\phi_i}/B_{zi}=0$ and $10^{-3}$ overlap with each other.}
    \label{5abckz0.5t}
\end{figure}
%----------------------------------------------------------------------------------------------------
\begin{figure}
    \centering
    \begin{subfigure}[b]{0.5\textwidth}
        \centering
        \includegraphics[width=\textwidth]{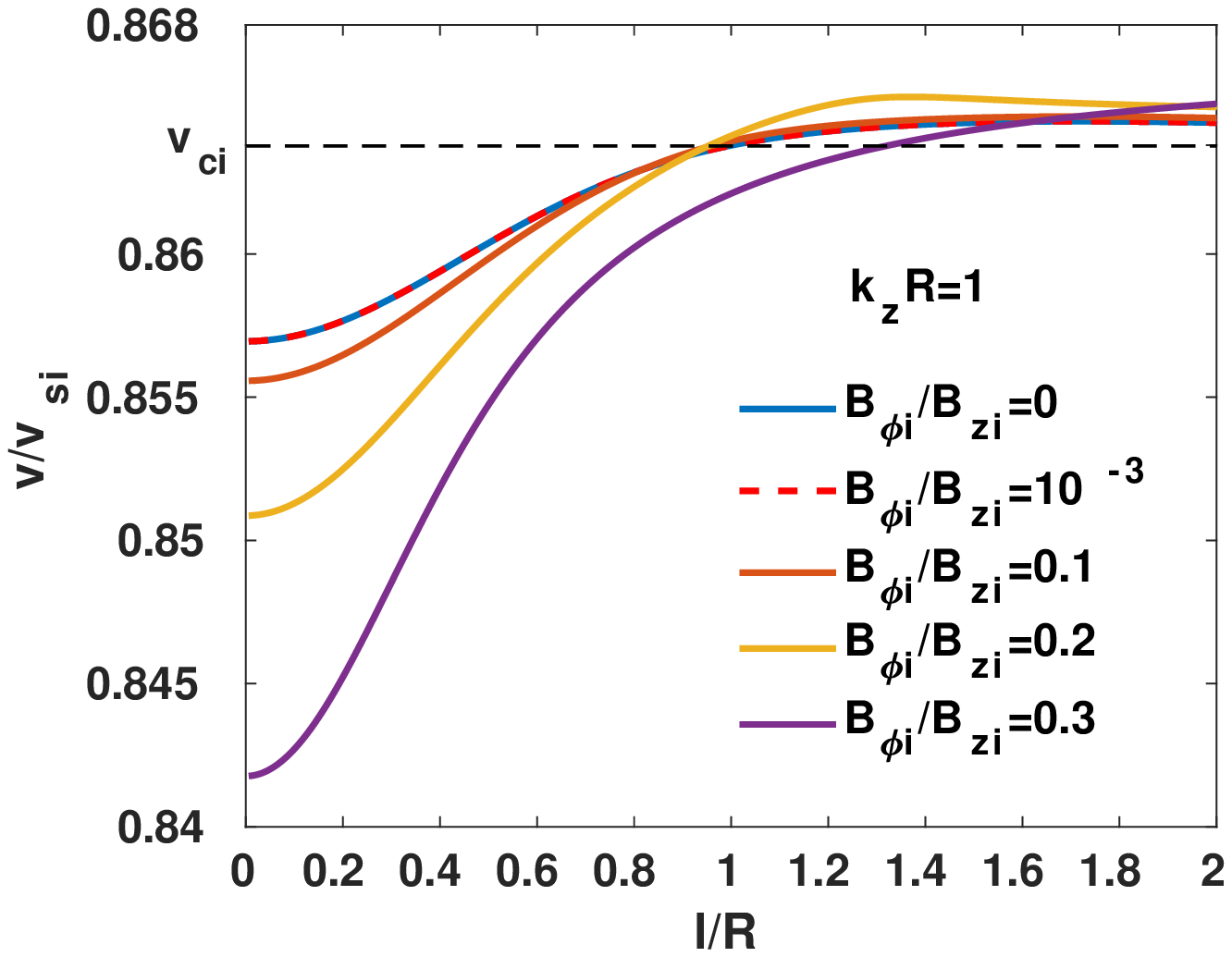}
      \caption{}
        \label{5akz1t}
    \end{subfigure}
    \vfill
    \begin{subfigure}[b]{0.5\textwidth}
        \centering
        \includegraphics[width=\textwidth]{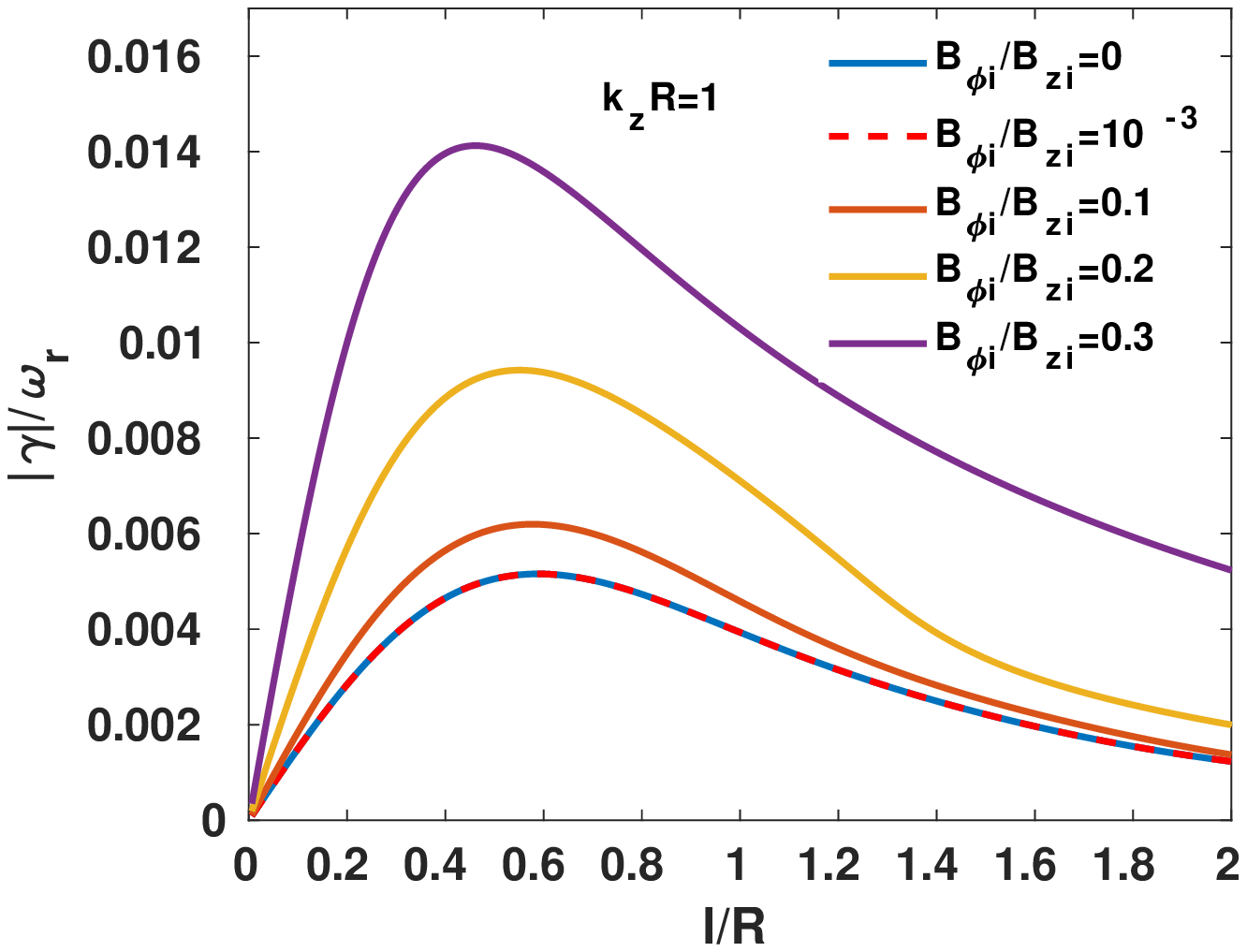}
       \caption{}
        \label{5bkz1t}
    \end{subfigure}
    \vfill
    \begin{subfigure}[b]{0.5\textwidth}
        \centering
        \includegraphics[width=\textwidth]{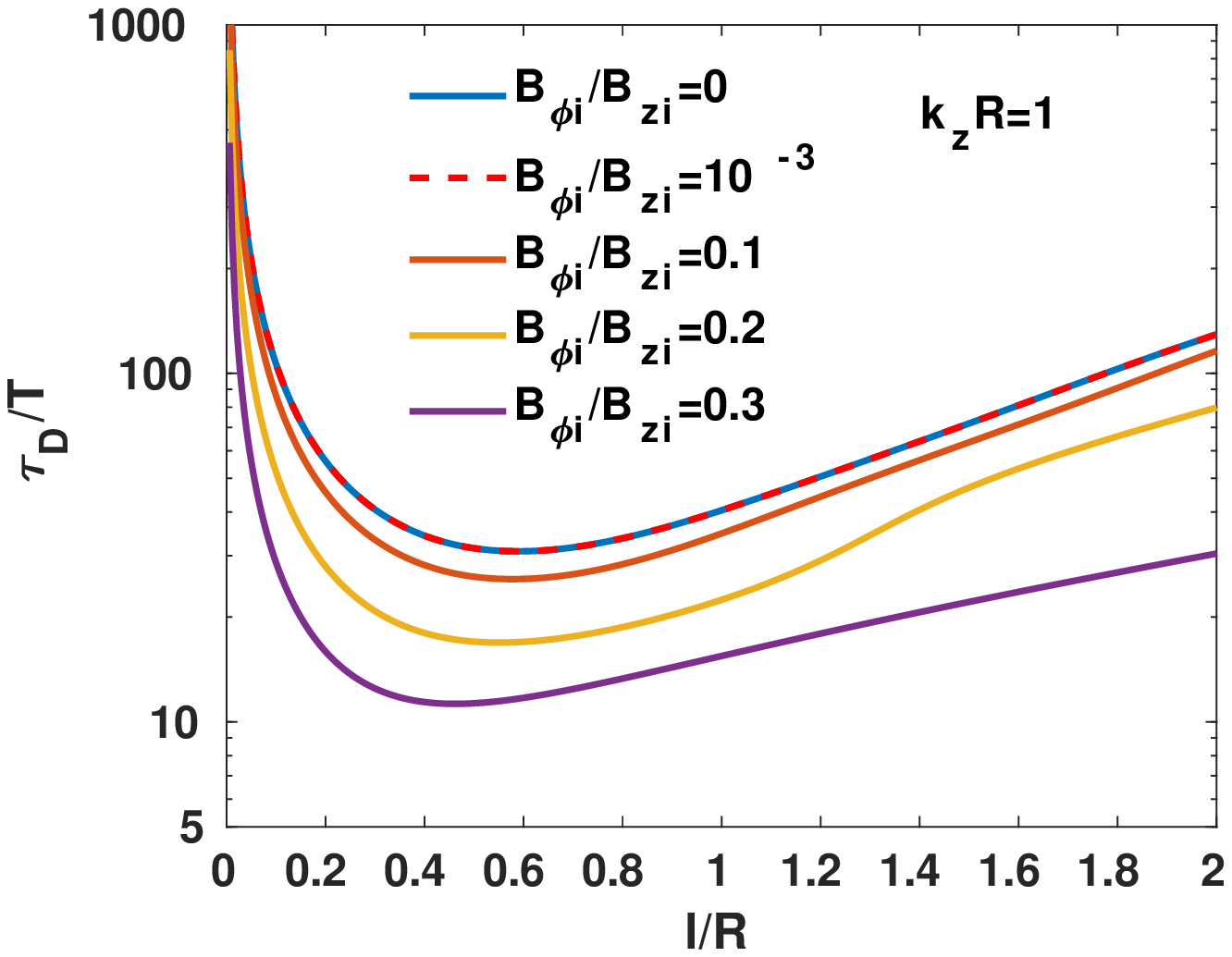}
        \caption{}
        \label{5ckz1t}
    \end{subfigure}
    \caption{Same as Fig. \ref{5abckz0.5t}, but for $k_zR=1$.}
    \label{5abckz1t}
\end{figure}
%----------------------------------------------------------------------------------------------------
\begin{figure}
    \centering
    \begin{subfigure}[b]{0.5\textwidth}
        \centering
        \includegraphics[width=\textwidth]{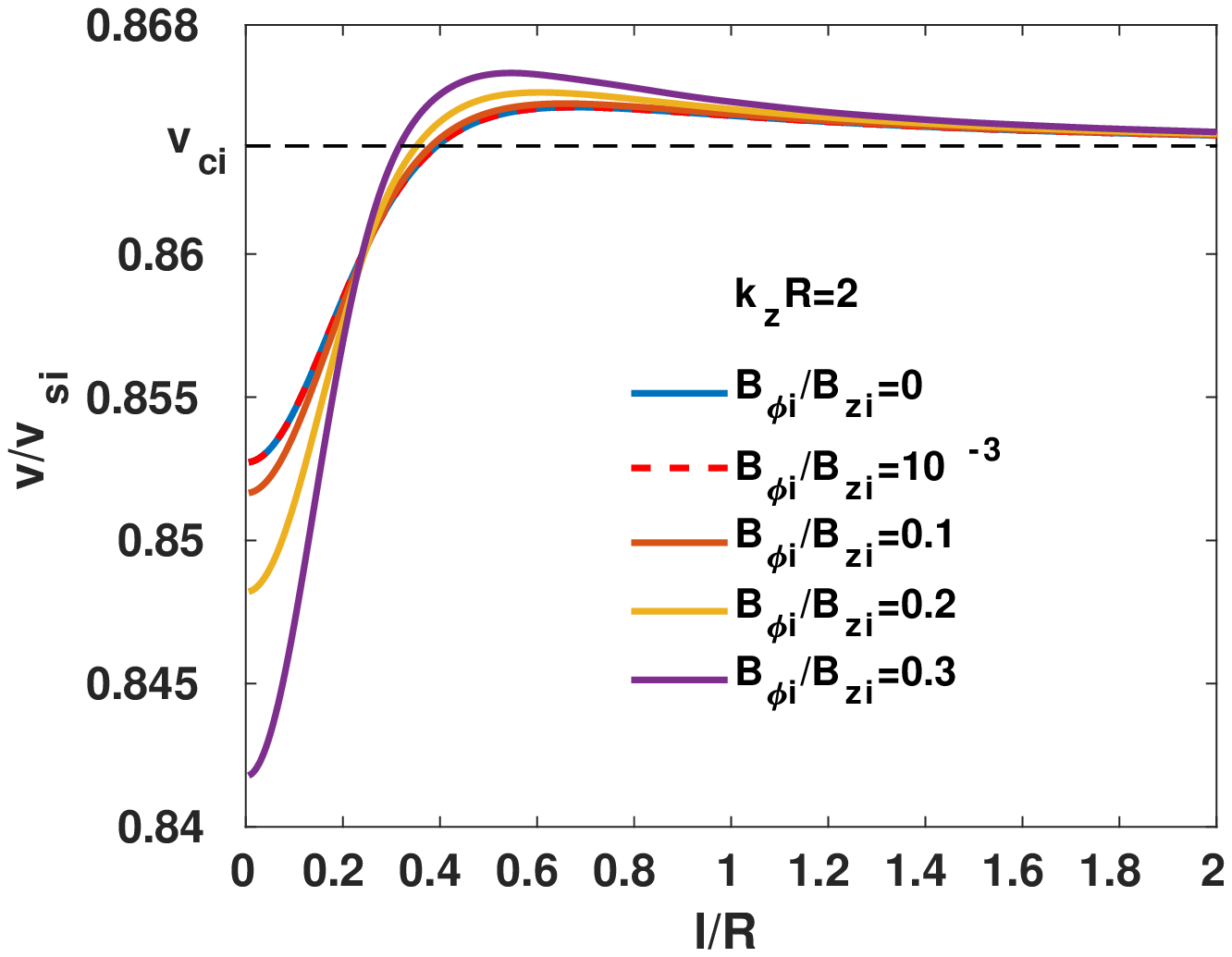}
      \caption{}
        \label{5akz2t}
    \end{subfigure}
    \vfill
    \begin{subfigure}[b]{0.5\textwidth}
        \centering
        \includegraphics[width=\textwidth]{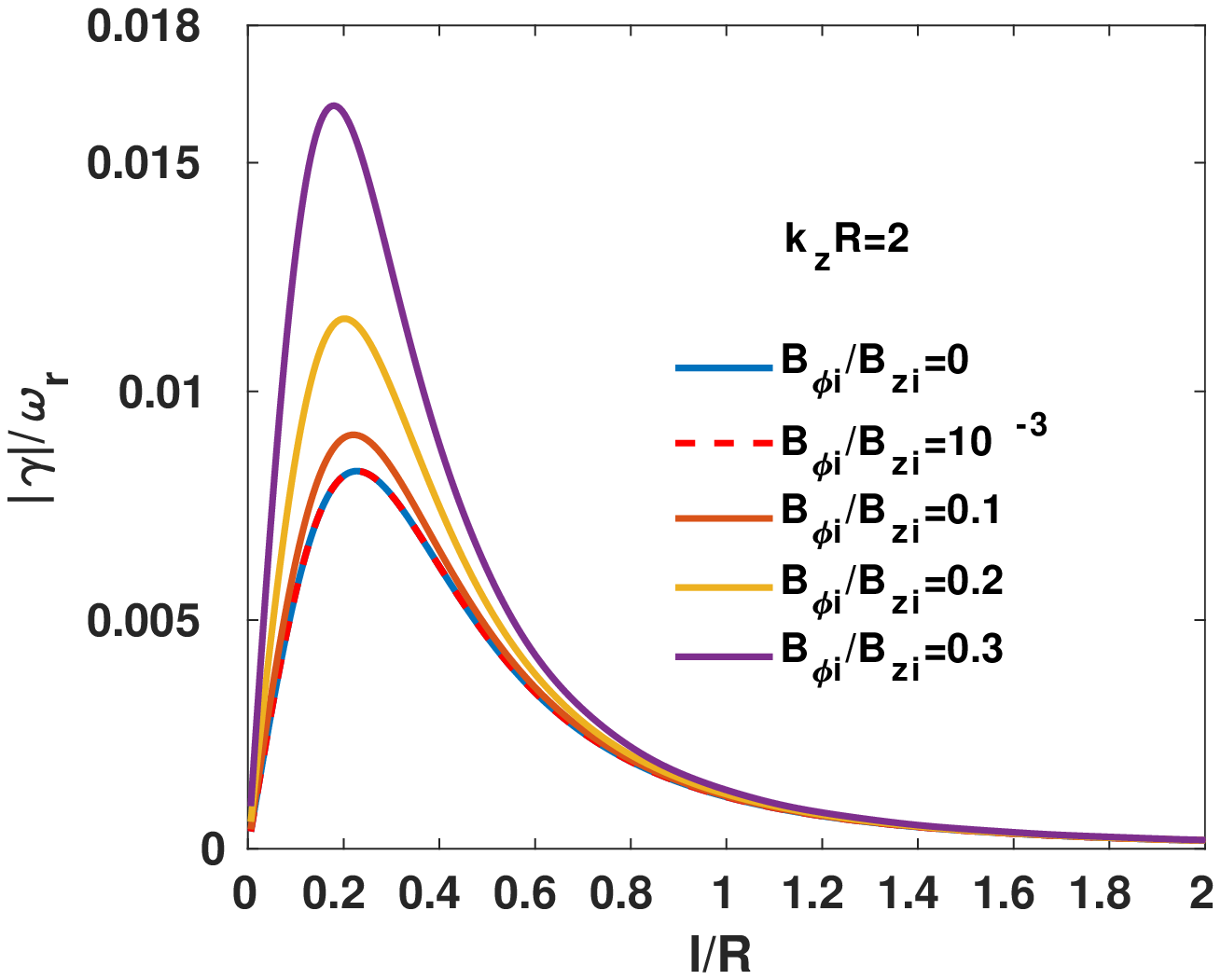}
       \caption{}
        \label{5bkz2t}
    \end{subfigure}
    \vfill
    \begin{subfigure}[b]{0.5\textwidth}
        \centering
        \includegraphics[width=\textwidth]{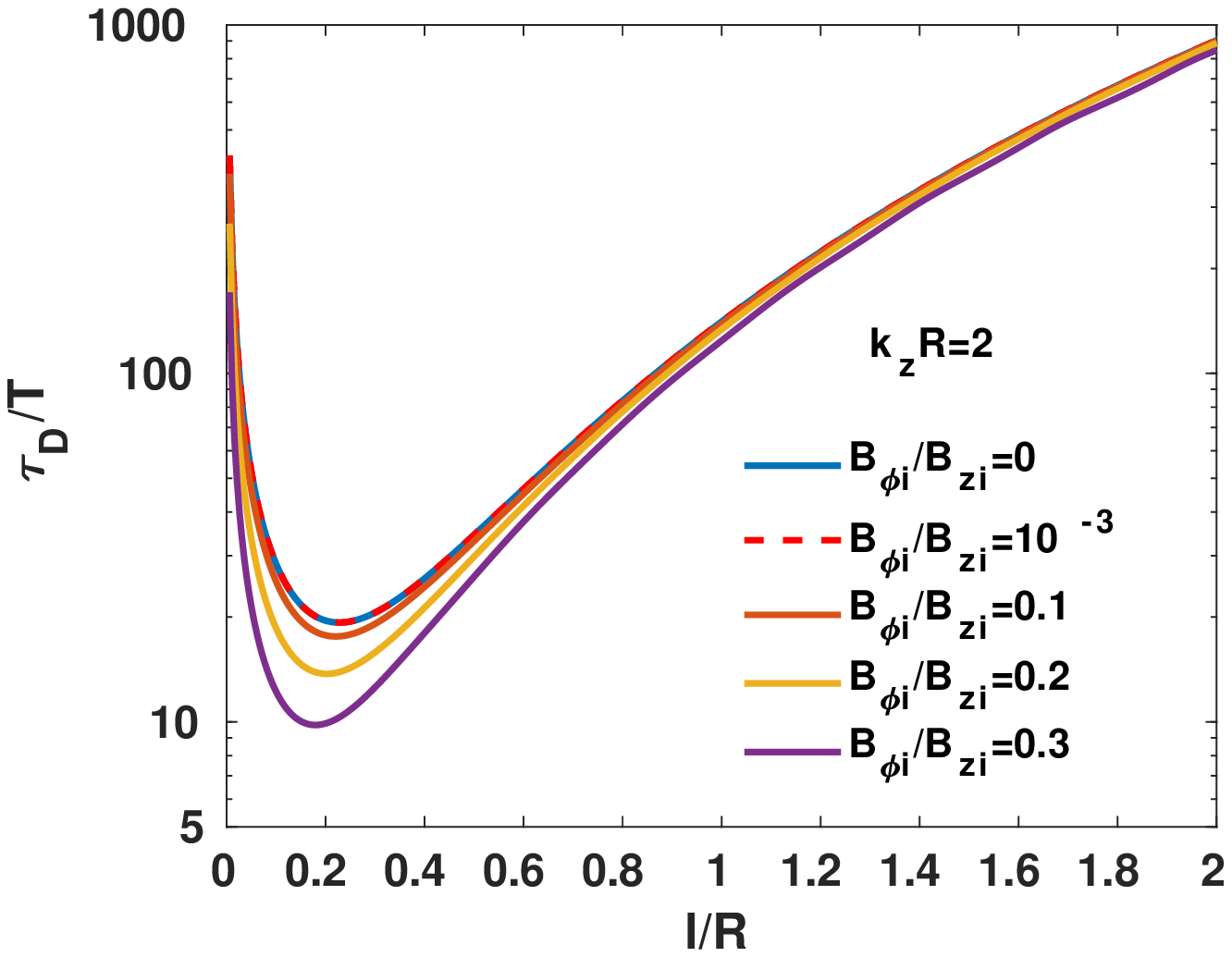}
        \caption{}
        \label{5ckz2t}
    \end{subfigure}
    \caption{Same as Fig. \ref{5abckz0.5t}, but for $k_zR=2$.}
    \label{5abckz2t}
\end{figure}
%----------------------------------------------------------------------------------------------------
\begin{figure}
    \centering
    \begin{subfigure}[b]{0.5\textwidth}
        \centering
        \includegraphics[width=\textwidth]{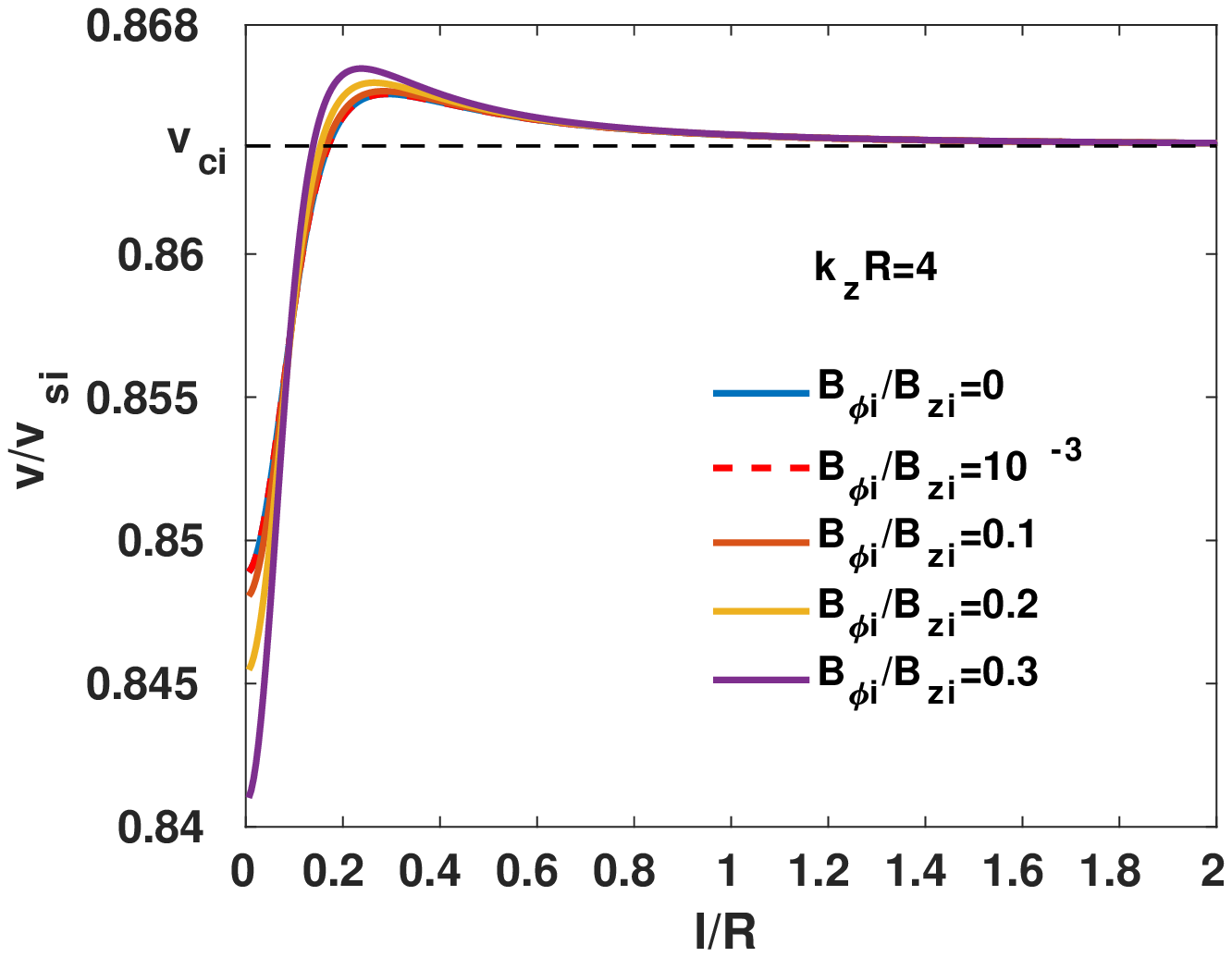}
      \caption{}
        \label{5akz4t}
    \end{subfigure}
    \vfill
    \begin{subfigure}[b]{0.5\textwidth}
        \centering
        \includegraphics[width=\textwidth]{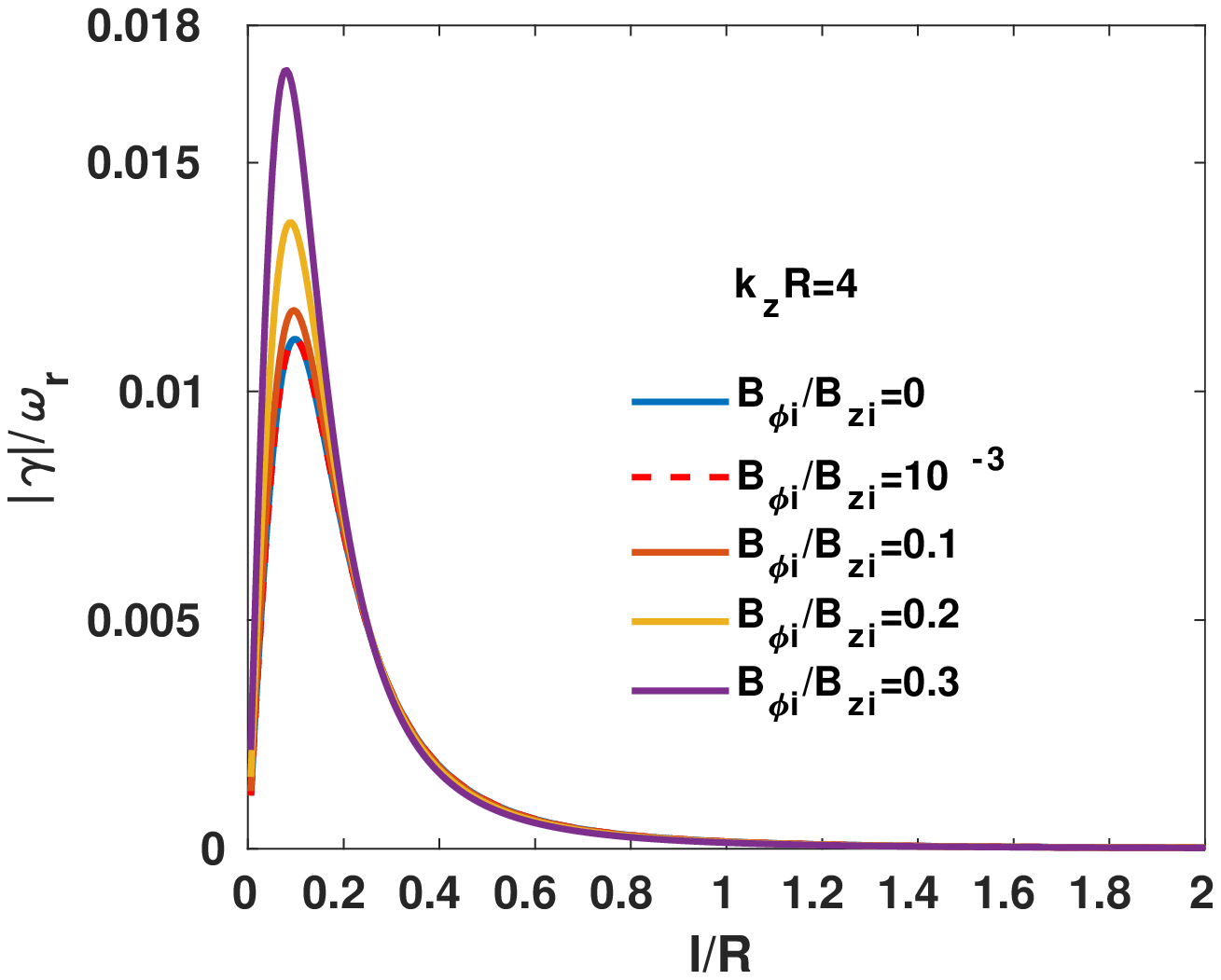}
       \caption{}
        \label{5bkz4t}
    \end{subfigure}
    \vfill
    \begin{subfigure}[b]{0.5\textwidth}
        \centering
        \includegraphics[width=\textwidth]{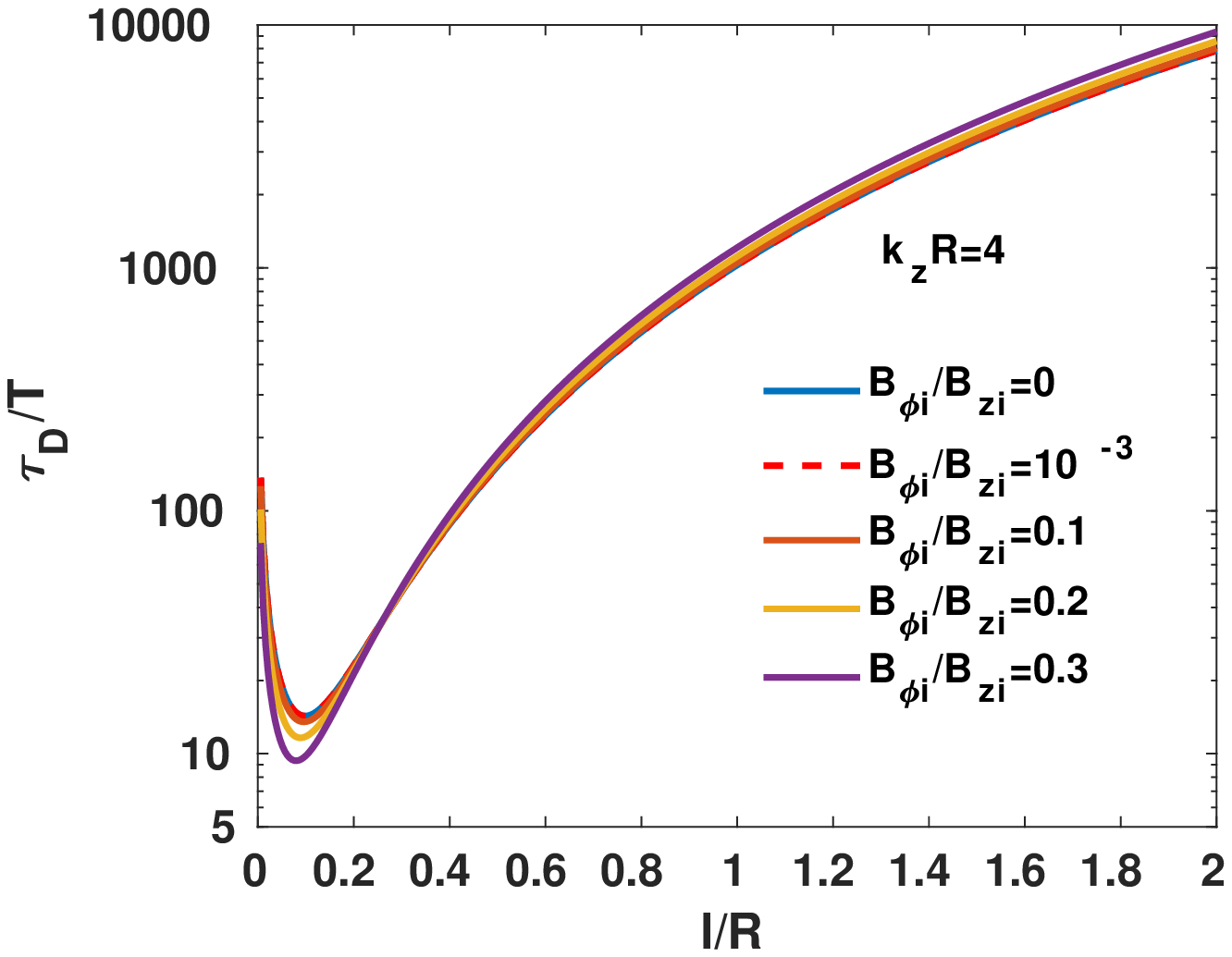}
        \caption{}
        \label{5ckz4t}
    \end{subfigure}
    \caption{Same as Fig. \ref{5abckz0.5t}, but for $k_zR=4$.}
    \label{5abckz4t}
\end{figure}
%----------------------------------------------------------------------------------------------------
\begin{figure}[h]
    \centering
    \begin{subfigure}[b]{0.5\textwidth}
        \centering
        \includegraphics[width=\textwidth]{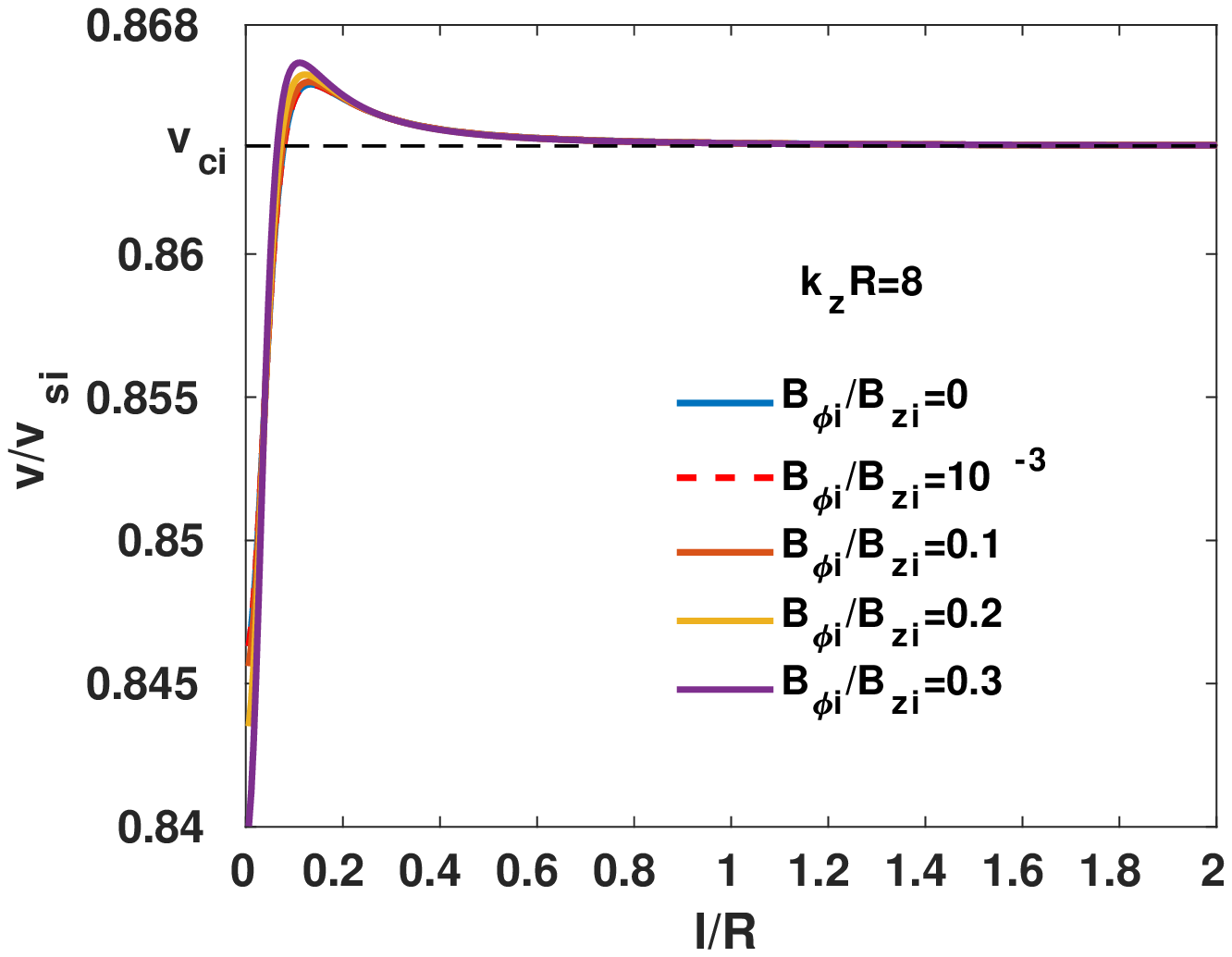}
      \caption{}
        \label{5akz8t}
    \end{subfigure}
    \vfill
    \begin{subfigure}[b]{0.5\textwidth}
        \centering
        \includegraphics[width=\textwidth]{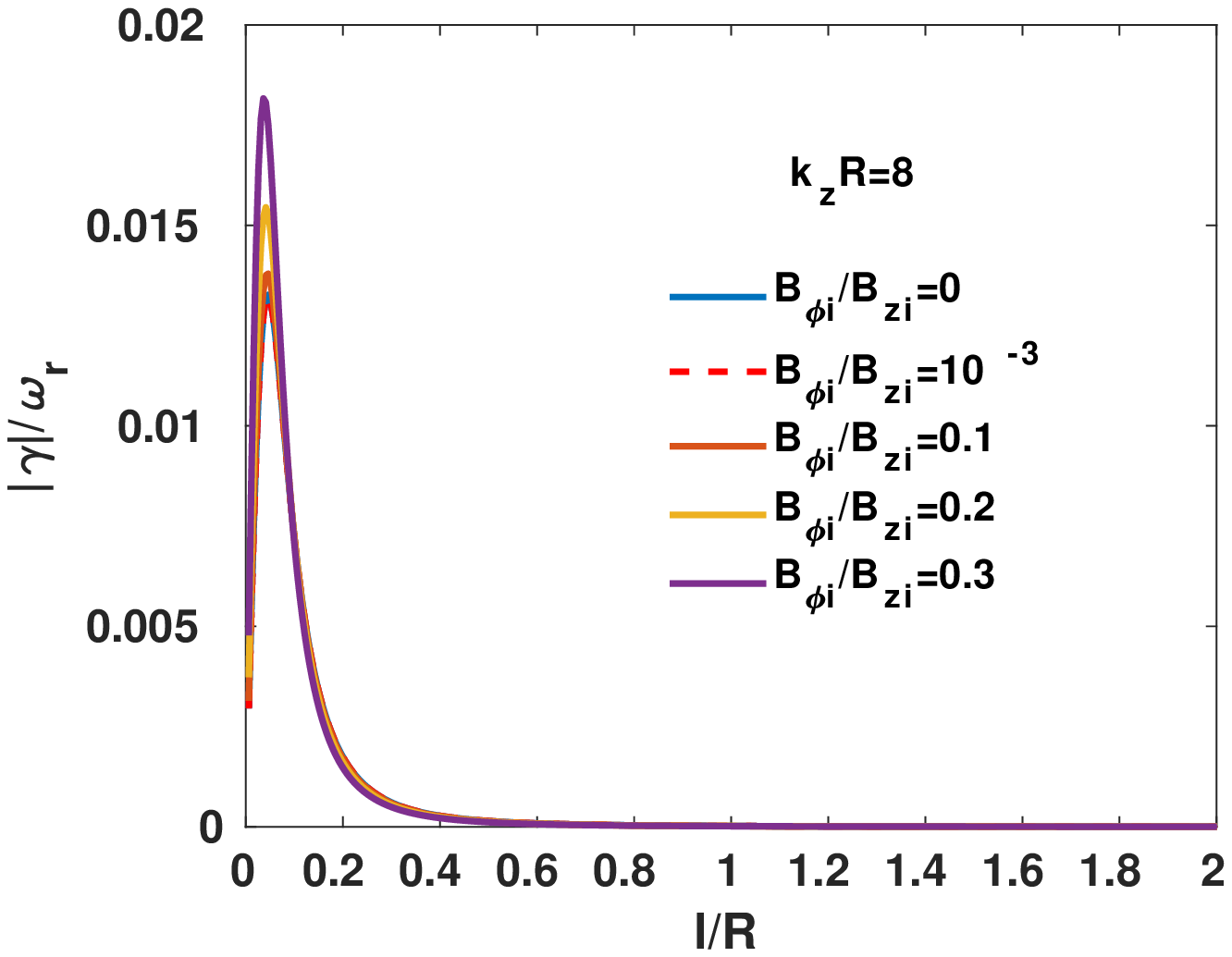}
       \caption{}
        \label{5bkz8t}
    \end{subfigure}
    \vfill
    \begin{subfigure}[b]{0.5\textwidth}
        \centering
        \includegraphics[width=\textwidth]{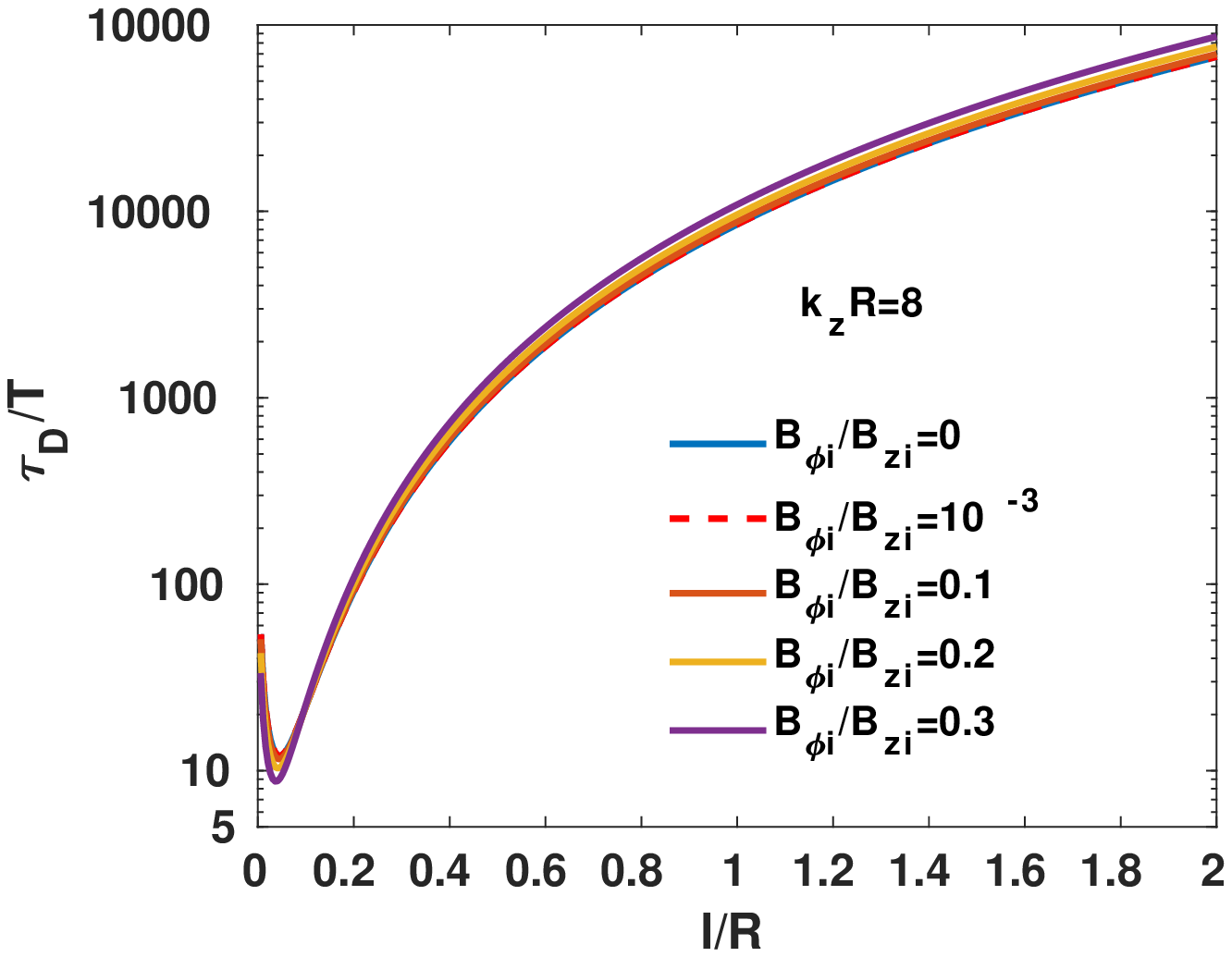}
        \caption{}
        \label{5ckz8t}
    \end{subfigure}
    \caption{Same as Fig. \ref{5abckz0.5t}, but for $k_zR=8$.}
    \label{5abckz8t}
\end{figure}
%----------------------------------------------------------------------------------------------------
%\begin{figure}
%    \centering
%    \begin{subfigure}[b]{0.5\textwidth}
%        \centering
%        \includegraphics[width=\textwidth]{5aphil0.2.eps}
%      \caption{}
%        \label{5aphi0.2t}
%    \end{subfigure}
%    \vfill
%    \begin{subfigure}[b]{0.5\textwidth}
%        \centering
%        \includegraphics[width=\textwidth]{5bphil0.2.eps}
%       \caption{}
%        \label{5bphi0.2t}
%    \end{subfigure}
%    \vfill
%    \begin{subfigure}[b]{0.5\textwidth}
%        \centering
%        \includegraphics[width=\textwidth]{5cphil0.2.eps}
%        \caption{}
%        \label{5cphi0.2t}
%    \end{subfigure}
%    \caption{(a) The phase speed $v/v_{si}\equiv\omega_r/\omega_{si}$, (b) the damping rate to frequency ratio $\left|\gamma\right|/\omega_r$, and (c) the damping time to period ratio $\tau_D/T=\omega_r/(2\pi|\gamma|)$ of the slow surface sausage modes versus $l/R$ for $B_{\phi_i}/B_{zi}=0.2$ and different $k_zR=(0.5,1,2,4,8)$. Auxiliary parameters are as in
%Fig. \ref{5bl0tn}.}
%    \label{5abcphil0.2t}
%\end{figure}
%----------------------------------------------------------------------------------------------------

%----------------------------------------------------------------------------------------------------
\begin{figure}
    \centering
    \begin{subfigure}[b]{0.5\textwidth}
        \centering
        \includegraphics[width=\textwidth]{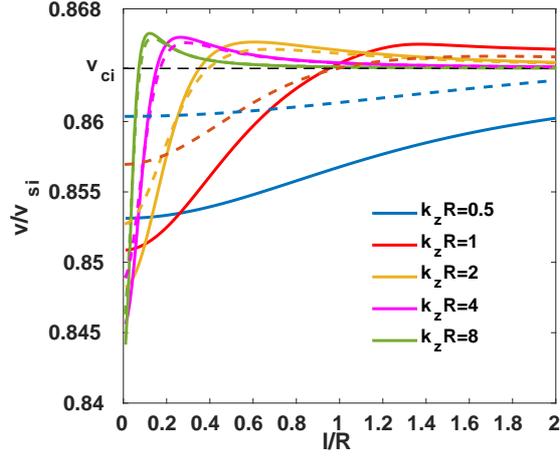}
      \caption{}
        \label{5aphi0.2t}
    \end{subfigure}
    \vfill
    \begin{subfigure}[b]{0.5\textwidth}
        \centering
        \includegraphics[width=\textwidth]{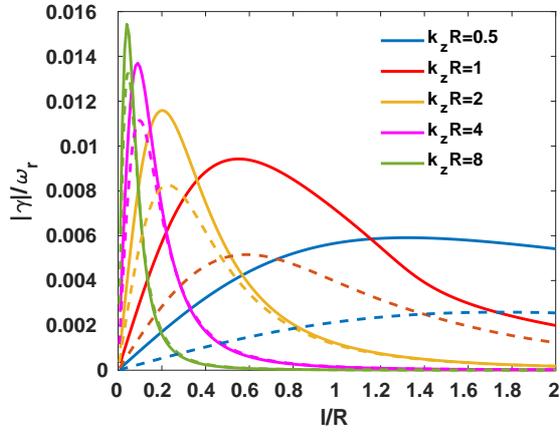}
       \caption{}
        \label{5bphi0.2t}
    \end{subfigure}
    \vfill
    \begin{subfigure}[b]{0.5\textwidth}
        \centering
        \includegraphics[width=\textwidth]{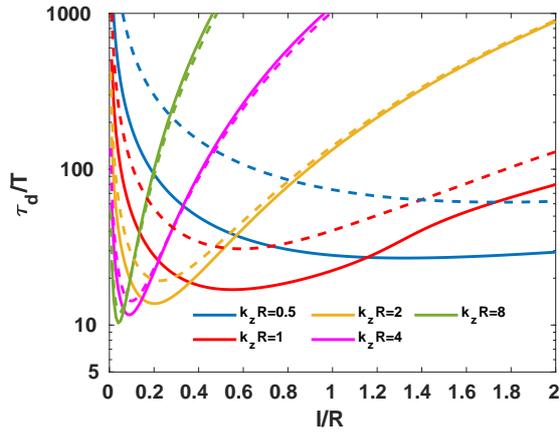}
        \caption{}
        \label{5cphi0.2t}
    \end{subfigure}
    \caption{(a) The phase speed $v/v_{si}\equiv\omega_r/\omega_{si}$, (b) the damping rate to frequency ratio $\left|\gamma\right|/\omega_r$, and (c) the damping time to period ratio $\tau_D/T=\omega_r/(2\pi|\gamma|)$ of the slow surface sausage modes versus $l/R$ for different $k_zR=(0.5,1,2,4,8)$. Here, the dashed and solid line curves, respectively, are related to $B_{\phi_i}/B_{zi}=0$ and $B_{\phi_i}/B_{zi}=0.2$. Auxiliary parameters are as in
Fig. \ref{5bl0tn}.}
    \label{5abcphil0.20t}
\end{figure}
%----------------------------------------------------------------------------------------------------

\end{document}